\documentclass[12pt]{article}
\title{Perturbative quantum E7(7) symmetry in N=8 supergravity}
\usepackage{mathrsfs}
\usepackage{amsmath}

\global\arraycolsep=1pt
\usepackage[colorlinks=true,backref=true,linkcolor=black,anchorcolor=black,citecolor=black,filecolor   =black,menucolor=black,pagecolor=black,urlcolor=black]{hyperref}
\usepackage[Symbol]{upgreek}
\usepackage{amsthm}
\usepackage{amssymb}
\usepackage[Symbol]{upgreek}
\usepackage{dsfont}
\usepackage{textcomp}

\usepackage{latexsym}
\usepackage{graphicx}

\usepackage{pstricks}
\usepackage{pst-node}
\usepackage{calc}
\usepackage{ifthen}

\def\bea{\begin{eqnarray}}
\def\eea{\end{eqnarray}}
\def\be{\begin{equation}}
\def\ee{\end{equation}}

\renewcommand{\Im}{{\rm Im}}
\renewcommand{\Re}{{\rm Re}}

\newcommand{\pa}{\partial}

\newcommand{\nn}{\nonumber}

\newcommand{\IN}{\mathbb{N}}
\newcommand{\IR}{\mathbb{R}}
\newcommand{\IC}{\mathbb{C}}

\newcommand{\IZ}{\mathbb{Z}}

\newcommand{\cA}{\mathcal{A}}

\newcommand{\cX}{\mathcal{X}}
\newcommand{\cR}{\mathcal{R}}

\usepackage{mathrsfs}
\usepackage[Symbol]{upgreek}
\usepackage{bbm}
\usepackage{dsfont}
\usepackage{amssymb}
\usepackage{textcomp}
\usepackage{wasysym}
\usepackage{caption}
\usepackage{graphicx}

\def\un{{\mathpzc{1}}}
\def\deux{{\mathpzc{2}}}
\def\trois{{\mathpzc{3}}}
\def\quatre{{\mathpzc{4}}}

\newcommand{\Scal}[1]{\Bigl ({#1} \Bigr )}
\newcommand{\scal}[1]{\bigl ({#1} \bigr )}
\newcommand{\CR}{\nonumber \\*}

\newcommand{\trace}{\hbox {Tr}~}

\def\L{{\mathcal{L}}}

\DeclareMathAlphabet{\mathpzc}{OT1}{pzc}{m}{it}

\DeclareMathOperator{\td}{Td}
\DeclareMathOperator{\ad}{ad}

\DeclareMathOperator{\res}{res}

\newcommand{\gra}[2]{{\scriptscriptstyle (#1 , #2 )}}
\newcommand{\ord}[1]{{\scriptscriptstyle (#1)}}

\def\ie{{\it i.e.}\ }
\def\eg{{\it e.g.}\ }
\def\nn{\nonumber}

\def\N{\mathcal{N}}

\def\V{{\mathcal{V}}}
\def\L{{\mathcal{L}}}
\def\w{{\scriptstyle W}}

\def\m{{\fontsize{10.35pt}{8pt}\selectfont {\mbox{$\mathpzc{m}$}} \fontsize{12.35pt}{12pt}\selectfont }}
\def\n{{\fontsize{10.35pt}{8pt}\selectfont {\mbox{$\mathpzc{n}$}} \fontsize{12.35pt}{12pt}\selectfont }}

\def\bm{{\fontsize{10.35pt}{8pt}\selectfont {\mbox{$\bar {\mathpzc{m}}$}} \fontsize{12.35pt}{12pt}\selectfont }}
\def\bn{{\fontsize{10.35pt}{8pt}\selectfont {\mbox{$\bar {\mathpzc{n}}$}} \fontsize{12.35pt}{12pt}\selectfont }}

\def\bigm{{\fontsize{13.35pt}{12pt}\selectfont {\mbox{$\mathpzc{m}$}} \fontsize{12.35pt}{12pt}\selectfont }}
\def\bign{{\fontsize{13.35pt}{12pt}\selectfont {\mbox{$\mathpzc{n}$}} \fontsize{12.35pt}{12pt}\selectfont }}
\def\bbigm{{\fontsize{13.35pt}{12pt}\selectfont {\mbox{$\bar {\mathpzc{m}}$}} \fontsize{12.35pt}{12pt}\selectfont }}
\def\bbign{{\fontsize{13.35pt}{12pt}\selectfont {\mbox{$\bar {\mathpzc{n}}$}} \fontsize{12.35pt}{12pt}\selectfont }}

\def\i{{\fontsize{8.35pt}{8pt}\selectfont {\mbox{\texttt{i}}} \fontsize{12.35pt}{12pt}\selectfont }}
\def\j{{\fontsize{8.35pt}{8pt}\selectfont {\mbox{\texttt{j}}} \fontsize{12.35pt}{12pt}\selectfont }}
\def\k{{\fontsize{9.35pt}{9pt}\selectfont {\mbox{\texttt{k}}} \fontsize{12.35pt}{12pt}\selectfont }}
\def\l{{\fontsize{9.35pt}{9pt}\selectfont {\mbox{\texttt{l}}} \fontsize{12.35pt}{12pt}\selectfont }}
\def\h{{\fontsize{9.35pt}{9pt}\selectfont {\mbox{\texttt{h}}} \fontsize{12.35pt}{12pt}\selectfont }}

\def\dd{{\ddagger}}

\def\zero{{\mathpzc{0}}}
\def\un{{\mathpzc{1}}}
\def\deux{{\mathpzc{2}}}
\def\trois{{\mathpzc{3}}}
\def\quatre{{\mathpzc{4}}}
\def\DJo{$\;$\kern-.4em \hbox{D\kern-.8em\raise.15ex\hbox{--}\kern.35em okovi\'c}}

\newcommand{\eprint}[1]{{\href{http://arxiv.org/abs/#1}{\texttt{[#1}]}}}
\newcommand{\eprintN}[1]{{\href{http://arxiv.org/abs/#1}{\texttt{#1 [hep-th]}}}}

\def\e{\boldsymbol{e}}

\def\w{{\scriptstyle W}}

\def\g{\mathfrak{g}}
\def\ka{\mathfrak{k}}
\def\pa{\mathfrak{p}}

\def\sl{\mathfrak{sl}}
\def\so{\mathfrak{so}}
\def\su{\mathfrak{su}}

\def\e{\mathfrak{e}}

\def\SU{SU_{\scriptscriptstyle \rm c}(8)}

\def\nn{\nonumber}
\def\N{\mathcal{N}}

\def\DJo{$\;$\kern-.4em \hbox{D\kern-.8em\raise.15ex\hbox{--}\kern.35em okovi\'c}}

\newcommand{\ba}{/ \hspace{-1.2ex}} 

\newcommand{\baaa}{\, / \hspace{-1.6ex}}

\def\susy{{\delta^{\mathpzc{Susy}}}}
\def\sept{{\delta^{\e_{7(7)}}}}

\def\Afi{\mathpzc{a}}
\def\Bfi{\mathpzc{b}}
\def\Cfi{\mathpzc{c}}
\def\Dfi{\mathpzc{d}}

\def\beg{\begin{equation*}}
\def\eeg{\end{equation*}}
\def\X{\mathscr{E}}
\def\ve{\varepsilon}

\newcommand{\cF}{\mathcal{F}}

\newcommand{\cV}{\mathcal{V}}
\newcommand{\dR}{R}

\allowdisplaybreaks[1]

\begin{document}

\renewcommand{\thefootnote}{\fnsymbol{footnote}}
\numberwithin{equation}{section}

\begin{titlepage}
\begin{flushright}
\
\vskip -2.5cm
{\small AEI-2010-118}\\
\vskip 1cm
\end{flushright}
\begin{center}
{\Large \bf
$E_{7(7)}$ symmetry in perturbatively \\[3mm] quantised $\N=8$ supergravity}
\\
\lineskip .75em
\vskip 3em
\normalsize
{\large  Guillaume Bossard\footnote{email: bossard@aei.mpg.de},
Christian Hillmann\footnote{email: hillmann@ihes.fr} and 
Hermann Nicolai\footnote{email: Hermann.Nicolai@aei.mpg.de}}\\
\vskip 1 em
$^{\ast\ddagger}${\it AEI, Max-Planck-Institut f\"{u}r Gravitationsphysik\\
Am M\"{u}hlenberg 1, D-14476 Potsdam, Germany}
\\
\vskip 1 em
$^{\dagger}${\it Institut des Hautes Etudes Scientifiques\\ 35, route de Chartres, 91440 Bures-sur-Yvette, France}\\

\vskip 3 em
\end{center}
\begin{abstract}
{
We study the perturbative quantisation of $\N=8$ supergravity in a 
formulation where its $E_{7(7)}$ symmetry is realised {\em off-shell}.
Relying on the cancellation of $SU(8)$ current anomalies we show that 
there are no anomalies for the non-linearly realised $E_{7(7)}$ either;
this result extends to all orders in perturbation theory. As a 
consequence, the $\e_{7(7)}$ Ward identities can be consistently 
implemented and imposed at all orders in perturbation theory, and 
therefore potential divergent counterterms must in particular respect the full 
non-linear $E_{7(7)}$ symmetry. 
}
\end{abstract}


\vspace{1cm}
\end{titlepage}
\renewcommand{\thefootnote}{\arabic{footnote}}
\setcounter{footnote}{0}

\pagebreak
\tableofcontents
\setcounter{page}{1}

\section{Introduction}

Maximally extended $\N=8$ supergravity \cite{CremmerJulia,deWitNicolai} 
is the most symmetric field theoretic extension of Einstein's theory
in four space-time dimensions. Although long thought to diverge at
three loops \cite{KalloshCT,Superactions}, spectacular computational 
advances have recently shown 
that, contrary to many expectations, the theory is finite at least up to 
and including four loops \cite{Bern,Bern2}, and thereby fuelled speculations 
that the theory may actually be finite to {\em all orders} in perturbation 
theory. It appears doubtful whether maximal supersymmetry alone could 
suffice to explain such a far reaching result  \cite{BHS}, if true. Rather, 
it seems plausible that the possible finiteness of $\N=8$ supergravity 
will hinge on known or unknown `hidden symmetries' of the theory. 
Indeed, already the construction of the $\N=8$ Lagrangian 
itself was only possible thanks to the discovery of the non-linear duality 
symmetry $E_{7(7)}$ of its equations of motion \cite{CremmerJulia}. This 
symmetry is expected to be a symmetry of perturbation theory, and to be 
broken to an arithmetic subgroup of $E_{7(7)}$ by non-perturbative effects
when the theory is embedded into string theory (see \eg \cite{Green,Pioline} 
for a recent update, and also the comments below). Nevertheless, the status of 
the non-linear duality symmetry at the level of quantised perturbation theory 
has remained rather unclear, because $E_{7(7)}$ is not a symmetry of 
the original $\N=8$ Lagrangian and the corresponding non-linear functional 
Ward identities therefore have not been worked out so far.

Inspired by earlier work devoted to the definition of an action for 
self-dual form fields \cite{HenneauxChiral}, one of the authors recently 
was able to set up a formulation of $\N=8$ supergravity in which the 
Lagrangian is {\em manifestly} $E_{7(7)}$-invariant \cite{Christian}.~\footnote{The
formalism had been applied earlier to the definition of 
 a manifestly $SL(2,\IR)$ bosonic action for $\N=4$ supergravity 
 \cite{Schwarz}.}  The main peculiarity of the formalism is to replace
the 28 vector fields $A^\m_\mu$ of the original formulation by 
$56 = 28 \! + \! 28$ vector fields $A_\i^m \equiv ( A_\i^\m , A_\i^{\bar \m})$ 
{\em with spatial components only}, whose conjugate 
momenta are determined by second class constraints in the canonical 
formulation, in such a way that they represent the same number of physical 
degrees of freedom as the original 28 vector fields in the conventional 
formulation of the theory. Although not manifestly diffeomorphism invariant, 
the theory still admits diffeomorphism and local supersymmetry gauge 
invariance \cite{Christian}. By virtue of its manifest off-shell 
$E_{7(7)}$ invariance, the theory possesses a {\em bona fide} $E_{7(7)}$ 
Noether current, unlike the covariant formulation \cite{GaillardZumino},
and this is the feature which permits to write down functional Ward 
identities for the non-linear duality symmetry. 

In this paper we will consider the perturbative quantisation of $\N=8$ supergravity
in this duality invariant formulation. As our main result, we will prove 
that there exists a renormalisation scheme which maintains the full 
{\em non-linear} (continuous) $E_{7(7)}$ duality symmetry at all orders 
in a perturbative expansion of the theory in the gravitational coupling $\kappa$.
A key element in this proof is the demonstration of the absence 
of linear $SU(8)$ and non-linear $E_{7(7)}$ anomalies. 

As is well known \cite{PiguetSorella}, the proper definition of any quantum 
field theory relies on the {\em quantum action principle}, according to which
the ultra-violet divergences of the 1PI generating functional are 
always {\em local} functionals of the 
fields. Only thanks to this property can one carry out the renormalisation 
program by consistently modifying the local bare action order by order 
to eliminate both divergences and trivial anomalies. Because of the 
non-conventional character of our reformulation of $\N=8$ supergravity, 
and its lack of manifest Lorentz invariance in particular, the validity 
of the quantum action principle is however not automatically guaranteed.

To deal with this problem, we will in a first step prove that the 
duality invariant path integral of the theory is equivalent to the conventional 
formulation by means of a Gaussian integration. In order to ensure the 
validity of the quantum action principle, we will require the existence of 
a local regularisation scheme in the two formulations of the theory, which 
are equivalent modulo a Gaussian integration (but note that the
Gaussian integration reduces the manifest $E_{7(7)}$ invariance to
an on-shell symmetry). The validity of the quantum action principle in 
the conventional formulation of the theory then ensures 
its validity in the duality invariant formulation. We will 
define a Pauli--Villars regularisation of the theory satisfying these
criteria. Although this regularisation would break Lorentz invariance in 
the covariant formulation as well, it is local and invariant with respect 
to abelian gauge invariance in the two formulations. We will exhibit the 
consistency of this regularisation in the explicit computation of the 
one-loop vector field contribution to the $\su(8)$-current anomaly. 

With a consistent duality invariant formulation at hand, we can address 
and answer the question of whether the $\e_{7(7)}$ current Ward identities 
are anomalous or not in perturbation theory. According to \cite{Ferrara}, 
the local $\su(8)$ gauge invariance in the version of $\N=8$ supergravity 
with {\em linearly realised} $E_{7(7)}$ is anomalous at one-loop. However,
as shown in \cite{deWit} this anomaly can be cancelled by an $SU(8)$ 
Wess--Zumino term which in turn breaks the manifest $E_{7(7)}$ invariance,
whereby the local $SU(8)$ anomaly is converted into an anomaly of the global
$E_{7(7)}$ --- unless there appear new contributions to the latter, as 
happens to be the case for $\N =8$ supergravity. According to \cite{deWit}
one thus has the option of working either with the locally $SU(8)$ invariant 
version of $\N=8$ supergravity, or with its gauge-fixed version where 
$E_{7(7)}$ is realised non-linearly. Here we prefer the second option, 
that is, we will consider an explicit parametrisation of the scalar manifold 
$E_{7(7)} / \SU$~\footnote{Where throughout the notation $\SU$ will be 
 used as a shorthand for the quotient of $SU(8)$ by the $\mathds{Z}_2$ 
 kernel of the representations of even rank.} in terms of 70 scalar fields
$\Phi \in \e_{7(7)} \ominus \su(8)$ which coordinatise the coset 
manifold. A consistent anomaly must then be a non-trivial solution to 
the Wess--Zumino consistency condition. We will prove that the associated 
cohomology problem reduces to the cohomology problem associated to the 
current $\su(8)$ Ward identities. It follows from this result that, although 
the non-linear character of the $\e_{7(7)}$ symmetry is such that the 
associated anomalies involve infinitely many correlation functions 
with arbitrarily many scalar field insertions, the Wess--Zumino 
consistency condition implies that the corresponding coefficients are 
all determined in function of the linear $\su(8)$ anomaly coefficient
--- thereby saving us the labour of having to determine an infinitude of 
anomalous diagrams! Now, thanks to a crucial insight of \cite{Marcus}, 
it is known that for $\N=8$ supergravity, the anomalous contributions to 
the current (rigid) $\su(8)$ Ward identities from the fermions cancel against the
contributions from the vector fields  because the latter are also {\em chiral}
under $SU(8)$. Therefore the non-linear  $\e_{7(7)}$  Ward identities 
are likewise free of anomalies. Moreover, the cohomological arguments 
of  section~\ref{Anomaly1loop} show that this results extends to all loop orders.

The fact that the consistent $\e_{7(7)}$ anomalies are in one-to-one 
correspondence with the set of consistent $\su(8)$ anomalies can 
also be understood more intuitively, and in a way that makes the result
almost look trivial. Namely, in differential geometric terms, this 
correspondence is based on the homotopy equivalence 
\be\label{E7homotopy}
E_{7(7)} \cong \SU \times \IR^{70} \ , 
\ee
which implies that the two group manifolds have the same De Rham cohomology. 
We will show how to extend the algebraic proof of this property by means 
of equivariant cohomology to the cohomology problem of classifying
the $\e_{7(7)}$ anomalies in $\N=8$ supergravity, and in this way 
arrive at a very explicit derivation of the {\em non-linear $\e_{7(7)}$ 
anomaly} from the corresponding linear $\su(8)$ anomaly.

$\N=8$ supergravity is a gauge theory, and its first class constraints 
(associated to diffeomorphisms, local supersymmetry, abelian gauge invariance, 
and Lorentz invariance) must be taken care of by means of the BRST 
formalism. This likewise requires the explicit parametrisation of the coset 
manifold $E_{7(7)} / \SU$, such that there are no first class constraints 
associated to $SU(8)$ gauge invariance in the formulation. For the validity 
of the proof of the $E_{7(7)}$ invariance of the theory, one must therefore 
establish the compatibility of the latter with the BRST invariance. We will 
demonstrate in the last section that the theory can be quantised in its 
duality invariant formulation within the Batalin--Vilkovisky formalism, 
as it does in the ordinary formulation. It is not difficult to see that 
one can define a consistent $E_{7(7)}$-invariant fermionic gauge
fixing-functional (or `gauge fermion').  We will explain how the $E_{7(7)}$ 
Noether current can be coupled consistently to the theory, despite 
its lack of gauge invariance. 

In summary, the proof of the duality invariance of the quantised perturbation
theory relies on establishing the following results:

\begin{enumerate}
\item Existence of a local action $\Sigma$ depending on the physical fields
      and sources, well suited for Feynman rules, and satisfying 
      consistent functional identities 
   associated to both $\e_{7(7)}$ current Ward identities and BRST invariance. 
\item Existence of a regularisation prescription consistent with the 
      quantum action principle; as dimensional regularisation appears
      unsuitable in the present formulation, we will employ a Pauli--Villars 
      regulator.
\item Existence of a unique non-trivial solution to the $E_{7(7)}$ 
    Wess--Zumino consistency condition associated to the one-loop anomaly.
\item Vanishing of the coefficient of the unique anomaly, which implies
   the absence of any obstruction towards implementing the full
   nonlinear $E_{7(7)}$ symmetry at each order in perturbation
   theory via an associated $\e_{7(7)}$ master equation.
\end{enumerate} 

However, our exposition will not follow these steps in this order, \ie as a 
successive proof of each of these points. Instead, we chose to postpone the 
discussion of the first point, \ie the consistency with BRST invariance, 
to the end and to first discuss other components of the proof that we 
consider to be more interesting (and perhaps also more easily accessible). 
As one of our main results we separately derive the master equations 
(or ``Zinn--Justin equations'') for both $\N=8$ supersymmetry and 
non-linear $E_{7(7)}$. Using standard textbook results (see e.g.
\cite{Weinberg,Weinberg2}) readers may then directly deduce from these 
any (non-linear) Ward identity of interest if they wish.

Our results confirm the expectation that any divergent counterterm 
must respect the full non-linear  $E_{7(7)}$ symmetry. They may thus 
be taken as further evidence that divergences of $\N=8$ supergravity, 
if any, will not make their appearance before seven loops. The strongest evidence so far of the 6-loop finiteness was the absence of logarithm in the string effective action threshold \cite{Pierre}. The chiral invariants associated to potential logarithmic divergences at three, five and six loops are only known in the linear 
approximation \cite{SusyInvariants}, and if they are invariant with respect 
to the linearised duality transformations, there is no reason to believe that 
their non-linear completion would be duality invariant. Indeed, it has recently been exhibited through the study of on-shell tree amplitudes in type II string theory that the 1/2 BPS invariant corresponding to the potential 3-loop divergence {\em is not} $E_{7(7)}$ invariant \cite{BroedelDixon,Elvang}. The same argument applies to the invariants associated to potential 5 and 6-loop divergences. The manifestly $E_{7(7)}$ invariant 7-loop counterterm is the full superspace 
integral of the supervielbein determinant. This is known to vanish for
lower $\N$ supergravities, suggesting that the first $E_{7(7)}$ invariant
counterterm may actually not appear before {\em eight} loops. As a 
corollary of our results, we may also point out that $\N\leq 4$ 
supergravities whose R-symmetry group $K$ possesses a $U(1)$ factor, 
do exhibit anomalies, and therefore possible divergences need not respect 
the non-linear duality invariance.

It is important to emphasise that the preservation of the continuous 
duality symmetry in perturbation theory is not in contradiction with the 
string theory expectation that only its arithmetic subgroup remains a 
symmetry at the quantum level. Within supergravity, we expect that 
only $E_{7(7)}(\mathds{Z})$ will be preserved by non-perturbative 
corrections in $\exp(- \kappa^{-2}{S^{\scriptscriptstyle \rm Instanton}} )$. 
Although the status of instanton corrections in $\N=8$ supergravity is not clear 
by any means, we will provide some evidence relying on the classical 
breaking of the $E_{7(7)}$ current conservation in non-trivial gravitational
backgrounds, see section~\ref{ClassicalCurrent}. On the other hand, considering $\N=8$ 
supergravity as a limit $\ell_s \rightarrow 0$ (decoupling the massive string 
states) of type II string theory compactified on a product of circles of radii $r_i$
(to be taken $\rightarrow 0$ to decouple massive Kaluza--Klein states), 
one cannot avoid non-perturbative string corrections in the four-dimensional 
effective string coupling constant 
\be 
g_\quatre^{\; 2}  \equiv  \frac{{\ell_s^{\; 6} g_s^{\; 2}}}{  \prod_{i=1}^6 r_i} \; , 
\ee
while keeping the gravitational coupling constant 
$ \kappa^2 = 8 \pi g_\quatre^{\; 2}Ê\ell_s^{\; 2}Ê$ fixed, since necessarily
$g_\quatre^{\; 2} \rightarrow \infty$ in this limit. It is therefore clear that 
the supergravity limit of string theory must involve string theory non-perturbative 
states \cite{GOS}, and thus defines some non-perturbative completion of 
the supergravity field theory. If the supergravity limit of the string theory 
effective action is the effective action in field theory, the latter must necessarily 
include non-perturbative contributions associated to field theory instantons. 
The $E_{7(7)}(\mathds{Z})$ `Eisenstein series' that multiplies 
the Bel--Robinson square $R^4$ term in the string theory effective 
action is defined in string theory as an expansion in 
$\exp(-1/{g_\quatre^{\;   2}})$ 
\cite{Green,Pioline}. This expansion diverges as $g_\quatre^{\; 6}$ in the supergravity limit 
$g_\quatre^{\; 2} \rightarrow \infty$ \cite{Pioline}, see also \cite{Pierre} for an explicit resummation of the eight-dimensional $SL(2,\IZ) \times SL(3,\IZ)$ invariant threshold in the supergravity limit. The result of the present paper suggests that if this limit 
makes sense in field theory, it should be defined as an expansion in $e^{-1/\kappa^2}$, 
and that the perturbative contribution would vanish. 

The paper is organised as follows. We will first recall the duality invariant 
formulation of the classical theory defined in \cite{Christian}, and exhibit 
its equivalence with the conventional formulation of the theory 
\cite{CremmerJulia,deWitNicolai} by means of a Gaussian integration. Then 
we will recall the definition of the $E_{7(7)}$ Noether current. In order 
to deal with the non-linear realisation of the $E_{7(7)}$ symmetry in the 
symmetric gauge, it will be convenient to define the non-linear 
transformations in terms of formal power series in $\Phi$ in the adjoint representation. 
We derive such formulas in Section \ref{SymmetricGauge}, and we exhibit 
the commutation relations between local supersymmetry and the $\e_{7(7)}$ 
symmetry. More generally, we show that {\em the BRST operator commutes 
with the non-linear $\e_{7(7)}$ symmetry}, cf. (\ref{E7BRST}), hence 
is $E_{7(7)}$ invariant.

Section \ref{Anomaly1loop} exhibits the well definedness and consistency
of the formalism (and particular the validity of the quantum action principle),
through the explicit computation of the one-loop vector field contribution 
to the $\su(8)$ anomaly. It will therefore provide answers to both 2 and 4. 
In this section we discuss the Feynman rules  for the vector fields in 
detail, exhibiting the equivalence with the conventional formulation in 
terms of free photons. It has been shown in \cite{WittenGaume} that 
self-dual form fields contribute to (gravitational) anomalies, just like 
chiral fermion fields, by means of a formal Fujikawa-like path integral 
derivation. This result can be understood geometrically from the family's  
index theorem \cite{Singer}, and it has been used in \cite{Marcus} to 
establish the absence of anomalies for the $\su(8)$ current Ward identities 
in $\N=8$ supergravity. Here we will exploit the duality invariant 
formulation to provide a full fledged Feynman diagram computation of 
the vector field contribution which confirms the expected result, and 
therefore the absence of anomalies in the theory. In this section we also
set up the Pauli--Villars regularisation for the vector fields, and 
exhibit its (non-trivial) compatibility with the quantum action principle.

Section \ref{WessZumino} is also very important: it will provide the 
definition of the non-linear $\e_{7(7)}$ Slavnov--Taylor identities for 
the current Ward identities, and define and solve the Wess--Zumino 
consistency condition, incidentally answering 3. 

The last section finally provides an answer to the first point of the above list.
We there discuss the solution of the Batalin--Vilkovisky master equation in the 
duality invariant formulation, including the coupling to the $E_{7(7)}$ 
Noether current. Using the property that the BRST operator commutes with 
the $\e_{7(7)}$ symmetry and considering a duality invariant gauge-fixing, 
we are able to define consistent and mutually compatible master equations 
for BRST invariance and $\e_{7(7)}$ symmetry. In this section we also discuss 
the `energy Coulomb divergences' in the one-loop insertions of 
$E_{7(7)}$ currents, which constitute a special subtlety of the formalism. We will exhibit that these divergences can be consistently removed within the Pauli--Villars regularisation.

As this paper is rather heavy on formalism, we here briefly summarise our 
notational conventions for the reader's convenience. (Curved) space-time 
indices are $\mu, \, \nu,\, ...$, (curved) spatial indices are 
\texttt{i}, \texttt{j}, \texttt{k}, $...$, and space-time Lorentz indices are 
$a,\, b,\, c,\, ....$. Indices in the fundamental representation 
$\bf{56}$ of $E_{7(7)}$ are
$m,n,...= 1,...,56$;  when split into 28+28 they become $\bigm ,\, \bign,\, ..$ 
and $\bbigm,\, \bbign,\, ..$. Rigid $SU(8)$ indices are $I,\, J,\, K ,\, ... $ 
such that the $E_{7(7)}$  adjoint representation $\bf{133}$ decomposes as
$\bf{63}\oplus\bf{70}$ with generators $X^{IJ}{}_{KL} \equiv \ 
2 \delta^{[I}_{[K} X^{J]}{}_{L]}, \, X^{IJKL}\equiv \frac{1}{2} X^{[IJKL]} 
+ \frac{1}{48} \varepsilon^{IJKLPQMN}ÊX_{PQMN}Ê$, etc. Local $SU(8)$  
indices are $i,\, j,\, k\, ... = 1,\, ...,\, 8$, and raising or lowering 
them corresponds to complex conjugation. Space-time indices are 
lowered with the metric $g_{\mu\nu}$, and the tensor densities $\ve^{\i\j\k}$ and 
$\ve^{\mu\nu\rho\sigma}$ are normalised as $\ve^{123}=\ve^{0123} =1$. 
Finally, we will use the letters $S$ for the classical action, $\Sigma$
for the classical action with sources, ghost and antifield terms included. While 
both $S$ and $\Sigma$ are local, the full quantum effective action 
$\Gamma$ is not, but obeys $\Gamma = \Sigma + {\cal{O}}(\hbar)$.

\section{$\N=8$ supergravity with off-shell $E_{7(7)}$ invariance}
\subsection{Manifestly duality invariant formulation}
We start from the usual ADM decomposition of the 4-metric 
\be 
g_{\mu\nu} dx^\mu dx^\nu  = - N^2 dt^2 + h_{\i\j} ( dx^\i + N^\i dt ) 
( dx^\j + N^\j dt ) \; ,
\ee
with the lapse $N$ and the shift $N^\i$; $h_{\i\j}$ is the metric on the
spatial slice. The vector fields $A_\i^m$
of the theory appear only with {\em spatial} indices, and are labeled 
by internal indices $m,n,...$ which transform in a given representation 
of the internal symmetry group $G$ with maximal compact subgroup $K$
(for $\N=8$ supergravity $G \cong E_{7(7)}$ and $K \cong \SU$, with the vector fields
transforming in the $\bf{56}$ of $E_{7(7)}$). In comparison with the usual 
on-shell formalism this implies a {\em doubling} of the vector fields,
such that the multiplet $A_\i^m$ comprises both the (spatial
components of the) electric and their dual magnetic vector 
potentials. To formulate an action we also need the field dependent 
$G$-invariant metric $G_{mn}$  on the vector space on which 
the electromagnetic fields are defined (\ie the $E_{7(7)}$ invariant 
metric on $\IR^{56}$ for $\N=8$ supergravity; this metric
is explicitly given in (\ref{VTV}) below). In addition we 
need the symplectic invariant $\Omega_{mn}= -\Omega_{nm} = 
\Omega^{mn}$,~\footnote{Hence, with our conventions $\Omega_{mp} 
  \Omega^{pn}= -\delta_m^n$.} which is always present, because 
the generalised duality symmetry is generally a subgroup of a 
symplectic group acting on the electric and magnetic vector 
potentials \cite{GaillardZumino} (the group $Sp(56,\IR)\supset E_{7(7)}$ for 
$\N=8$ supergravity). Duality invariance implies the following relation
for the inverse metric $G^{mn}$ 
\be\label{GOO}
G^{mn} = \Omega^{mp} \Omega^{nq}G_{pq} \; ,  \qquad
(G^{mp} G_{pn} = \delta^m_n) \; \;  . 
\ee 
For later purposes we also define the `complex structure' tensor
\be\label{J}
J^m{}_n \equiv G^{mp} \Omega_{pn} \quad \Rightarrow\;\;
J^m{}_p J^p{}_n = - \delta^m_n \; \; . 
\ee
Note that $J^m{}_n$ depends on the scalar fields via the metric $G_{mn}$.
The maximal compact subgroup $K$ can be characterised 
as the maximal subgroup in $G$ which commutes with $J^m{}_n(\mathring{\Phi})$ (for some background value $\mathring{\Phi}$ of the scalar fields).

After these preparations we can write down the part of the action 
containing the vector fields
\begin{multline}\label{Action1} 
S_{\scriptscriptstyle \rm vec} = \frac{1}{2}Ê  \int d^4x \Bigl(  
\frac{1}{2}\Omega_{mn}  \varepsilon^{\i\j\k} \scal{ \partial_\zero A_\i^m  
+  N^\l F^m_{\i\l}  } F_{\j\k}^n 
-  \frac{1}{2} N \sqrt{h} \, G_{mn}  
h^{\i\k}h^{\j\l} F_{\i\j}^m F_{\k\l}^n \\*
- N \sqrt{h} h^{\i\k}h^{\j\l}  F^m_{\i\j} W_{\k\l\, m}  
- \frac{1}{2}N \sqrt{h} \, G^{mn}  h^{\i\k}h^{\j\l} 
W_{\i\j\, m}W_{\k\l\, n}   \Bigr)   \;\;.
\end{multline}
Here $W_{\i\j\, m}$ is a bilinear function of the fermion fields, 
which will be discussed in more detail shortly (see (\ref{Wuv}) below). 
We also consider the $W^2$ term which define the non-manifestly 
diffeomorphism covariant quartic terms in the fermions. For quantisation,
the above action must be supplemented by further terms depending on
the ghost fields as well as the anti-fields; this will be discussed in more 
detail below.

As shown in \cite{Christian}, the main advantage of the above reformulation 
is that it incorporates both the electric and the dual magnetic vector 
potentials {\em off-shell}, at the expense of manifest space-time 
diffeomorphism invariance. In particular, the equation of motion of the
$56$ vector fields $A_\i^m$ can be expressed as a twisted self-duality 
constraint \cite{CremmerJulia} for the supercovariant field strength 
$\hat{F}_{\mu\nu}^m$ (see \cite{Christian} for further details)
\be\label{EOM2}
\hat{F}_{\mu\nu}^m=-\frac{1}{2\sqrt{\mbox{-}g}}\ve_{\mu\nu}{}^{\sigma\rho}J^m{}_n 
\hat{F}_{\sigma\rho}^n \; \; , 
\ee
where the tensor $J$ takes the place of an imaginary unit.
We briefly explain this procedure and why the time-components $A_\zero^m$ of 
the vector fields naturally enter this equation, although they are absent 
in the original Lagrangian (\ref{Action1}). The variation of the action 
functional $S_{\scriptscriptstyle \rm vec}$ (\ref{Action1}) with respect to 
the $56$ vector fields $A_\i^m$ leads to the second order equation of motion \footnote{Do not confuse the equation of motion function $\X_\i^m$ with the electric potential $\mathcal{E}_\i^m$ introduced in \cite{HenneauxChiral,Christian}.}
\be\label{EOM1}
\ve^{\i\j\k}\partial_\j\X_\k^m = 0 \ , 
\ee
with the abbreviation
\be\label{Xk}
\X_\i^m \equiv  \partial_\zero A_\i^m  
+  N^\j F^m_{\i\j}   -  \frac{N}{2\sqrt{h}}h_{\i\j}\ve^{\j\k\l}
\Bigl(J^m{}_n F_{\k\l}^n +\Omega^{mn} W_{\k\l\, n}\Bigr) \ .
\ee
This equation is equivalent to the statement that the one-form 
$\X_k^m\,dx^k $ is closed. On any contractible open set of the $d=4$ 
space-time manifold, every closed form is exact by Poincar\'e's lemma, 
which implies the existence of a zero-form $A_\zero^m$
satisfying
\be\label{EOM3}
\X_\i^m = \partial_\i A_\zero^m \ .
\ee
It is straightforward to verify that this equation of motion is completely 
equivalent to the twisted self-duality constraint of equation (\ref{EOM2}). 
Furthermore, only the identification of the zero-form with the time-component 
$A_\zero^m$ gives rise to an equation of motion that is diffeomorphism 
covariant in the usual sense. 

Before we prove that the action functional (\ref{Action1}) and the usual 
second order form of the action are equivalent, and related by functional 
integration, we briefly explain the realisation of the diffeomorphism 
algebra on the vector fields. To this aim we recall that the Lie derivative 
on the vector field in the covariant formulation can be rewritten as
\be
\delta A_\mu^m = \partial_\mu \xi^\nu A_\nu^m + \xi^\nu \partial_\nu A_\mu^m
  = \partial_\mu(\xi^\nu A_\nu^m) + \xi^\nu F_{\nu\mu}^m \; . 
\ee
Considering the vector fields $A^m$ as abelian connections, the geometrical action of diffeomorphism is defined via the horizontal lift of the vector $\xi^\mu$ to the principle bundle, and is modified by a gauge transformation. We will consider this {\em covariant (or `horizontal') diffeomorphism} 
\be
\delta A_\mu^m = \xi^\nu F_{\nu\mu}^m \; . 
\ee 
Splitting indices into time and space indices, we get 
\be\label{dAi1}
\delta A_\i^m = \xi^\zero F_{\zero\i}^m\ + \xi^\j F_{\j\i}^m \; . 
\ee
The recipe for obtaining the correct formula in the present formulation 
then consists simply in replacing
\be\label{Recipe}
F_{\zero\i}^m \;\rightarrow\; \partial_\zero A_\i^m - \X_\i^m
\ee
everywhere according to (\ref{Xk}), such that (\ref{dAi1}) becomes
\be\label{DiffA}
\delta_\xi A_\i^m \equiv  \xi^\mu\partial_\mu A_\i^m -
\xi^\j\partial_\i A_\j^m  -   \xi^\zero \X_\i^m \; . 
\ee
We note that the recipe (\ref{Recipe}) also yields the correct formulas
for all other transformations in the manifestly duality invariant formalism,
including the modified supersymmetry transformations and the BRST transformations of the ghosts.

The non-standard representation of the diffeomorphism algebra (\ref{DiffA})
on the vector fields is consistent, because it closes {\it off-shell} up to 
a gauge transformation with parameter $\Lambda^m$, which cannot be separated 
from the diffeomorphism action:
\bea
\bigl[\delta_{\xi_1},\delta_{\xi_2}\bigr]A_\i^m &=& 
\delta_{[\xi_2,\xi_1]} A_\i^m +\partial_i\Lambda^m
\CR
\text{with}\quad \Lambda^m&=&\xi_2^\i\xi_1^\j F_{\i\j}^m +
(\xi^\zero_2\xi^\j_1-\xi^\zero_1\xi^\j_2)( \partial_\zero A_\j^m -\X_\j^m ) \, \, .
\eea
The gauge transformation $\Lambda^m$ can be obtained from the one that would appear in the covariant formulation by the substitution (\ref{Recipe}).

To sum up: although the equations of motion are covariant under 
the diffeomorphism action in both formulations of maximal supergravity, 
the representations of the diffeomorphism algebra on the vector fields do 
not coincide {\it off-shell}. Agreement can {\it a priori}Ê   be achieved 
only {\it on-shell}, if we impose the equations of motion in their first order 
form (\ref{EOM3})  with the introduction of the time-component of 
the $56$ vector fields. Nevertheless, the two formulations are also formally 
equivalent at the quantum level, as we are going to see.

\subsection{Equivalence with the covariant formalism}	
\label{Equivalence}		
To establish the link with the manifestly  diffeomorphism covariant 
formalism, we must in a first step decompose the electromagnetic fields 
into Darboux components associated to the symplectic form
\be \label{Omega}
\Omega_{\m\n}  = \Omega_{\bar \m \bar \n}= 0 \qquad 
\Omega_{\m\bar \n} = - \Omega_{\bn \m} = \delta_{\m \bar \n} 
\ee 
where the indices $m,n,...$ are split into pairs $(\bigm,\bbigm)$ each running 
over half the range of $m,n$. For the vector fields this entails
the split
\be
A_\i^m \;\rightarrow\; (A_\i^\m , A_\i^{\bar\m})
\ee
into electric and magnetic vector potentials. With the above split,
the manifest off-shell $E_{7(7)}$ symmetry will be lost after the 
Gaussian integration to be performed below, and is thus reduced to
the on-shell symmetry of the standard version. Extending (\ref{Action1}) 
by a gauge-fixing term, the action functional becomes
\begin{multline}\label{Action2} 
S_{\scriptscriptstyle \rm vec} = \frac{1}{2}Ê  \int d^4x \Bigl(  
\frac{1}{2}\Omega_{mn}  \varepsilon^{\i\j\k} \scal{ \partial_\zero A_\i^m  
+  N^\l F^m_{\i\l}  } F_{\j\k}^n - \frac{1}{2} N \sqrt{h} \, G_{mn}  
h^{\i\k}h^{\j\l} F_{\i\j}^m F_{\k\l}^n \\*
- N \sqrt{h} h^{\i\k}h^{\j\l}  F^m_{\i\j} W_{\k\l\, m}  
- \frac{1}{2}N \sqrt{h} \, G^{mn}  h^{\i\k}h^{\j\l} 
W_{\i\j\, m}W_{\k\l\, n} +  2 b_m \partial_\i A_\i^m  \Bigr) \; . 
\end{multline}
Sums over repeated indices are understood even when they are both down, 
which only reflects the property that the corresponding terms are not 
invariant with respect to diffeomorphisms. Performing the split, and up 
to an irrelevant boundary term, we arrive at the following Lagrange density 
\begin{multline} 
\hspace{-3mm} S_{\scriptscriptstyle \rm vec} = \frac{1}{2}Ê\int d^4x \biggl(   \Bigl( 
\delta_{\m\bar \n }  \varepsilon^{\i\j\k} \scal{ \partial_\zero A_\i^\m  +  
N^\l F^\m_{\i\l}  }  - N \sqrt{h} \, G_{\m\bar \n}  h^{\i\k}h^{\j\l} 
F_{\i\j}^\m   - N \sqrt{h} \,   h^{\i\k}h^{\j\l} W_{\i\j\, \bar \n}   \Bigr)
F_{\k\l}^{\bar \n} \\*    - \frac{1}{2} N \sqrt{h} \, G_{\bar \m\bar \n}  
h^{\i\k}h^{\j\l} F_{\i\j}^{\bar \m} F_{\k\l}^{\bar \n}  -  
\frac{1}{2} N \sqrt{h} \, G_{ \m\n}  h^{\i\k}h^{\j\l} F_{\i\j}^{\m} 
F_{\k\l}^{\n} \\* -N \sqrt{h} \,   h^{\i\k}h^{\j\l} F_{\i\j}^{\m} 
W_{\k\l\, \m} -\frac{1}{2}N \sqrt{h} \, G^{mn}  h^{\i\k}h^{\j\l} 
W_{\i\j\, m}W_{\k\l\, n} +   2 b_\m \partial_\i A_\i^\m +  
2 b_{\bar \m} \partial_\i A_\i^{\bar \m}  \biggr) \; . \label{PreGauss}
\end{multline}
Integrating out the auxiliary field $b_{\bar \m}$ enforces the constraint 
$ \partial_\i A_\i^{\bar \m} = 0 $, and the Lagrangian only depends on 
$A_\i^{\bar \m} $ through $F^{\bar \m} _{\i\j} = \partial_\i A_\j^{\bar \m} 
-  \partial_\j A_\i^{\bar \m} $ (note that this is the case even when 
considering the ghost field terms that we neglect in this discussion). 
One has then an isomorphism between the square integrable fields 
$A_\i^{\bar \m}$ satisfying $ \partial_\i A_\i^{\bar \m} = 0 $, and the 
square integrable fields $\Pi^{\i \, \bar \m}$ satisfying the same constraint 
$\partial_\i \Pi^{\i\, \bar \m} = 0 $, through
\be 
\Pi^{\i \, \bar \m} = \varepsilon^{\i\j\k}\partial_{\j} A_{\k}^{\bar \m} 
\; , \qquad A_\i^{\bar \m}= - [  \partial_\l \partial_\l ]^{-1} 
\varepsilon_{\i\j\k} \partial_\j \Pi^{\k\, \bar \m} \; \; ,  \label{VaMo}
\ee
where repeated indices are summed (and appropriate
boundary conditions assumed). This change of variables leads to a 
non-trivial functional Jacobian, but the latter does not depend on the fields 
and can therefore be disregarded.~\footnote{Note that this is only true in the 
  specific metric independent Coulomb gauge we used, in which the ghosts 
  decouple. For a metric dependent gauge, the functional Jacobian would 
  depend non-trivially on the metric, but this field dependence would be exactly 
  compensated by the functional determinant generated by the Gaussian 
  integration over the ghosts $\bar c_{\bar \m}$ and $c^{\bm}$, as is ensured 
  by BRST invariance.} Introducing a Lagrange multiplier 
$A_\zero^\m$ for the constraint  $\partial_\i \Pi^{\i\, \bar \m} = 0 $, 
one has the action 
\begin{multline}  
\hspace{-3.6mm} S_{\scriptscriptstyle \rm vec} =\frac{1}{2}Ê \int d^4x \biggl(   \Bigl( 2 
\delta_{\m\bar \n }  \scal{ \partial_\zero A_\i^\m - \partial_\i A_\zero^\m  
+  N^\l F^\m_{\i\l}  }  - N \sqrt{h} \,  
\varepsilon_{\i\l\h} h^{\l\j}h^{\h\k}  \scal{G_{\m\bar \n} F_{\j\k}^\m  + 
 W_{\j\k\, \bar \n} }   \Bigr)  \Pi^{\i\bar \n} \\*    -  
\frac{N}{\sqrt{h}} \, G_{\bar \m\bar \n}  h_{\i\j} \Pi^{\i\, \bar \m} 
\Pi^{\j\, \bar \n}  -  \frac{1}{2} N \sqrt{h} \, G_{ \m\n}  h^{\i\k}h^{\j\l}
F_{\i\j}^{\m} F_{\k\l}^{\n} \\* -N \sqrt{h} \,   h^{\i\k}h^{\j\l} 
F_{\i\j}^{\m} W_{\k\l\, \m} - \frac{1}{2}N \sqrt{h} \, 
G^{mn}  h^{\i\k}h^{\j\l} W_{\i\j\, m}W_{\k\l\, n}+  
2 b_\m \partial_\i A_\i^\m \biggr) \; , \label{PosGauss}
\end{multline}
where we normalised $A_\zero^\m$ such that it can be identified as 
the time component of the vector field, and
\be 
F_{\zero\i}^\m  = \partial_\zero A_\i^\m - \partial_\i A_\zero^\m \; \; . 
\ee
Note that this is the form of the action that one would obtain by deriving 
the path integral formulation from the Hamiltonian quantisation in the 
Coulomb gauge, such that $\Pi^{\bar \m\, \i} $ define the momentum conjugate 
to the vector fields $A_\i^\m$. One then sees that (\ref{VaMo}) actually
corresponds to a {\em second class constraint}, as one would expect in
a first order formalism. We also emphasise that, when the equations of motion are satisfied, the Lagrange multiplier
field $A_\zero^\m$ in the path integral can be identified with the corresponding component of 
 $A_\zero^m$ appearing in (\ref{EOM3}), which is the classical field resulting from rewriting a given expression $\X_\i^m$ as a curl.

One can now integrate the momentum variables $\Pi^{\i \bar \m}$ through 
formal Gaussian integration, the remaining action is 
\begin{multline}\label{Sv}  
S_{\scriptscriptstyle \rm vec} = 
\frac{1}{2} \int d^4x  \biggl(    \frac{\sqrt{h}}{N}  
\delta_{\m\bar\m} \delta_{\n\bar \n} H^{\bar \m \bar \n} 
h^{\i\j}\scal{ F_{\zero\i}^\m  +  N^\k F^\m_{\i\k}  }  
\scal{ F_{\zero\j}^\n  +  N^\l F^\n_{\j\l}  }  \\* -  
\frac{1}{2} N \sqrt{h} \, \scal{ G_{ \m\n} - G_{\m\bar\m} 
H^{\bar \m \bar \n} G_{\bar \n \n}}   h^{\i\k}h^{\j\l} F_{\i\j}^{\m} 
F_{\k\l}^{\n} - \varepsilon^{\i\j\k}\delta_{\m\bar \m } 
H^{\bar \m \bar \n}G_{\bar \n \n}\scal{ F_{\zero\i}^\m  +  
N^\l F^\m_{\i\l}  } F_{\j\k}^{\n} \\* 
- \varepsilon^{\i\j\k}\delta_{\m\bar \m } H^{\bar \m \bar \n} 
\scal{ F_{\zero\i}^\m +  N^\l F^\m_{\i\l}  }  W_{\k\l \bar \n}  + 
H^{\bar \n \bar \m}G_{\bar \m \m}   N \sqrt{h} \, h^{\i\k}h^{\j\l}  
F_{\i\j}^\m   W_{\k\l \bar \n}        -N \sqrt{h} \,   
h^{\i\k}h^{\j\l} F_{\i\j}^{\m} W_{\k\l  \m} \\* + 
\frac{1}{2}N \sqrt{h} \, H^{\bar \m\bar \n} h^{\i\k}h^{\j\l}  
W_{\i\j\, \bar \m}   W_{\k\l\, \bar \n}- \frac{1}{2}N \sqrt{h} \, 
G^{mn}  h^{\i\k}h^{\j\l} W_{\i\j\, m}W_{\k\l\, n}+  
2 b_\m \partial_\i A_\i^\m \biggr) \; \; , 
\end{multline}
where $H^{\bar \m \bar \n} $ is the inverse of $G_{\bar \m \bar \n}$
(not to be confused with the component $G^{\bm\bn}$ of the inverse 
metric $G^{mn}$). We will discuss the functional determinant afterward. 
First note that, by (\ref{GOO}), duality invariance implies 
\be 
G_{ \m\n} - G_{\m\bar\m} H^{\bar \m \bar \n} G_{\bar \n \n} = 
\delta_{\m\bar\m} \delta_{\n\bar \n} H^{\bar \m \bar \n} \; ,  \qquad  \delta_{\m\bar \m } 
H^{\bar \m \bar \n}G_{\bar \n \n}  = \delta_{\n\bar \m } 
H^{\bar \m \bar \n}G_{\bar \n \m} \; \; , 
\ee
and therefore the bosonic component is manifestly diffeomorphism invariant 
\be 
S_{\scriptscriptstyle \rm vec} = \frac{1}{4}\int d^4x  \biggl( -  \sqrt{-g}   
\delta_{\m\bar\m} \delta_{\n\bar \n} H^{\bar \m \bar \n} g^{\mu\sigma} 
g^{\nu\rho} F_{\mu\nu}^\m F_{\sigma\rho}^\n  - \frac{1}{2}Ê 
\varepsilon^{\mu\nu\sigma\rho} \delta_{\m\bar \m } 
H^{\bar \m \bar \n}G_{\bar \n \n}F^\m_{\mu\nu}  F_{\sigma\rho}^{\n}  + 
\mathcal{O}(W) \biggr) \; \; . 
\label{Svector}
\ee 
The formal Gaussian integration over the momentum variables 
$\Pi^{i\, \bar \m}$ also produces a functional determinant 
\be 
\mbox{Det}^{-\frac{1}{2}}\left[\frac{N}{\sqrt{h}} \, 
G_{\bar \m\bar \n}  h_{\i\j} \delta^4(x-y) \right]  = 
\prod_x \Scal{\mbox{det}^{-\frac{3}{2}}[ G_{\bar \m\bar \n} ]  
N^{-42} h^{7} } \label{Det} \; \; , 
\ee
which defines a one-loop local divergence quartic in the cutoff 
$\sim \Lambda^4$. This determinant defines in particular the modification of the diffeomorphism invariant measure of the metric field from the duality invariant formulation to the conventional one \cite{Fradkin}, and respectively for the $E_{7(7)}$ invariant scalar field measure. This kind of volume divergence is in fact a general property of (super)gravity theories \cite{Measure}.

\subsection{$\N=8$ supergravity}
The discussion was rather general so far, and we now turn to the  
specific case of maximal $\N=8$ supergravity, where the formalism
developed in the foregoing section leads to a formulation of the 
theory with {\em manifest and off-shell $E_{7(7)}$ invariance.}
Here we show that the formalism reproduces the vector Lagrangian
as well as the couplings of the vector fields to the fermions 
and the scalar field dependent quartic fermionic terms in the 
form given in \cite{deWitNicolai} (the remaining quartic terms in 
the Lagrangian are manifestly $E_{7(7)}$ invariant). In this case the 
choice of Darboux coordinates amounts to decomposing the 28 complex 
vector fields $A^{IJ}_\i$ into  imaginary and real (or `electric'
and `magnetic') components~\footnote{With our usual convention 
  $A^{IJ}_\i = (A_{\i IJ})^*$. Recall that the standard formulation 
  of $\N=8$ supergravity has 28 {\em real} vectors, for which there 
  is no need to distinguish between upper and lower indices.}
\be 
A_\i^{\m} \, \, \hat{=} \,\,  \Im[ A_{\i\, IJ} ] \; , 
\qquad A_\i^{\bar \m} \, \, \hat{=} \, \, \Re[ A_{\i\, IJ} ] \; \; . \label{Darboux}
\ee
For the coset representative $E_{7(7)}/\SU$, this corresponds to the 
passage from the $SU(8)$ basis in which 
\be \label{VDefi}
\V \,   \, \hat{=} \, \,  \left(\begin{array}{cc} \, {u_{ij}}^{IJ}\;  & \; v_{ijKL}\, 
\vspace{2mm} \\ \, v^{klIJ} \; & \; {u^{kl}}_{KL} \, \end{array}\right) 
\ee
to an $SL(8,\IR)$ basis in which~\footnote{This transformation is analogous
to the M\"obius transformation mapping the unit (Poincar\'e) disk to the
upper half plane, and relating $SU(1,1)$ to $SL(2,\mathbb{R})$.}
\be
\widetilde\V =
\left(\begin{array}{cc} \; \frac1{\sqrt{2}} \; & \; \frac{1}{\sqrt{2}} \; 
\vspace{2mm} \\ \; \frac{-i}{\sqrt{2}} \; & \frac{i}{\sqrt{2}}\  \end{array}\right) 
{\cal V}
\left(\begin{array}{cc} \; \frac1{\sqrt{2}} \;  &\;  \frac{i}{\sqrt{2}} \;Ê
\vspace{2mm} \\ \frac{1}{\sqrt{2}} \; &\;Ê \frac{-i}{\sqrt{2}} \; \end{array}\right) \; \; , 
\ee
or, written out in components,
\be 
\widetilde\V \, \, \hat{=} \, \, \left(\begin{array}{cc}\;  \Re\scal{{u_{ij}}^{IJ} +
v_{ijIJ} } \;  & \; \Im\scal{-{u_{ij}}^{KL} +v_{ijKL} } \;  \vspace{4mm} \\ 
\; \Im\scal{{u_{kl}}^{IJ} +v_{klIJ} }  \; & \; \Re\scal{{u_{kl}}^{KL} -v_{klKL} } \; 
\end{array}\right) \; \; . 
\ee
Then one computes that 
\bea\label{VTV}
G &=& \widetilde\V^T \widetilde\V \, \, \, \hat{=} \, \,   \left(\begin{array}{cc}  \; \; \scal{{u^{ij}}_{IJ} + v^{ijIJ} } \scal{{u_{ij}}^{KL} + 
v_{ijKL} }   \; & \; 
2 \Im\bigl[  \scal{Êu_{ij}{}^{IJ}  + v_{ijIJ} } u^{ij}{}_{KL}  \bigr]  \; \; 
\vspace{4mm} \\ 
2 \Im \bigl[ u^{ij}{}_{IJ} \scal{{u_{ij}}^{KL} + v_{ijKL} }  \bigr]  &  
\scal{{u^{ij}}_{IJ} - v^{ijIJ} } \scal{{u_{ij}}^{KL} - 
v_{ijKL} }  \end{array}\right) \CR &&\CR && \CR
&=& \left(\begin{array}{cc} \Scal{ \Re\bigl[2 S - \mathds{1} \bigr]} ^{-1} 
&\Scal{\Re\bigl[2 S - \mathds{1} \bigr]} ^{-1}\Im\bigl[2 S\bigr]  
\vspace{4mm} \\  \Im\bigl[2 S\bigr] \Scal{\Re\bigl[2 S - \mathds{1} 
\bigr]}^{-1} \hspace{4mm}& \hspace{2mm} \Scal{\Re\bigl[2 S - \mathds{1} \bigr]} ^{-1}\hspace{-1mm} 
+ \Im\bigl[2 S\bigr]\Scal{\Re\bigl[2 S - \mathds{1} \bigr]} ^{-1}    
\Im\bigl[2 S\bigr] \end{array}\right) \CR \label{GS} 
\eea
where we used \be 
\Im\bigl[ \scal{{u^{ij}}_{IJ} + v^{ijIJ} } \scal{{u_{ij}}^{KL} + 
v_{ijKL} } \bigr] = 0 \; \; , \label{RealUplusV}
\ee
to compute the first matrix, and where the symmetric matrix $S$ is defined such that 
\be 
\scal{u^{ij}{}_{IJ} + v^{ijIJ} } S^{IJ,KL} = u^{ij}{}_{KL} \; \; . 
\ee 
To prove the equality of the two matrices in (\ref{VTV}), one uses again (\ref{RealUplusV}) to show that 
\be 
2 \, \Im \bigl[ u^{ij}{}_{IJ} \scal{{u_{ij}}^{KL} + v_{ijKL} } \bigr] =  
\Im\, [ 2S]^{IJ,PQ}\,  \scal{{u^{ij}}_{PQ} + v^{ijPQ} } \scal{{u_{ij}}^{KL} 
+ v_{ijKL} } \; , 
\ee
and 
\bea 
& &  \!\!\!\!\!\!\!\!\!\!\!\!\!\!\!\!\!\!\!\!\!\!\!\!\!\!\!\!\!\!
\Re\bigl[2S- \mathds{1} \bigr]^{IJ,PQ}   \, \scal{{u^{ij}}_{PQ} + 
v^{ijPQ} } \scal{{u_{ij}}^{KL} + v_{ijKL} } \, = \CR
&=& \Re\bigl[ \scal{u^{ij}{}_{IJ}  - v^{ijIJ} } \scal{{u_{ij}}^{KL} + 
v_{ijKL} } \bigr] = \delta_{IJ}^{KL} \label{SR} \; , 
\eea
which establishes the equality for the first column in (\ref{GS}). The 
equality in the second column then follows by using the property that 
the matrix  is symmetric and symplectic. Identifying
\be 
H^{\bar \m\bar \n} \, \, \hat{=}\,  \, \Re\bigl[2S - \mathds{1}]^{IJ,KL} 
\; , \qquad  H^{\bar \m \bar \n}G_{\bar \n \m} \, \,  \hat{=} \, \, 
\Im\bigl[2S]^{IJ,KL} \; \; ,
\ee
one recovers the conventional form of the action (\ref{Svector})
as given in \cite{deWitNicolai}.

To investigate the couplings of the vectors to the fermions, we recall
from \cite{Christian} that the fermionic bilinears $W_{\i\j m}$ in
(\ref{PreGauss}) are determined by~\footnote{Readers should 
  keep in mind the different meanings of the letters $\i,\j,...$ and
  $i,j,...$ in this and other equations of this section (with apologies
  from the authors for the proliferation of different fonts!).}
\be\label{Wuv} 
W_{\i\j}^{IJ} = e^a_\i e^b_\j \Scal{u_{ij}{}^{IJ} 
O^{\scriptscriptstyle +}_{ab}{}^{ij} + v^{ijIJ}  
O^{\scriptscriptstyle -}_{ab \, ij} } \; \; , 
\ee
via the identification (analogous to (\ref{Darboux}))
\be 
W_{\i\j\, \m} \, \, \hat{=} \, \, \Im[ W_{\i\j}^{IJ} ] \; \; , 
\qquad W_{\i\j\, \bar \m} \, \, \hat{=} \, \, \Re[ W_{\i\j}^{IJ} ] \; \; . 
\label{Darboux1}
\ee
Here, $O^{\scriptscriptstyle +}_{ab}{}^{ij}$ and its complex conjugate
$O^{\scriptscriptstyle -}_{ab \, ij}$ are the fermionic bilinears 
defined in \cite{deWitNicolai} 
\be 
O^{\scriptscriptstyle +}_{ab}{}^{ij} =  
\bar \psi^i_{c} \gamma^{[c}Ê\gamma_{ab} \gamma^{d]}Ê \psi^j_{d} - 
\frac{1}{4}Ê  \bar \psi_{kc} \gamma_{ab} \gamma^c  \chi^{ijk}  - 
\frac{1}{(4!)^2}Ê\varepsilon^{ijklmnpq} \bar \chi_{klm} \gamma_{ab} 
\chi_{npq} \ \ .  
\ee
modulo normalisations (our coefficients here are chosen to agree 
with \cite{Christian}). By complex self-duality they satisfy
\be\label{csd}
O^{\scriptscriptstyle +}_{ab}{}^{ij} =  \frac{i}2  \varepsilon_{ab}{}^{cd} O^{\scriptscriptstyle +}_{cd}{}^{ij}\;\; , \quad 
O^{\scriptscriptstyle -}_{ab\,  ij} = - \frac{i}2  \varepsilon_{ab}{}^{cd} O^{\scriptscriptstyle -}_{cd\, ij}\;\;  . 
\ee  
These relations allow us to express the `timelike' components 
$W^{IJ}_{\zero\i}$ in terms of the purely spatial components 
$W^{IJ}_{\i\j}$, and thereby to recover the full fermionic Lagrangian
of the covariant formulation in terms of just the purely spatial 
components $W^{IJ}_{\i\j}$. 

After these preparations we return to the Lagrangian (\ref{Sv}), from which 
we read off the couplings of the vector fields to the fermions  
\begin{multline} 
   \varepsilon^{\i\j\k}\Im\bigl[ F_{\zero\i}^{IJ}  +  N^\l F^{IJ}_{\i\l}  
\bigr]\,  \Re\bigl[2 S - \mathds{1} \bigr]^{IJ,KL} \,
 \Re\bigl[W_{\k\l}^{KL} \bigr]\\*  + N \sqrt{h} \,   h^{\i\k}h^{\j\l} 
\, \Im\, [ F_{\i\j}^{IJ}]  \, \Scal{\Im \bigl[ W_{\k\l}^{IJ} \bigr] - 
\Im\bigl[2S\bigr]^{IJ,KL} \Re\bigl[W_{\k\l}^{KL} \bigr] }  \; \; . 
\end{multline}
Using the properties of $S^{IJ,KL}$ one computes that
\bea
&& \!\!\!\!\!\!\!\!\!\!\!\!\!\!\!\!
\,\Re\bigl[ 2S - \mathds{1} \bigr]^{IJ,KL} \, \Re\bigl[ u_{ij}{}^{KL} 
O^{\scriptscriptstyle +}_{ab}{}^{ij} + v^{ijKL}  
O^{\scriptscriptstyle -}_{ab \, ij} \bigr] \CR
&=& \, \Re\bigl[( 2 S - \mathds{1} )^{IJ,KL} \scal{u^{ij}{}_{KL} + v^{ijKL} }
O^{\scriptscriptstyle -}_{ab\, ij}\bigr]+ \Im[2S \bigl]^{IJ,KL} \,
\Im\bigl[  \scal{u_{ij}{}^{KL} + v_{ijKL} }O^{\scriptscriptstyle +}_{ab}{}^{ij} \bigr]\CR
&=& \, \Re\bigl[ \scal{ u_{ij}{}^{IJ} - v_{ijIJ}} 
O^{\scriptscriptstyle +}_{ab}{}^{ij} \bigr]+ \Im[2S \bigl]^{IJ,KL} \,
\Im\bigl[  \scal{u_{ij}{}^{KL} + v_{ijKL} }O^{\scriptscriptstyle +}_{ab}{}^{ij} \bigr] \; \; .
\eea
Invoking the complex self-duality of $O^{\scriptscriptstyle +}_{ab}{}^{ij}$ 
one recovers the manifest diffeomorphism invariant coupling 
\bea 
& &  e \,   e^{a\, \mu} e^{b\, \nu}  \Im [F_{\mu\nu}^{IJ} ]
\Scal{\Im \bigl[ \scal{u_{ij}{}^{IJ} - v_{ijIJ} } 
O^{\scriptscriptstyle +}_{ab}{}^{ij}  \bigr] - 
\Im\bigl[2S\bigr]^{IJ,KL} \Re\bigl[ \scal{u_{ij}{}^{KL} + v_{ijKL} } 
O^{\scriptscriptstyle +}_{ab}{}^{ij}  \bigr] }  \CR
 &&  \quad\qquad = \, e \, e^{a\, \mu} e^{b\, \nu} \Im [ F_{\mu\nu}^{IJ}]  \, 
\Re\bigl[2S- \mathds{1} \bigr]^{IJ,KL}\Im \bigl[\scal{u_{ij}{}^{KL} + 
v_{ijKL}}O^{\scriptscriptstyle +}_{ab}{}^{ij} \bigr]\CR
 &&  \quad\qquad = \, e \,   e^{a\, \mu} e^{b\, \nu} \, 
\Im\, [F_{\mu\nu}^{IJ} ] \,
\Im \bigl[S^{IJ,KL} (u^{-1})^{KL}{}_{ij} 
O^{\scriptscriptstyle +}_{ab}{}^{ij} \bigr]\; \; . 
  \eea
Next we consider the quartic terms in the fermions. They read 
\bea\label{WW1}
&& \frac{1}{2}N \sqrt{h} \, H^{\bar \m\bar \n} h^{\i\k}h^{\j\l}  
    W_{\i\j\, \bar \m}   W_{\k\l\, \bar \n}         \\
 && =  \,   \frac{1}{2} e \, 
h^{\i\k} h^{\j\l} e^a_\i e^b_\j e^c_\k e^d_\l \,  
\Re\bigl[\scal{u_{ij}{}^{IJ} + v_{ijIJ} } O^{\scriptscriptstyle +}_{ab}{}^{ij}
 \bigr] \Re\bigl[ 2S - \mathds{1} \bigr]^{IJ,KL}  \Re\bigl[  
 \scal{ u_{kl}{}^{KL} + v_{klKL} }  O^{\scriptscriptstyle +}_{cd}{}^{kl} 
\bigr] \nn
\eea
and 
\be\label{WW2}
- \frac{1}{2} N \sqrt{h} \, G^{mn}  h^{\i\k}h^{\j\l} W_{\i\j\, m}W_{\k\l\, n} 
\,  =  \,  - \frac{1}{4}e  \, h^{\i\k} h^{\j\l} 
 e^a_\i e^b_\j e^c_\k e^d_\l \, O^{\scriptscriptstyle -}_{ab\, ij} 
 O^{\scriptscriptstyle +}_{cd}{}^{ij} \; \; , 
\ee 
where in the last equation the dependence of $W_{\i\j m}$ on scalar fields 
in (\ref{Wuv}) is eliminated through the contraction with $G^{mn}$. 
Using (\ref{SR}) and
\begin{multline}
\Re \bigl[Ê2S-\mathds{1}\bigr]^{IJ,KL} \,
\scal{Êu_{ij}{}^{IJ} + v_{ij IJ} }\scal{Êu_{kl}{}^{KL} + v_{kl KL}} \\*
= \; \;  (u^{-1})^{IJ}{}_{ij} \Scal{ S^{IJ,KL} + u^{pq}{}_{IJ} v_{pq IJ} } 
  (u^{-1})^{KL}{}_{kl} \; \; , 
\end{multline}
we obtain
\begin{multline} 
\frac{1}{2}N \sqrt{h} \, H^{\bar \m\bar \n} h^{\i\k}h^{\j\l}  
    W_{\i\j\, \bar \m}   W_{\k\l\, \bar \n}   \;  = \,  \,\frac{1}{4}e \,h^{\i\k} h^{\j\l} e^a_\i e^b_\j e^c_\k e^d_\l 
\, \biggl( O^{\scriptscriptstyle -}_{ab\, ij}  O^{\scriptscriptstyle +}_{cd}{}^{ij} \,  \biggr . \\* \biggl .  
 + \; \frac{1}{2}Ê  \Big[ O^{\scriptscriptstyle +}_{ab}{}^{ij} 
 (u^{-1})^{IJ}{}_{ij} \Big(S^{IJ,KL} + u^{pq}{}_{IJ} v_{pqKL} \Bigr) 
  (u^{-1})^{KL}{}_{kl}  O^{\scriptscriptstyle +}_{cd}{}^{kl} \,   + \, \mbox{c.c.} \Big] \biggr) \; \; .
\end{multline}
The first term in parentheses cancels the (manifestly $E_{7(7)}$ invariant) 
expression (\ref{WW2}) --- as must be the case because any Lorentz
invariant extension of type $O^{\scriptscriptstyle +}{}^{ij} O^{\scriptscriptstyle -}_{ij}$ must necessarily vanish because of the opposite duality phases.
Altogether we have shown that the relevant part of the Lagrangian
agrees with the corresponding one from \cite{deWitNicolai} which
reads, in the present notations and conventions~\footnote{The conventions
  of \cite{deWitNicolai} are recovered with the identifications $\hspace{2mm} A_\mu^{\hspace{-4.7mm} {Ê\scriptscriptstyle \rm [2]} \hspace{ 2.5mm} IJ} \equiv \sqrt{2}\, Ê\Im[A_\mu^{IJ}] $, $\hspace{2.7mm}\psi_\mu^{\hspace{-5.2mm} {Ê\scriptscriptstyle \rm [2]} \hspace{ 3mm} i} \equiv \frac{1}{\sqrt{2}} \psi^i_\mu$, $\,~\hspace{2.5mm}~\chi^{\hspace{-4.9mm} {Ê\scriptscriptstyle \rm [2]} \hspace{2.2mm} ijk} \equiv \frac{1}{4} \chi^{ijk}$. The charge conjugation matrix of \cite{deWitNicolai} is 
  related to ours by,  $\hspace{2mm}Ê{\cal C}^{\hspace{-4.7mm} {Ê\scriptscriptstyle \rm [2]} \hspace{ 3mm} } \equiv i \, {\cal C}$, such that, for instance $O^{\scriptscriptstyle+}_{ab}{}^{ij} = - i 2 \sqrt{2}Ê\hspace{3.2mm}ÊO^{\hspace{-5.2mm} {Ê\scriptscriptstyle \rm [2]} \hspace{ 2.4mm}  \scriptscriptstyle+}_{ab}{}^{ij} $ and the complex self-duality convention is reversed.}
\begin{multline} \label{VF}
{\cal{L}}_{\scriptscriptstyle \rm VF} = \, \frac{e}{4}  \biggl( - \bigl[ 2S - \mathds{1} \bigr]^{IJ,KL} \Im[ F_{\mu\nu}^{IJ} ]^{\scriptscriptstyle -} \Im[F^{\mu\nu \, KL}]^{\scriptscriptstyle -}  - i Êe^{a \mu}Êe^{b \nu} O^{\scriptscriptstyle+}_{ab}{}^{ij} 
  (u^{-1})^{IJ}{}_{ij} S^{IJ,KL} \Im[Ê  F_{\mu\nu}^{KL} ] 
\biggl .    \\* \biggr .  
 +\; \frac{1}{8} O^{\scriptscriptstyle+}_{ab}{}^{ij} (u^{-1})^{IJ}{}_{ij} 
\big( S^{IJ,KL} + u^{pq}{}_{IJ} v_{pq KL} \big)
(u^{-1})^{KL}{}_{kl} O^{{\scriptscriptstyle+ \, } ab kl} + \mbox{c.c.} \biggr) \; \; . 
\end{multline}

Because the vector fields only appear through the field strength 
$F^{IJ}_{\i\j}$ in the BRST transformations of the fields, the Gaussian 
integration can be carried out for the complete Batalin--Vilkovisky action 
which will be discussed in the last section. The validity of the BRST master 
equation all along the process of carrying out the Gaussian path integrals 
to pass from one formalism to the other ensures the validity 
of the above formal argument, by fixing all possible ambiguities associated 
to the regularisation scheme.

\subsection{The classical $E_{7(7)}$ current}
\label{ClassicalCurrent}
A main advantage of the present formulation is that the $E_{7(7)}$ current
can be derived as a {\em bona fide} Noether current \cite{Christian}. 
It consists of two pieces
\be
J^\mu = J^{(1)\mu} + J^{(2)\mu} \; \; . 
\ee
Here the first piece $J^{(1)}$ does not depend on the vector fields and 
has the standard form as in any $\sigma$-model with fermions (see also
\cite{Kallosh}). The more important piece for our discussion here is
the second term $J^{(2)}$, which depends on the 56 electric and magnetic 
vector fields and is of Chern-Simons type; this part of the current does 
not exist off-shell in the usual formulation \cite{GaillardZumino}, 
where it would be given by a non-local expression on-shell. The 
current $J^\mu$ is an axial vector which defines the current 
three-form
\be
J = J^{(1)} + J^{(2)} \equiv \frac1{3!}\varepsilon_{\mu\nu\rho\sigma} J^\mu \, 
      dx^\nu_{\, \, \wedge }dx^\rho_{\, \,   \wedge} dx^\sigma \; \; ,
\ee
in terms of which the classical current conservation simply reads
$dJ=0$.

\vskip 3mm

Following the standard Noether procedure, the $E_{7(7)}$-current $J^\mu$ 
was computed in \cite{Christian} by an infinitesimal displacement along 
$\Lambda\in \e_{7(7)} $.  Under the $\SU$ subgroup of $E_{7(7)}$, 
$J^\mu$ decomposes into  $63$ components $\left( J^{\mu}\right)^{I}{}_{K}$ 
and $70$ components $\left( J^{\mu}\right)^{IJKL}$: 
\be\label{JDeco}
J^{\mu} (\Lambda) = \left( J^{\mu}\right)^{I}{}_{K} 
\Lambda_{I}{}^{K} + \left( J^{\mu}\right)^{IJKL}\Lambda_{IJKL}\; \; .
\ee
The easiest way to write the first piece $J^{(1)}$ is in terms of matrices:
\be\label{J1F}
 J^{(1)\mu} (\Lambda) =
				-\frac{1}{24}\text{tr}\scal{Ê\cV^{-1}\dR^\mu \cV \Lambda}Ê\; \; .
\ee
Here, we are using the matrix form of the scalar coset $\cV$ (\ref{VDefi}) 
and the matrices $\Lambda$ and $\dR^\mu$ in  $\e_{7(7)}$ that 
are defined as usual
\be
\Lambda \ \hat{=} \;  \left( \begin{tabular}{cc} $2 \delta_{[I}^{[M} \Lambda_{J]}{}^{N]}$ & $\Lambda_{IJOP} $\\  $\Lambda^{KLMN}$& $  - 2  \delta_{[O}^{[K} \Lambda_{P]}{}^{L]}
$\end{tabular}\right)  \, \, , \qquad 
\dR^\mu  \ \hat{=} \; \left(\begin{tabular}{cc} $ -  2 \delta_{[i}^{[m} \dR^{\mu\, n]}{}_{j]} $ & $\dR^{\mu}{}_{ijop}$\\  $\dR^{\mu\, klmn}$& $2 \delta^{[k}_{[o} \dR^{\mu\, l]}{}_{p]}$\end{tabular}\right) \, \, Ê.
\ee
The components $R^{\mu\, i}{}_j$ and $\dR^\mu{}_{ijkl}=\frac{1}{4!}
\ve_{ijklmnop} \dR^{\mu\, mnop}$ have the form
\bea\label{Rmu}
\dR^{\mu\, i}{}_j&\equiv&
2i\ve^{\mu\nu\sigma\rho} \Scal{Ê\bar{\psi}_{\nu}^i\gamma_\sigma \psi_{\rho j}  - \frac{1}{8}ÊÊ\delta^i _j  \, \bar{\psi}_{\nu}^k\gamma_\sigma \psi_{\rho k} }Ê
 +\frac{\sqrt{-g}}{8} \Scal{Ê\bar{\chi}^{ikl}\gamma^\mu \chi_{jkl}
-\frac{1}{8}Ê\delta^i_j \, \bar{\chi}^{klm}\gamma^\mu \chi_{klm} }Ê \CR
\dR^\mu{}_{ijkl}&\equiv&
\sqrt{-g} \hat{\cA}^\mu_{ijkl}-\frac{i}{2}\ve^{\mu\nu\sigma\rho}\Big(
\bar{\chi}_{[ijk}\gamma_{\sigma\rho}\psi_\nu{}_{l]}-\frac{1}{4!}\ve_{ijklmnop}\bar{\chi}^{mno}\gamma_{\sigma\rho}\psi_\nu{}^{p}
\Big) \; \; , 
\eea
where $\hat{\mathcal{A}}_\mu^{ijkl}$ is the supercovariant derivative of the scalar coset
\be
\hat{\mathcal{A}}_\mu^{ijkl} \equiv u^{ij}{}_{IJ}\partial_\mu v^{kl IJ} -v^{ijIJ}\partial_\mu u^{kl}{}_{IJ} - \bar{\psi}_{\mu}^{[i}\chi^{jkl]}-\frac{1}{4!}Ê\ve^{ijklmnop}\bar{\psi}_{\mu\, m}\chi_{nop} \ .
\ee
Since the second part $ J^{(2)\mu}$ of the current contains the $56$ vector fields 
$A_\i^m$, it necessarily lacks manifest covariance. With spatial indices $\i,\j=1,2,3$, 
it has the form:~\footnote{Note that the normalisation of the vector fields here
  differs from the one in \cite{Christian} by a factor 2.} 
\bea\label{JFirst}
  J^{(2) \k  }  &=&  -\frac12  \ve^{\i\j\k}A_\i^m \left(\partial_\zero A_\j^n -2N^\l F_{\l\j}^n\right)
\Omega_{pm}\Lambda_n{}^p + \sqrt{-g} h^{\j\k}h^{\i\l}A_\i^n\Big(
 G_{pm}  F^m_{\j\l} + W_{\j\l\, p} \Big)\Lambda_n{}^p \CR
 J^{(2)\zero } &=&   \frac14  \ve^{\i\j\k}A_\i^m F_{\j\k}^n  \Omega_{pm}\Lambda_n{}^p  \ .
\eea
Like for  $J^{(1)}$ in eq. (\ref{J1F}), the independent components 
of $J^{(2)}$ are provided by the $133$ independent 
components of $\Lambda$ within the $56\times 56$-matrix 
$\Lambda_n{}^p$. For instance, the 
time-like components in the $\su(8)$ basis are given by 
\bea J^{(2)\zero\, I }{}_J &=& \frac{i}{4} \varepsilon^{\i\j\k} \Scal{ÊA^{IK}_\i F_{\j\k\, JK} + A_{\i\, JK} F_{\j\k}^{JK} - \frac{1}{8} \delta^I_J \scal{ÊÊA^{KL}_\i F_{\j\k\, KL} + A_{\i\, KL} F_{\j\k}^{KL} }} \CR
J^{(2)\zero\, IJKL} &=& - \frac{i}{2}Ê\varepsilon^{\i\j\k}Ê\Scal{ÊA^{[IJ}_{\i} F_{\j\k}^{KL]} - \frac{1}{4!} \ve^{IJKLMNOP} A_{\i\, MN} F_{\j\k\, OP} } \eea
The space-like components admit a similar form that can straightforwardly be 
obtained from (\ref{JFirst}). However, the explicit expressions are rather 
complicated, and would not provide any further insight in this discussion.

As a next step, we want to rewrite the vector field part (\ref{JFirst}) 
in a way that allows a direct comparison with the current constructed 
in \cite{GaillardZumino}. A simple computation reveals the identity 
\cite{Christian}
\be \label{Noether23}
 J^{(2)\mu} =\Big(
\frac14\ve^{\mu\nu\rho\sigma}F_{\nu\rho}^m A_\sigma^n
+\frac12\delta^\mu_\k\ve^{\i\j\k}\partial_\i\left(A_\j^m A_\zero{}^n\right)
-\delta^\mu_\k \ve^{\i\j\k} A_\i^n \left(\X_j^m -\partial_\j A_\zero^m\right)
\Big)\Lambda_n{}^p \Omega_{pm} \, \, Ê.
\ee
where a spurious dependence in the component $A_\zero^m$ has been 
introduced, such that all the $A_\zero^m$ dependent terms add up to zero. 
This form of the current decomposes into three terms:
\begin{enumerate}
\item The first term in $J^{(2)\mu}$ is a Chern--Simons three-form. 
      It is manifestly diffeomorphism covariant in the usual sense.
\item The second term is a `curl',  and thus does not affect 
      current conservation.\footnote{Although the Noether procedure only 
      determines the current $J$ up to a `curl', this term 
  cannot be avoided in (\ref{Noether23}), because $A_\zero^m$ is not a 
  fundamental field in the duality invariant formulation.}
\item  The third term is proportional to the integrated equation of motion 
  of the vector field (\ref{EOM3}) with $\X_\k^m$ defined in (\ref{Xk}).
\end{enumerate}
Let us now recall the procedure of \cite{GaillardZumino} for obtaining the 
conserved current associated to the duality invariance of the equations 
of motion. The idea
is to supplement the manifestly covariant part of the current 
$ J^{(1)\mu}(\Lambda)$ by a further term $J_{\scriptscriptstyle \rm GZ}^{(2)\mu}$ in such
a way that the complete current (which we will henceforth refer to as 
the {\em Gaillard--Zumino current}) is conserved 
\be
\partial_\mu  \left(J^{(1)\mu}(\Lambda)+J_{\scriptscriptstyle \rm GZ}^{(2)\mu}(\Lambda)\right)=0 
\, \, ,
\ee
if the equations of motion are enforced. Therefore, it is clear that 
$J_{\scriptscriptstyle \rm GZ}^{(2)\mu}(\Lambda)$ is only defined up to a curl, and modulo 
terms proportional to the equations of motion. From the complete 
Noether current (\ref{Noether23}), we thus deduce 
\be\label{Noether3}
 J_{\scriptscriptstyle \rm GZ}^{(2)\mu}(\Lambda) =
\frac{1}{4}\ve^{\mu\nu\rho\sigma} A_\nu^m \, F_{\rho\sigma}^n  \, 
\Lambda_m{}^p \Omega_{pn}\, \, .
\ee
This Chern-Simons three-form exhibits manifest diffeomorphism 
covariance and it depends {\em only} on the $56$ vector fields,
unlike the current (\ref{JFirst}). The explicit form of $J_{\scriptscriptstyle \rm GZ}^{(2)\mu}(\Lambda)$
as given in \cite{Kallosh} is indeed equivalent to the decomposition of
(\ref{Noether3}) in a Darboux basis for the $56$ electromagnetic fields $A_\mu^m$ into $A_\mu^{\m}$ and $A_\mu^{\bm}$ (\ref{Darboux}). The usual covariant formulation of 
\cite{CremmerJulia} contains only the $28$ vector fields $A_\mu^{\m}$ off-shell, whereas
the dual fields $A_\mu^{\bm}$ are non-local functionals of
all the other fields satisfying the equations of motion.

In a non-trivial background, the Chern--Simons like component (\ref{Noether3})
is not globally defined in general. For a non-trivial connection, one must 
introduce a reference connection $\mathring{A}^m$, such that the one-form  
$A^m-\mathring{A}^m$ is gauge invariant (and so globally defined), and 
$\mathring{F}^m$ represents a non-trivial cohomology class in the given 
background. The background dependent extension of (\ref{Noether3}) is 
given from the Cartan homotopy formula as
\be \label{BackNoeth}
 J_{\scriptscriptstyle \rm GZ}^{(2)}(\Lambda) = - 
\frac{1}{2} \scal{ÊA^m - \mathring{A}^m}{}_{\wedge} 
\scal{ÊF^n + \mathring{F}^n }
\Lambda_m{}^p \Omega_{pn}\; \; .
\ee
By definition of the Cartan homotopy formula, it follows that the globally 
defined $E_{7(7)}$ current then suffers from a {\em classical anomaly}
\be 
d  J(\Lambda) = \frac{1}{2}\mathring{F}^m_{\hspace{3mm} \wedge}  \mathring{F}^n \, \, 
\Lambda_m{}^p \Omega_{pn}\; \; . \label{CSB}
\ee
Even without a general classification of the instanton backgrounds
that may occur in $\N=8$ supergravity, this result by itself already
shows how the continuous $E_{7(7)}$ symmetry can be broken in a non-trivial
background. When the gravity background is such that there is a non-trivial 
cohomology group
\be 
H^2(\mathds{Z}) \wedge H^2(\mathds{Z}) \rightarrow H^4(\mathds{Z}) 
\ee
and both $\mathring{F}^m$ and $\mathring{F}^m_{\hspace{3mm} \wedge}\mathring{F}^n$
define non-trivial cohomology classes in $H^2(\mathds{Z})$ and
$H^4(\mathds{Z})$, respectively,~\footnote{This is {\em not}
  the case for dyonic solutions in an asymptotically Minkowskian space-time: 
  even though $\mathring{F}^m$ is non-trivial for such solutions of Maxwell's
  equations, the product $\mathring{F}^m_{\hspace{2.5mm} \wedge}\mathring{F}^n$ is trivial.} 
the $\e_{7(7)}$ Ward identities will be broken in the background. 
In this case the 1PI generating functional $\Gamma$ evaluated on $E_{7(7)}$ 
transformed fields varies as (with appropriate normalisation)
\be 
\Gamma[g] = \Gamma[\mathds{1}] + 2\pi \,  \Omega_{mp}\,  g^p{}_n \, q^m q^n \; \; , 
\ee
with integer charges $q^m = \frac{1}{2\pi} \int \mathring{F}^m$. This `classical anomaly' is not affected by the Legendre transform, and the generating functional $W$ of connected diagrams transforms as $\Gamma$ with respect to $E_{7(7)}$ transformations. 
As a consequence, the generating functional $Z = \exp[\, Êi W \, ]$ will no longer be invariant 
under continuous $E_{7(7)}$ transformations, but only with respect to 
transformations $g \in E_{7(7)}(\mathds{Z})$. Such backgrounds appear 
for example in the classification of  \cite{GHinstantons}  as $\mathds{C}P^2$ 
and $S^2 \times S^2$ type spaces. One might therefore anticipate that 
$E_{7(7)}$ gets broken to a discrete subgroup when the path integral 
also includes a sum over such instanton contributions. 

However, we should caution readers that the status of `instanton solutions' 
in $\N=8$ supergravity is not clear by any means. Unlike the usual self-duality
constraint (which requires a Euclidean metric) the twisted self-duality 
constraint (\ref{EOM2}) contains an additional `imaginary unit' $J$,
and any $E_{7(7)}$ invariant Euclidean theory must therefore involve scalar 
fields parameterising a pseudo-Riemmanian symmetric space 
$E_{7(7)} / SU^*(8)_{\scriptscriptstyle \rm c} $ or $E_{7(7)} / 
SL(8)_{\scriptscriptstyle \rm c}$ such that the $\bf 28$ representation is 
real.\footnote{A positive definite `kinetic term' could then be recovered
   by decomposing $E_{7(7)} / SU^*(8)_{\scriptscriptstyle \rm c}  \cong \mathds{R}_+^* \times E_{6(6)} / Sp(4) \times \mathds{R}^{27}$ (respectively $E_{7(7)} / SL(8)_{\scriptscriptstyle \rm c}  \cong SL(8) / SO(8)  \times \mathds{R}^{35}$), and dualising 27 axionic scalars (respectively 35) into 2-forms, in analogy with the type IIB D-instantons \cite{GreenInstantons}.}
 It is thus doubtful whether a `Wick rotation' really makes sense, or 
whether one should instead look for {\em real} saddle points in a Lorentzian 
path integral. The second approach would still require to define the action 
in a non-trivial non-globally hyperbolic background. It is rather 
straightforward to modify the classical action similarly as (\ref{CSB}) 
such that the equations of motion are not modified, and such that the 
Lagrangian density is gauge invariant and transforms covariantly with respect 
to spatial diffeomorphisms. Nonetheless, this Lagrangian density 
transforms covariantly with respect to $D=4$ diffeomorphisms only up to 
terms linear in the equations of motion.

\subsection{Transformations in the symmetric gauge} 
\label{SymmetricGauge}
Under the combined action of local $SU(8)$ and rigid $E_{7(7)}$ the
56-bein transforms as 
\be\label{V'}
\V(x) \; \rightarrow \; \V'(x) = h(x) \V(x) g^{-1}  \; \; , \qquad
h(x)\in SU(8) \;, \;\; g\in E_{7(7)} \; \; . 
\ee
For the classical theory, one has the option of either keeping the
local $SU(8)$ with linearly realised $E_{7(7)}$, or fixing a gauge for 
the local $SU(8)$, retaining only the 70 physical scalar fields, whereby the 
rigid $E_{7(7)}$ becomes realised non-linearly. However, we are here 
concerned with the {\em quantised theory}, where the compatibility 
and mutual consistency of these two descriptions is not immediately
evident. Indeed, the $SU(8)$ gauge-invariant formulation of the theory 
may appear {\em not} to be well defined at the quantum level because the 
gauge $\su(8)$ Ward identity is anomalous at one loop due to the
contribution from the spin-$\frac12$ and spin-$\frac32$ fermions
\cite{Ferrara}. On the other hand, as shown by Marcus \cite{Marcus}, 
the rigid $SU(8)\subset 
E_{7(7)}$ left after gauge-fixing is {\em non-anomalous}, implying
the absence of anomalies for the {\em rigid $\su(8)$} current Ward identities 
in the gauge-fixed formulation of the theory. This is because the rigid 
$\su(8)$ symmetry acts linearly on the vector fields, whose chiral nature 
under $SU(8)$ implies that there is an extra contribution to the anomaly 
from the vector fields which precisely compensates the contribution from 
the fermion fields. From the path integral perspective, the main 
difference between those two kinds of $\su(8)$ Ward identities can 
be viewed as resulting from a redefinition of the 56 vector fields as 
\be 
A_\i^{IJ} \, \rightarrow \, \check{A}{}_\i^{ij} \equiv 
  u^{ij}{}_{IJ} A_\i^{IJ} + v^{ij\,IJ} A_{\i\, IJ}
\ee
that is, to the passage between objects transforming under rigid 
$E_{7(7)}$ and local $SU(8)$, respectively. According to the 
family's index theorem this change of variables does not leave the 
path integral measure for the vector fields invariant (because the action of $E_{7(7)}$ on the vector fields is chiral), and thus generates 
an anomaly. The results of \cite{Ferrara} and \cite{Marcus} are 
therefore perfectly consistent with each other, because the associated 
sets of Ward identities cannot be both free of anomalies. In the following
section we will present an explicit Feynman diagram computation of the 
vector field contribution to the $\su(8)$ anomaly. This explicit computation 
was not given in \cite{Marcus}, which relied on the formulation of 
$\N=8$ supergravity with only on-shell $E_{7(7)}$ and on 
arguments based on the family's index theorem.

We emphasize that the $\su(8)$ anomaly for the local $SU(8)$ gauge invariance 
is somewhat artificial because it can be compensated by the addition of 
an appropriate Wess--Zumino term for the $\SU$ components of the 
$E_{7(7)} / \SU$ vielbein $\V(x)$ \cite{deWit}. This procedure replaces 
the gauge $\su(8)$ anomaly by a corresponding anomaly of the $\su(8)$ 
current Ward identities (with the same coefficient). While restoring
local $SU(8)$, the latter by itself would break the rigid $E_{7(7)}$ 
symmetry, but for $\N=8$ supergravity this anomaly is cancelled in turn 
by the contribution from the vector fields! Consequently we anticipate
that our results can be re-obtained for the version of $\N=8$ supergravity
with local $SU(8)$ and linearly realised $E_{7(7)}$ such that both 
descriptions of the quantised theory are consistent, but a detailed 
verification of this claim remains to be done.

In order to set up the perturbative expansion of the quantised theory, we 
will nevertheless parameterise the symmetric space $E_{7(7)} / \SU$ with
explicit coordinates. We will consider as coordinates the scalar fields 
$\phi^{ijkl}$ in the ${\bf 70}$ of $SU(8)$, which parameterise a 
representative $\V(x)$ in the {\em symmetric gauge}, {\it viz.}
\be\label{gauge}
\V(x) \equiv  \exp \Phi(x)\; \hat{=} \; 
\exp\left( \begin{array}{cc}  0  &  \phi_{ijkl}(x) \vspace{0mm}  \\
  \phi^{ijkl}(x)  & 0 \end{array} \right) 
\ee
with $\Phi\in\e_{7(7)}\ominus\su(8)$ and the standard convention 
$\phi^{ijkl} = (\phi_{ijkl})^*$, (having fixed the $SU(8)$ gauge there is no need any more to distinguish
between $SU(8)$ and $E_7$ indices). After this gauge choice we are left with a 
{\em rigid} $E_{7(7)}$ symmetry, whose $SU(8)$ subgroup is realised 
linearly. The remaining rigid $E_7$ transformations require field 
dependent compensating $SU(8)$ rotation in order to maintain the chosen 
gauge (\ref{gauge}), and are therefore realised {\em non-linearly} 
on the 70 scalar fields. In this section, we work out these non-linear
transformations in more detail to set the stage for the implementation 
of the full nonlinear $E_7$ symmetry at the quantum level. For this 
purpose we adopt the following notational convention: for any two 
Lie algebra elements $X$ and $Y$ and any function $f(X)$ that is 
analytic at $X=0$, we abbreviate 
the adjoint action of $f(X)$ on $Y$ by 
\be
f(X) * Y \equiv  f({\rm ad}(X)) (Y)
\ee
Here, the right hand side is to be evaluated term by term in the Taylor
expansion, where the $n$-th order term $(\ad)^n(X)(Y)$ is the 
$n$-fold commutator $[X,[X,...[X,Y]...]]$. It is easy to check that
$f(X)*g(X)*Y= (fg)(X)*Y$. For the evaluation of the non-linear 
transformations the main tool is the Baker--Campbell--Hausdorf formula 
\bea\label{BCH}
\exp(X) \exp(Y) &=& 
\exp\Big( X + \td(X) * Y + \mathcal{O}( Y^2) \Big) \nn\\ 
&=& \exp\Big( Y + \td(-Y)*  X + \mathcal{O}(X^2)\Big) 
\eea
with 
\be 
\td(x) \equiv  \frac{x}{1 - e^{-x}} = 1 + \frac12 x + {\cal O}(x^2) \; \; . 
\ee
Accordingly we now consider an $E_7$ transformation with parameter $\Lambda$
in the $\bf{70}$ of $SU(8)$, {\it viz.}
\be
g \equiv \exp {\Lambda} \; \hat{=} \;  \left( \begin{array}{cc}  0  &  \Lambda_{ijkl}(x)    \vspace{0mm}  \\
  \Lambda^{ijkl}(x)  & 0 \end{array} \right) \; \; .
\ee
Then, by use of (\ref{BCH}),
\be\label{BCH1}
\exp \Phi \exp (-{\Lambda}) =
\exp\left( \Phi - \frac{\Phi/2}{\tanh(\Phi/2)}*  {\Lambda}
          - \frac12 \big[ \Phi , {\Lambda}\big] + {\cal O}({\Lambda}^2)
     \right)
\ee
where the odd piece is $[\Phi,\Lambda]\equiv \Phi*\Lambda\in\su(8)$.
Now we must choose the compensating $SU(8)$ transformation from the left 
so as to cancel the third term in the exponential. Using (\ref{BCH1})
and the second line of (\ref{BCH}), we obtain
\bea
\exp\big(\Phi + \delta\Phi\big) &=&
\exp\Big( \tanh(\Phi/2) *  {\Lambda} \Big)
\, \exp\Phi \, \exp(-{\Lambda}) \nn\\
&=&
\exp\left( \Phi - \frac{\Phi}{\tanh\Phi} * {\Lambda} 
         + {\cal O}({\Lambda}^2)\right)
\eea
or
\be
\sept\Phi \equiv \sept (\Lambda) \Phi = - \frac{\Phi}{\tanh\Phi} * {\Lambda} \; \; . 
\ee
In the same way one computes the supersymmetry transformation of the 
scalar fields and the non-linear modifications due to the compensating
$SU(8)$ rotations. Infinitesimally, local supersymmetry acts on the 
scalar fields by a shift along the non-compact directions with parameter
\be 
\cX  \, = \,  \left( \begin{array}{cc}  0  & \ { \overline{\epsilon}_{[i} 
\chi_{jkl]}} + \frac{1}{24} \varepsilon_{ijklmnpq} 
{ \overline{\epsilon}^{m} \chi^{npq}  } \ \vspace{2mm} \\
\ { \overline{\epsilon}^{[i} \chi^{jkl]}} + \frac{1}{24} 
\varepsilon^{ijklmnpq} { \overline{\epsilon}_{m} \chi_{npq}} & 0 \end{array} 
\right) 
\ee
Observing that this shift acts on $\V$ {\em from the left} (unlike the 
$E_7$ transformation in (\ref{V'}) which acts from the right), one
computes that, again using (\ref{BCH}), 
\be 
\exp\Big(\cX + \tanh(\Phi/2) *\cX \Big)\,\exp\Phi =
\exp\left(\Phi + \frac{\Phi}{\sinh\Phi} *  \cX
             + {\cal O}(\cX^2)\right) \; \; , 
\ee
whence the supersymmetry transformation of $\Phi$ is  
\be 
\susy \Phi \equiv\susy(\cX)\Phi = \frac{\Phi}{\sinh\Phi} * \cX    \; \; . 
\ee
By elementary algebra, this can be re-written in terms of an $E_{7(7)}$
transformation with parameter $\cX$,
\be
\susy(\cX)\Phi = - \sept(\cX)\Phi - \Phi\tanh(\Phi/2) * \cX \; \; . 
\ee

We can now check the commutation rules between supersymmetry and $E_{7(7)}$. 
The expectation is that the commutator of two such transformations gives 
rise to a supersymmetry transformation whose parameter $\epsilon'$ is 
obtained from the original supersymmetry parameter $\epsilon$ by acting 
on it with the compensating $SU(8)$ transformations induced by the 
action of $\e_{7(7)}$ on the fermions, \ie 
\be\label{E7S} 
[\sept({\Lambda})  , \susy(\epsilon) ] = \susy( \epsilon^\prime) \ , \qquad 
\epsilon^\prime \equiv\delta^{\su(8)}
\left(-  \tanh(\Phi/2) * {\Lambda} \right) \epsilon \; \; . 
\ee
At this point it is convenient to modify the supersymmetry variation 
by requiring the spinor parameter $\epsilon$ also to transform 
with respect to the induced $SU(8)$ transformation as
\be\label{E7epsilon} 
\sept \epsilon =  \delta^{\su(8)}\Big( \tanh(\Phi/2) *{\Lambda} \Big) \epsilon 
\ee
so $\epsilon$ transforms in the same way as the gravitino field $\psi$  under
the compensating $SU(8)$. As a consequence, the parameter $\cX$ 
in the adjoint representation simply transforms as
\be\label{septSigma}
\sept (\Lambda)\,\cX = \Big[ \tanh(\Phi/2) * {\Lambda} \, , \, \cX \Big]
\ee
which correctly reproduces the corresponding $\su(8)$ action 
in the $\bf 70$. As we will show below, with this extra compensating 
transformation we obtain
\be\label{E7SUSY} 
\big[\sept , \susy \big] \, \Phi = 0 \; ,
\ee
If (\ref{E7SUSY}) holds on the scalar fields, this commutator will also 
vanish on functions of $\Phi$ as well as on all other fields. Indeed, the  
only transformation that could still appear is a local Lorentz 
transformation which does not act on $\Phi$. To check the absence of
the latter we simply evaluate the above commutator on the vierbein field. 
While $\sept$ does not act on the vierbein, $\susy$ produces a term 
$\bar\epsilon^i\gamma^a\psi_{\mu i} +\bar\epsilon_i\gamma^a\psi^i_{\mu }$. However, with the extra
compensating transformation (\ref{E7epsilon}), this expression becomes a
singlet under the induced $SU(8)$ transformation, and therefore the 
commutator also vanishes on the vierbein. The main advantage 
of defining the transformation such that (\ref{E7SUSY}) is satisfied 
will become apparent when we discuss the quantum theory, because 
with (\ref{E7SUSY}) the BRST transformation will commute with $E_{7(7)}$, 
and this enables us to directly formulate the Ward identities
for the full non-linear $E_{7(7)}$ symmetry. 

In the remainder of this section we prove the key formula (\ref{E7SUSY}). 
It is more convenient to evaluate the commutator on $\exp\Phi$ rather 
than on $\Phi$ itself, because then all the non-linear terms appear
via the compensating $SU(8)$ transformation
\bea\label{dexp}
\sept (\Lambda) \exp\Phi &=& \big(\tanh (\Phi/2) * \Lambda\big)\exp\Phi 
         - \big(\exp\Phi \big) \, \Lambda \nn\\
\susy(\cX) \exp\Phi &=& 
\Big(\cX + \tanh(\Phi/2) * \cX\Big)\exp\Phi
\eea
from which we read off that
\be\label{SusyE7}
\susy(\cX) \exp\Phi = \sept(\cX)\exp\Phi + \cX\exp\Phi
                          + \big( \exp\Phi\big) \cX \;\;,
\ee
a relation that will be useful below. Using (\ref{septSigma}) we get
\bea\label{E7S1}
\big[ \sept(\Lambda)\; ,\; \susy(\cX)\big] \exp\Phi &=& \nn\\
&& \!\!\!\!\!\!\!\!\!\!\!\!\!\!\!\!\!\!\!\!\!\!\!\!\!
\!\!\!\!\!\!\!\!\!\!\!\!\!\!\!\!\!\!\!\!\!\!\!\!\!
\!\!\!\!\!\!\!\!\!\!\!\!\!\!
= \left(\big( \sept(\Lambda) \tanh(\Phi/2)\big)* \cX \right) \exp\Phi -
 \left(\big( \susy(\cX) \tanh(\Phi/2)\big)* \Lambda \right) \exp\Phi \nn\\
&& \!\!\!\!\!\!\!\!\!\!\!\!\!\!\!\!\!\!\!\!\!\!\!\!\!
 + \; \Big[ \tanh(\Phi/2)*\cX \; ,\, \tanh(\Phi/2)*\Lambda \Big] \exp\Phi
\nn\\
&& \!\!\!\!\!\!\!\!\!\!\!\!\!\!\!\!\!\!\!\!\!\!\!\!\!
 + \; \left( \tanh(\Phi/2)* \Big[ \tanh(\Phi/2)*\Lambda \,,\, \cX \Big]\right)
  \exp\Phi \; \; . 
\eea
To evaluate these terms further we need to make use of the closure property
\be\label{E7comm}
\big[ \sept (\Lambda_1), \sept (\Lambda_2) \big] \Phi =
   \Big[ \Phi \, ,\, \big[\Lambda_1, \Lambda_2 \big]\Big] \; \; , 
\ee
that is, the fact that the commutator of two compensated $E_{7(7)}$ 
transformations must close properly into $\su(8)$. This formula 
obviously extends to all functions $f(\Phi)$ which are expandable
in a power series. Observe that without the compensating $\su(8)$ 
transformation, the commutator $[\Lambda_1,\Lambda_2]$ in (\ref{E7comm})
would only act on $\Phi$ from the right (corresponding to the 
uncompensated $E_{7(7)}$ action), while its action from the left
is due to the compensating $\su(8)$. Using (\ref{dexp}) we now apply
this formula to $\exp\Phi$ to obtain
\bea
&& \left(\Big(\sept(\Lambda_1) \tanh(\Phi/2)\Big) * \Lambda_2 \right)\exp\Phi -
\left(\Big(\sept(\Lambda_1) \tanh(\Phi/2)\Big) * \Lambda_2 \right) \exp\Phi 
= \nn\\
&& \qquad = \, 
\Big[ \tanh(\Phi/2)*\Lambda_1 \, , \, \tanh(\Phi/2)*\Lambda_2 \Big] \exp\Phi
   - \big[ \Lambda_1 , \Lambda_2 \big] \exp\Phi \; \; . 
\eea
Modulo the difference between $\susy(\cX)$ and $\sept(\cX)$,
cf. (\ref{SusyE7}), this formula allows us to rewrite the right 
hand side of (\ref{E7S1}) as
\be\label{rhs}
 \big[\cX \, , \, \Lambda \big] \exp\Phi
 + \; \left(\tanh(\Phi/2)* \Big[ \tanh(\Phi/2)*\Lambda \,,\, \cX \Big]
   \right)\exp\Phi \; \; . 
\ee
Now exploiting (\ref{SusyE7}) in the form
\bea
\susy(\cX) \exp(n\Phi) &=& \sept(\cX) \exp(n\Phi) + 
        \cX \exp(n\Phi) + \exp(n\Phi) \cX + \nn\\
&& + \; 2\, \sum_{1\leq m\leq n-1} \exp(m\Phi)\cX \exp((n-m)\Phi) 
\eea
and expanding $\tanh(\Phi/2)$ as a formal power series in $\exp\Phi$
we get 
\be
\susy(\cX)\tanh(\Phi/2) = \sept(\cX)\tanh(\Phi/2) + \cX 
   - \tanh(\Phi/2)\,\cX\tanh(\Phi/2) \; \; , 
\ee
(as can also be checked by expanding the formal power series around
$\Phi = 0$). Therefore
\bea
- \susy(\cX) \tanh(\Phi/2) * \Lambda &=&
- \sept(\cX) \tanh(\Phi/2) * \Lambda  \; +\nn\\
&& \!\!\!\!\!\!\!\!\!\!\!\!\!\!\!\!\!\!\!\!\!\!\!\!\!\!\!\!\!\!\!\!\!\!\!
- \; \big[ \cX \, ,\, \Lambda\big]  
 + \; \tanh(\Phi/2)* \Big[\cX \, ,\,  \tanh(\Phi/2)*\Lambda \Big] \; \; . 
\eea
Acting with this expression on $\exp\Phi$ we see that these terms cancel
the ones in (\ref{rhs}), which proves the key formula (\ref{E7SUSY}).

Finally, it is straightforward to see that the remaining gauge
symmetries trivially commute with the non-linear action of $\sept$. 
Combining all gauge symmetries into a single BRST transformation
with generator $s$ in the usual way we therefore see that the relation 
(\ref{E7SUSY}) extends to the more general statement~\footnote{For
   anti-commuting $\e_{7(7)}$ parameters, this relation becomes an
   anti-commutator: $\{ \sept, s\} =0$.}
\be\label{E7BRST}
[ \sept , s ] = 0 \; \; . 
\ee
Consequently, at the classical level, the BRST (gauge) transformations 
can be completely disentangled from the non-linear action of 
$E_{7(7)}$. In the remaining sections it will be our task to elevate
this statement to the full quantum theory.

\section{The $SU(8)$ anomaly at one loop}
\label{Anomaly1loop}

As a first application of the formalism developed in the foregoing
sections, we now present a Feynman diagram computation of the 
$SU(8)$ anomaly considered long ago by very different methods.
In \cite{Marcus}, N.~Marcus pointed out the absence of rigid $SU(8)$ 
anomalies for $\N=8$ supergravity at one loop; the cancellation is based 
on the following identity 
\be\label{Marcus}
3 \times {\rm tr}_{\bf 8} X^3   - 2 \times {\rm tr}_{\bf 28}  X^3 + 
1 \times {\rm tr}_{\bf 56} X^3 = 
\Scal{ 3 \times 1 - 2 \times 4 + 1 \times 5 }Ê{\rm tr}_{\bf 8} X^3  = 0\,\,, 
\ee 
where $X$ is any $\su(8)$ generator, and  
where the first and third contributions are due to the eight gravitinos
and the 56 spin-$\frac12$ fermions of $\N=8$ supergravity, while the 
middle contribution is due to the  28 chiral vectors. 
In this section we will not consider the fermionic triangle diagrams which 
can be obtained by standard methods, but concentrate on the vector fields,
that is, the middle term in (\ref{Marcus}). The formalism of this paper 
makes possible 
(for the first time) a full fledged Feynman diagram calculation because it 
allows for an off-shell realisation of the chiral properties of the 
vector fields and their interactions under $SU(8)$. At the end of this 
section and in the following sections we will extend these considerations 
to the full $E_{7(7)}$ current, where we will encounter a non-linear 
variant of the familiar linear anomaly, with three currents and (in
principle) any number of scalar field insertions. We will also present 
arguments showing that this results extends to all loop orders.

Accordingly, our first aim will be to compute correlators with the insertion 
of three $SU(8)$ current operators obtained by restricting the $E_{7(7)}$ 
current to its $SU(8)$ subgroup. Now it is known (for linearly realised 
symmetries) that the anomaly involves a trace of the form (with Lie
algebra generators $X_\un, X_\deux, X_\trois$)
\be\label{rank3inv} 
\trace \{ X_\un , X_\deux \} X_\trois  \; \; . 
\ee
However, there is no invariant symmetric tensor of rank three in the $\bf{56}$ 
or the adjoint of $E_{7(7)}$, and hence {\em a priori} also none for its
$SU(8)$ subgroup (in these representations) so readers may wonder how one 
could get an anomaly at all. It is here that the distinction between a 
linearly realised symmetry and a non-linearly realised one makes all 
the difference. Namely, as the explicit calculation below will show, the 
relevant trace involves the complex structure tensor $J^m{}_n$ 
as an extra factor, so (\ref{rank3inv}) is replaced by
\be\label{rank3invJ} 
\trace J \{ X_\un , X_\deux \} X_\trois \; \; . 
\ee
This extra factor (which one might think of as being analogous to the 
insertion of a $\gamma_5$) breaks the manifest symmetry from $E_{7(7)}$ to
$SU(8)$, and at the same time allows for the appearance of chirality,
and hence a non-vanishing trace (effectively replacing the vector-like 
${\bf{56}}= {\bf{28}}\oplus \overline{\bf{28}}$ by the chiral $\bf{28}$ in 
the trace). Nevertheless, the $E_{7(7)}$ symmetry is still present, 
but necessarily non-linear.

\subsection{Feynman rules}

With these remarks we can now proceed to the actual computation.~\footnote{We 
  shall occasionally point out similarities of the present computation 
  with the familiar $\gamma_5$ anomaly; 
  readers may therefore find it useful to consult the textbooks 
  \cite{Ramond,Bertlmann,Weinberg,Weinberg2} for further information 
  on this well known topic.}   
We first work out the propagators by starting from the gauge-fixed 
kinetic term for the vector fields
\be\label{quadvect} 
\L_0 =  \frac{1}{2}Ê \Omega_{mn}  \varepsilon^{\i\j\k} \partial_\zero A_\i^m 
 \partial_\j A_\k^n - 
\frac{1}{2}G_{mn} ( \delta^{\i\k}\delta^{\j\l} - \delta^{\i\l} \delta^{\j\k}) 
\partial_\i A_\j^m \partial_\k A_\l^n + b_m \partial_\i A_\i^m \; \; , 
\ee
which is obtained from (\ref{Action2}) by retaining only the parts quadratic
in the fields. Furthermore, in the linearised approximation, we 
set $h_{\i\j} = \delta_{\i\j}$ in (\ref{quadvect}) and expand the 
scalar fields about a given background $\mathring{\Phi}$, so the 
metric $G_{mn}=G_{mn}(\mathring{\Phi})$ \footnote{In this section we 
  write $G_{mn}$ instead of using the (perhaps more appropriate)
  notation $\mathring{G}_{mn}\equiv G_{mn}(\mathring{\Phi})$, 
  since $G_{mn}(\Phi)$ does not appear and the notation is 
  therefore unambiguous. Except in (\ref{CommDirac}) and (\ref{FFcomm}), 
  we refrain from using boldface latters for the spatial components of 
  four-vectors, as it should be clear from the context which is meant.} 
also becomes constant 
(with $\mathring{\Phi} =0$ we have ${G}_{mn}= \delta_{mn}$). 
Going to momentum space, the quadratic operator to be inverted is 
\be 
\Delta^{-1}(p) = 
\left( \begin{array}{cc}\ \Omega_{mn} \varepsilon^{\i\j\k}  
p_\zero p_\k + G_{mn} ( \delta^{\i\j} p^2- p^\i p^\j ) \ & \ \ i p^\i  
\delta^n_m \ \ \vspace{2mm} \\ -  i p^\j  \delta^m_n & 0 \end{array}\right) \; \; , 
\ee
with $p^2 \equiv \delta^{\i\j}p_\i p_\j $. The vector propagator is
therefore 
\be\label{Propagator} 
\Delta(p) =  \frac{1}{p^2}\left( \begin{array}{cc}\ 
\frac{\Omega^{mn} \varepsilon_{\i\j\k}  p_\zero p^\k - 
G^{mn} ( \delta_{\i\j} p^2 - p_\i p_\j )}{{p_\zero}^2 - p^2 + i\varepsilon }  
\ & \ \ i p_\i  \delta^m_n \ \ \vspace{3mm} \\ -  i p_\j  \delta^n_m & 0 
\end{array}\right) 
\;\;   ;
\ee
it is a $(4\times56)$ by $(4\times 56)$ matrix, with three spatial 
directions and the fourth component corresponding to the Lagrange
multipliers $b_m$ which enforce the condition $\partial_\i A_\i^m=0$.
The propagating spin-1 degrees of freedom correspond to the residues of 
the poles of the propagator at $p_\zero = \pm | p | $. There is no pole in 
the off-diagonal components mixing $b_m$ and $A^m_\i$, and the  residue 
is given by 
\be 
2 |p| \res\, ( \Delta )\big|_{p_\zero= |p|} = 
\Omega^{mn} \varepsilon_{\i\j\k}  \hat{p}^\k  -  G^{mn} ( \delta_{\i\j}  
- \hat{p}_\i \hat{p}_\j ) \; \; , 
\ee
where $\hat{p}_\i \equiv p_\i/|p|$. An important difference between 
(\ref{Propagator}) and the usual covariant propagator in four dimensions 
is that (\ref{Propagator}) contains terms which are {\em odd} under parity 
(for which $p^\i\rightarrow -p^\i$ and $p_\zero \rightarrow p_\zero$). 
It is these terms, together with the parity odd vertices to be given below, 
which introduce the extra factor $J^m{}_n$ into the traces, and
hence can contribute to chiral anomalies, even if only vector fields 
circulate in the loop.

We can rephrase these results in canonical language.
Consider the free quantum field 
\be\label{QF} 
A_\i^m(x) \equiv \int \frac{d^3 p}{(2\pi)^\frac{3}{2}} 
\frac{1}{\sqrt{2 |p|}} \sum_{\sigma} \Scal{Êe^{-i x \cdot  p} 
e^*{}_\i^m(\sigma,p) a(\sigma,p) +  e^{i x \cdot  p} e_\i^m(\sigma,p) 
a^\dagger(\sigma,p)} \ ,  
\ee
where $a^\dagger(\sigma,p)$ and $a(\sigma,p)$ are creation and annihilation 
operators of asymptotic free particles of momentum $p$ and helicity $h(\sigma)
= \pm 1 $, and 56 $SU(8)$ quantum numbers $\sigma$ (we anticipate in this 
notation that $\sigma$ determines $h$ by (\ref{Jhelicity})),
\be 
\bigl[Êa(\sigma,p) , a^\dagger(\sigma',q) \bigr] = 
\delta_{\sigma \sigma'} \delta^\ord{3}(p-q) \ . 
\ee
In order for the operator algebra to reproduce the propagator 
(\ref{Propagator}) 
\be\label{Propagator1}
\Bigl< 0 \Big| T\Bigl\{ÊA_\i^m(x) \, A_\j^n(y) \Bigr\}Ê\Big| 0 \Bigr> = - i \int \frac{d^4 p}{(2\pi)^4}Ê\frac{ e^{i p \cdot ( x-y)} }{ÊÊ{p_\zero}^2 - p^2 + i\varepsilon } \Scal{Ê\Omega^{mn} \varepsilon_{\i\j\k}  \frac{p_\zero}{|p|} \hat{p}^\k - 
G^{mn} \scal{Ê \delta_{\i\j}  - \hat{p}_\i \hat{p}_\j }} \ ,  \ee
the polarisation vectors $e_\i^m(\sigma,p)$ and their complex conjugates 
$e^*{}_\i^m(\sigma,p)$ must satisfy 
\be 
\sum_{\sigma} e_\i^m(\sigma,p) e^*{}_\j^n(\sigma,p) = - \Omega^{mn} \varepsilon_{\i\j\k}  \hat{p}^\k  +  G^{mn} ( \delta_{\i\j}  - \hat{p}_\i \hat{p}_\j ) \ .  \label{ProjTwS} 
\ee
As usual, the polarisation vectors are transverse
\be 
\hat{p}^\i \, e_\i^m(\sigma,p)  = 0\; \; . 
\ee
With the convention 
\be
\varepsilon_{\i}{}^{\j\k} \hat{p}_\k \, e_\j^m(\sigma,p)    = i h(\sigma) e_\i^m(\sigma,p) \; \; , 
\ee
it follows from (\ref{ProjTwS}) that the polarisation vectors must satisfy
in addition
\be 
J^m{}_n e^n_\i( \sigma , p)  = i h(\sigma)  e^m_\i( \sigma , p) \ , 
\label{Jhelicity} 
\ee
with the `complex structure' tensor $J^m{}_n\equiv J^m{}_n(\mathring{\Phi})$, 
see (\ref{J}). With this extra constraint, there are only 56 independent 
polarisations, so $\sigma$ runs from $1$ to $56$. The linearised equations 
of motion are then satisfied with a zero gradient 
$\partial_\i A_\zero^m  = 0$ in (\ref{EOM3}),  
\be 
\partial_\zero A_\i^m  = \varepsilon_\i{}^{\j\k} J^m{}_n \partial_\j A^n_\k \ , \ee
such that the action of the Lorentz group on $A_\i^m$ is the same as in the standard formulation of the free theory in the Coulomb gauge. It follows that the 56 creation operators $a^\dagger(\sigma,p)$ are the same as in the standard formulation of the free theory, and the 28 states of helicity $h=1$ transform in the ${\bf 28}$ of $SU(8)$, whereas the 28 states of helicity $h=-1$ transform in the $\bf \overline{28}$ of $SU(8)$, as required by (\ref{Jhelicity}). 

Note that because of (\ref{ProjTwS}), the free quantum field $A_\i^m(x)$ 
does not commute with itself at equal time, but satisfies instead  
\be 
\bigl[ÊA_\i^m(x^\zero,{\bf{x}}) , A_\j^n(x^\zero,{\bf{y}}) \bigr]Ê= 
i \Omega^{mn} 
\varepsilon_{\i\j\k} \frac{\partial\, }{\partial x^\k} \, 
\frac{1}{4\pi | {\bf{x}}- {\bf{y}}|} \ . \label{CommDirac} 
\ee
This equal time commutator could be derived alternatively from the Dirac 
quantisation of the theory in the Coulomb gauge, with the second class 
constraints \footnote{As in the conventional formulation, the Poisson 
bracket of the first class Coulomb constraint $\partial_\i \Pi^\i_m \approx 0$
and the Coulomb gauge constraint $\partial^\i A_\i^m \approx 0 $ is 
non-degenerate, and they altogether define a set of second class constraints.}
\be 
\Pi_m^\i - \frac{1}{2} \Omega_{mn}  \varepsilon^{\i\j\k} \partial_\j A_\k^n \approx 0 \ , 
\qquad \partial^\i A_\i^m \approx 0 \ . 
\ee
The decomposition of the canonical momentum $\Pi_m^\i$ in the Darboux basis 
only coincides with the definition  (\ref{VaMo}) of the canonical momentum 
$\Pi^{\bar \m\, \i}$ in the conventional formulation of the theory 
(\ref{PosGauss}) up to a factor $2$. Although the canonical Poisson brackets 
therefore differ by a factor $2$ in the two formulations, the Dirac brackets 
are equivalent. The commutation relation (\ref{CommDirac}) is consistent 
with causality, because
\be\label{FFcomm}
\bigl[ F^m_{\i\j}(x^\zero,{\bf{x}}) , F^n_{\k\l}(x^\zero,{\bf{y}})\bigr]  
= 2i \Omega^{mn} \varepsilon_{\i\j[\k} \partial_{\l ]} 
\delta^\ord{3}( {\bf{x}} - {\bf{y}}) \ , 
\ee
as follows directly from (\ref{CommDirac}), and therefore gauge-invariant 
operators commute at space-like 
separation $( x^\zero - y^{\zero} )^2 < |{\bf{x}} - {\bf{y}}|^2 $. 

The cubic vertex defining the couplings of the $E_{7(7)}$ current to 
the vector fields can be obtained from the quadratic action by adding
to (\ref{quadvect}) terms with source fields $B_\mu^{\,m}{}_n$ coupling 
to the conserved $E_{7(7)}$ current, such that the latter is re-obtained by
taking the derivative with respect to the source fields and then setting 
them equal to zero. Here we will restrict attention to the $\su(8)$ part
of the full $E_{7(7)}$ current, for which the source $B_\mu^{\, m}{}_n$ 
leaves the background metric $G_{mn}$ invariant:  
\be 
B_\mu^{\, p}{}_m {G}_{pn} + 
B_\mu^{\, p}{}_n {G}_{pm} = 0  \label{antisymmetric}
\ee
(For ${G}_{mn}= \delta_{mn}$ this just means that $SU(8)$ is realised
by {\em anti-symmetric} matrices in the real basis of the $\bf{56}$
representation of $E_{7(7)}$). As is well known, the introduction of 
such sources corresponds to formally covariantising the action 
(\ref{quadvect}) with respect to {\em local} $SU(8)$, such that 
(\ref{quadvect}) is replaced by the density
\begin{multline} 
\L_0[B] = \frac{1}{2}Ê \Omega_{mn}  
\varepsilon^{\i\j\k} \scal{Ê\partial_\zero A_\i^m + 
B_\zero^{\, m}{}_p A_\i^p} \scal{ 
 \partial_\j A_\k^n + B_\j^{\, n}{}_q A^q_\k }Ê\\* -  
\frac{1}{2}G_{mn} ( \delta^{\i\k}\delta^{\j\l} - \delta^{\i\l} \delta^{\j\k}) 
\scal{Ê\partial_\i A_\j^m  + B_\i^{\, m}{}_p A_\j^p} 
\scal{Ê\partial_\k A_\l^n + B_\k^{\, n}{}_q A_\l^q }Ê+ b_m 
\scal{Ê\partial_\i A_\i^m + B_\i^{\, m}{}_n A_\i^n } 
\label{SourceLagrange}
\end{multline}
(in fact, dropping the restriction (\ref{antisymmetric}) this action  
becomes covariant with respect to local $E_{7(7)}$, as required for 
a study of the full $E_{7(7)}$ current, cf. section~\ref{BRSTCurrent}).
For the fermion fields, the $SU(8)$ tensor structure factorises out, 
and the vertex associated to one $SU(8)$ current insertion just has
the expected structure $\propto (1\pm i \gamma_5) \gamma^\mu$. For vector
fields, on the other hand, the Lorentz and $\su(8)$ tensor structures 
are {\it  a priori} entangled for the vertices computed from 
(\ref{SourceLagrange}).\footnote{The momentum dependence of the 3-point vertex can be derived 
  in the usual way \cite{Ramond} by writing the corresponding terms from 
  (\ref{SourceLagrange}) in momentum space and symmetrising in the
  internal legs involving the quantum fields $A_\i^m$ (not forgetting 
  the antisymmetry condition  (\ref{antisymmetric})).} 
Nevertheless, for correlation functions of 
$SU(8)$ currents only, the two tensor structures can be disentangled 
by using the property that the only tensors appearing in the trace are 
the $SU(8)$ invariant tensors $\delta^m_n$ and $J^m{}_n$; these can
be diagonalised according to the decomposition of the $E_{7(7)}$ 
representation ${\bf 56 }\cong {\bf 28}Ê\oplus \overline{\bf 28}Ê$ of $SU(8)$.
The calculation shows that all the $\su(8)$ Lie algebra generators $X$'s 
can be moved to the left  such that the vertex for linking an $\su(8)$-current $J^\mu$ and a chiral boson $A_\i^m$ with incoming momenta $p$ and $k$ respectively to a chiral boson $A_\j^n$ with outgoing momentum $p+k$
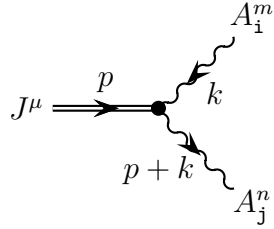
\begin{center}
\scalebox{1}{
\psset{xunit=5pt,yunit=5pt,linewidth=0.8pt,dotsize=2pt}
\def\zigzag{\psline[linearc=.1]{-}(0,0)(.5,1)(1.5,-1)(2,0)}

\begin{pspicture}(20,18)
\psline[doubleline=true]{-}(6,8)(10,8)
\psline[doubleline=true,arrowsize=0.24]{->}(2,8)(7,8)
\rput{*0}(0,8){$J^\mu$}
\rput{*0}(6,10){$p$}

\rput{45}(8.6,-5)
{
\multips(10,8)(2,0){4}{\zigzag}
\psline[arrowsize=0.24]{->}(14,8)(13,8)
\rput{*0}(20,8){$A_\i^m$}
\rput{*0}(14,6){$k$}
}

\rput[r]{315}(-2.7,9)
{
\multips(10,8)(2,0){4}{\zigzag}
\psline[arrowsize=0.24]{->}(13.5,8)(14.5,8)
\rput{*0}(20,8){$A_\j^n$}
\rput{*0}(13,5){$p+k$}
}

\qdisk(10,8){0.1}
\end{pspicture}
}
\end{center}
 is effectively given by 
\bea\label{Upsilon} 
& & \Upsilon^\zero(k+p,k)  = 
\left( \begin{array}{cc}\  i \Omega_{mn} \varepsilon^{\i\j\k}  
\scal{Êk_\k + \frac{1}{2} p_\k }  \ & \ \ 0 \ \ \vspace{2mm} \\ 
 0 & 0 \end{array}\right) ,
\CR &&
 \Upsilon^\k(k+p,k)  = \CR
  &&\hspace{-15mm} 
\left( \begin{array}{cc}\ i\Omega_{mn} \varepsilon^{\i\j\k}  
\scal{Êk_\zero + \frac{1}{2} p_\zero } +i  G_{mn} \Scal{ ( 2 k^\k + p^\k ) 
\delta^{\i\j} - \delta^{\k\i}Ê( k^\j + p^\j ) - \delta^{\k\j} k^\i }  \ & \ 
- \delta^{\k\i} \delta^n_m \ \ \vspace{2mm} \\ \delta^{\k\j}   
\delta^m_n & 0 \end{array}\right) \; \; , 
\eea
where the bottom-left component gets a positive sign because of 
(\ref{antisymmetric}).
 The notation we use here is {\em formally} very 
similar to the one used for the familiar fermionic vertices. The vertices 
$\Upsilon^\mu$ are analogous to the $(1\pm i \gamma_5) \gamma^\mu$ matrices 
that appear in the corresponding computation of the anomalous fermionic 
triangle diagram. This analogy is for instance reflected in the identity
\be 
- i p_\mu \Upsilon^\mu(k+p,k)  =  \Delta^{-1}(k+p)  
- \Delta^{-1}(k) \label{FeynWard} \; \; , 
\ee
which is analogous to the (trivial) identity $ \ba p = ( \ba k + \ba p ) 
- \ba k $, and will be similarly useful to cancel propagators in the
diagrams and thereby simplify them. However, in contradistinction to the 
case of fermion fields which are governed by a first order kinetic term, 
(\ref{SourceLagrange}) is quadratic in $B_\mu^{\, m}{}_n$ and thus the 
insertion of more than one current requires the consideration of 
contact terms absent in the fermionic triangle. The corresponding vertices $\cR^{\mu \nu }$
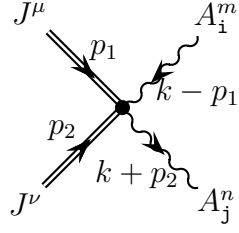
\begin{center}
\scalebox{1}{
\psset{xunit=5pt,yunit=5pt,linewidth=0.8pt,dotsize=2pt}
\def\zigzag{\psline[linearc=.1]{-}(0,0)(.5,1)(1.5,-1)(2,0)}

\begin{pspicture}(20,18)
\rput{315}(-2.7,9.5)
{
\psline[doubleline=true]{-}(6,8)(10,8)
\psline[doubleline=true,arrowsize=0.24]{->}(2,8)(7,8)
\rput{*0}(0,8){$J^\mu$}
\rput{*0}(6,10){$p_1$}
}

\rput{45}(8.3,-5)
{
\psline[doubleline=true]{-}(6,8)(10,8)
\psline[doubleline=true,arrowsize=0.24]{->}(2,8)(7,8)
\rput{*0}(0,8){$J^\nu$}
\rput{*0}(6,10){$p_2$}
}

\rput{45}(8.6,-5)
{
\multips(10,8)(2,0){4}{\zigzag}
\psline[arrowsize=0.24]{->}(14,8)(13,8)
\rput{*0}(20,8){$A_\i^m$}
\rput{*0}(15,5){$k-p_1$}
}

\rput[r]{315}(-2.7,9)
{
\multips(10,8)(2,0){4}{\zigzag}
\psline[arrowsize=0.24]{->}(13.5,8)(14.5,8)
\rput{*0}(20,8){$A_\j^n$}
\rput{*0}(14,5.5){$k+p_2$}
}

\qdisk(10,8){0.1}
\end{pspicture}
}
\end{center}
do not depend on the momenta:
\be\label{R} 
\cR^{\zero \k } = \cR^{\k \zero} = 
\left( \begin{array}{cc}\  \frac{1}{2} \Omega_{mn} 
\varepsilon^{\i\j\k}   \ & \ \ 0 \ \ \vspace{2mm} \\ 0 & 0 \end{array}\right) 
\ , \qquad 
\cR^{\k\l} = \left( \begin{array}{cc}  G_{mn} \Scal{ \delta^{\k\l}  
\delta^{\i\j} - \delta^{\k\j}Ê\delta^{\l\i}Ê}  \ & \  0  \ \ \vspace{2mm} 
\\ 0  & 0 \end{array}\right) \; \; . 
\ee
The vertices (\ref{Upsilon}) and (\ref{R}) satisfy 
\be\label{Vert}
\Upsilon^\mu(k+p,k)^T = - \Upsilon^\mu(-k,-k-p) \;\;, \quad 
\cR^{\mu\nu} = \big(\cR^{\nu\mu}\big)^T \; \; , 
\ee
where transposition is defined in the matrix notation, and includes the interchange of the index pairs
$(\i,m) \leftrightarrow (\j,n)$ of the top left component. Furthermore,
\be  
i p_\nu \cR^{\mu\nu} = \Upsilon^\mu(q , l + p  ) - 
\Upsilon^\mu(q,l) \label{ConWard}Ê\; \; , 
\ee
for any choice of momenta $l^\mu$ and $q^\mu$.
The contribution to the vacuum expectation value of three currents of 
the one-loop diagrams with vector fields circulating in the loop is  
encoded in the triangle diagram
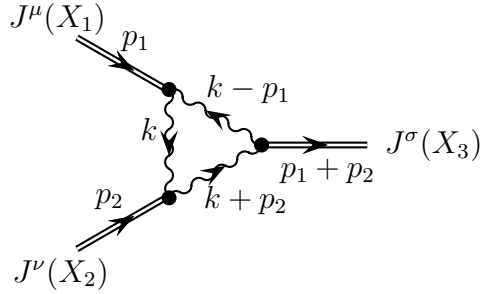
\begin{center}
\scalebox{1}{
\psset{xunit=5pt,yunit=5pt,linewidth=0.8pt,dotsize=2pt}
\def\zigzag{\psline[linearc=.1]{-}(0,0)(.5,1)(1.5,-1)(2,0)}
\begin{pspicture}(37,20)
\rput{330}(-2.7,11)
{
\psline[doubleline=true]{-}(6,8)(10,8)
\psline[doubleline=true,arrowsize=0.24]{->}(2,8)(7,8)
\rput{*0}(0,8.5){$J^\mu(X_1)$}
\rput{*0}(6,10){$p_1$}
}

\rput{30}(5.3,-7)
{
\psline[doubleline=true]{-}(6,8)(10,8)
\psline[doubleline=true,arrowsize=0.24]{->}(2,8)(7,8)
\rput{*0}(0,7){$J^\nu(X_2)$}
\rput{*0}(6,10){$p_2$}
}

\rput{90}(18,-5)
{
\multips(10,8)(2,0){4}{\zigzag}
\psline[arrowsize=0.24]{->}(14,8)(13,8)
\rput{*0}(15,9.5){$k$}
}

\rput{30}(5.3,-7.2)
{
\multips(10,8)(2,0){4}{\zigzag}
\psline[arrowsize=0.24]{->}(14,8)(15,8)
\rput{*0}(15,5){$k+p_2$}
}

\rput{330}(-2.5,11)
{
\multips(10,8)(2,0){4}{\zigzag}
\psline[arrowsize=0.24]{->}(14,8)(13,8)
\rput{*0}(15,11){$k-p_1$}
}

\psline[doubleline=true]{-}(21,8.8)(25,8.8) 
\psline[doubleline=true,arrowsize=0.24]{->}(17,8.8)(22,8.8)
\rput{*0}(30,8.8){$J^\sigma(X_3)$}
\rput{*0}(22,6.8){$p_1+p_2$}

\qdisk(10,4.8){0.1}
\qdisk(10,13){0.1}
\qdisk(17,8.8){0.1}
\end{pspicture}
}
\end{center}
and in the one with the orientation of the loop momenta reversed as well as in the six independent permutations of the bubble diagram
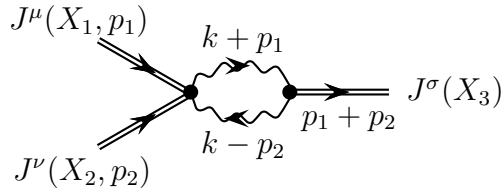
\begin{center}
\scalebox{1}{
\psset{xunit=5pt,yunit=5pt,linewidth=0.8pt,dotsize=2pt}
\def\zigzag{\psline[linearc=.1]{-}(0,0)(.5,1)(1.5,-1)(2,0)}
\begin{pspicture}(38,17)
\rput{330}(-2.7,6.8)
{
\psline[doubleline=true]{-}(6,8)(10,8)
\psline[doubleline=true,arrowsize=0.24]{->}(2,8)(7,8)
\rput{*0}(0,8.5){$J^\mu(X_1,p_1)$}
}

\rput{30}(5.3,-3.3)
{
\psline[doubleline=true]{-}(6,8)(10,8)
\psline[doubleline=true,arrowsize=0.24]{->}(2,8)(7,8)
\rput{*0}(0,7){$J^\nu(X_2,p_2)$}
}

\rput(10,8.8)
{
\psline[linearc=.06]
{-}(0,0)(.5,1.47)(1,1.32)(1.5,1.06)
(2,1.73)(2.5,2.35)(3,1.94)(3.5,1.48)
(4,2)(4.5,2.48)(5,1.94)(5.5,1.35)
(6,1.73)(6.5,2.06)(7,1.32)(7.5,0.47)(8,0)
\rput{*0}(4,4){$k+p_1$}
\psline[arrowsize=0.24]{->}(3.5,2)(4.5,2)
}

\rput(10,8.8)
{
\psline[linearc=.06]
{-}(0,0)(.5,-1.47)(1,-1.32)(1.5,-1.06)
(2,-1.73)(2.5,-2.35)(3,-1.94)(3.5,-1.48) 
(4,-2)(4.5,-2.48)(5,-1.94)(5.5,-1.35)
(6,-1.73)(6.5,-2.06)(7,-1.32)(7.5,-0.47)(8,0)
\rput{*0}(4,-4){$k-p_2$}
\psline[arrowsize=0.24]{->}(3.5,-2)(2.5,-2)
}

\psline[doubleline=true]{-}(21,8.8)(25,8.8) 
\psline[doubleline=true,arrowsize=0.24]{->}(17,8.8)(22,8.8)
\rput{*0}(30,8.8){$J^\sigma(X_3)$}
\rput{*0}(22,6.8){$p_1+p_2$}

\qdisk(10,8.8){0.1}
\qdisk(17.5,8.8){0.1}
\end{pspicture}
}
\end{center}
Summing all these contributions, we obtain the following expression
for the three-point function:
\begin{multline}\label{3point1}  
\Bigl< J^\mu(X_\un , p_\un ) J^\nu(X_\deux, p_\deux) J^\sigma(X_\trois, - p_\un - p_\deux ) \Bigr>_{\rm vec} =  i \int \frac{d^4k}{(2\pi)^4} \trace  X_\un X_\deux X_\trois  \biggl(  \\*  \Upsilon^\mu(k + p_\un  , k  ) \Delta(k) \Upsilon^\nu(k, k - p_\deux ) \Delta(k - p_\deux  )  \Upsilon^\sigma(k- p_\deux  , k + p_\un  ) \Delta(k+p_\un ) \\* 
+ \cR^{\mu\nu}  \Delta(k - p_\deux  )  \Upsilon^\sigma(k- p_\deux  , 
k + p_\un  ) \Delta(k+p_\un ) + \Upsilon^\mu(k + p_\un  , k  ) \Delta(k) 
\cR^{\nu\sigma} \Delta(k+p_\un ) \\* +  
\Delta(k) \Upsilon^\nu(k, k - p_\deux ) \Delta(k - p_\deux  ) \cR^{\sigma\mu}
  \biggr)\\   + i \int \frac{d^4k}{(2\pi)^4} \trace X_\deux X_\un X_\trois\biggl(  
   \\* \Upsilon^\nu(k + p_\deux  , k  ) 
\Delta(k) \Upsilon^\mu(k, k - p_\un ) \Delta(k - p_\un  )  
\Upsilon^\sigma(k- p_\un  , k + p_\deux  )  \Delta(k + p_\deux ) \\*
+ \cR^{\nu\mu}  \Delta(k - p_\un  )  \Upsilon^\sigma(k- p_\un  , 
k + p_\deux  ) \Delta(k+p_\deux )  + \Upsilon^\nu(k + p_\deux  , k  ) 
\Delta(k) \cR^{\mu\sigma} \Delta(k+p_\deux ) \\* +  \Delta(k) 
\Upsilon^\mu(k, k - p_\un ) \Delta(k - p_\un  ) \cR^{\sigma\nu} \biggr) \; \; . 
\end{multline}
Here, $X_\un,\, X_\deux,\, X_\trois$ are $\su(8)$ matrices,
 valued in the $\bf 28 \oplus \overline{\bf 28}$ and the trace is to be taken over 
$( 4 \times 56)^2$ matrices corresponding to components
of the vector propagator.

Let us compute the divergence of the third current in this expectation value. 
Using the formulas (\ref{FeynWard}, \ref{ConWard}), one computes that 
\begin{multline}  
i (p_{\un\, \sigma} + p_{\deux\, \sigma})  \Bigl< J^\mu(X_\un , p_\un ) J^\nu(X_\deux, p_\deux) J^\sigma(X_\trois, - p_\un - p_\deux ) \Bigr>_{\rm vec}  \\* =  i \int \frac{d^4k}{(2\pi)^4} \trace  X_\un X_\deux X_\trois  \biggl(    \Upsilon^\mu(k + p_\un  , k  ) \Delta(k) \Upsilon^\nu(k, k + p_\un ) \Delta(k+p_\un ) \\*
-   \Upsilon^\mu(k -  p_\deux  , k  ) \Delta(k) \Upsilon^\nu(k, k - 
p_\deux ) \Delta(k - p_\deux  ) 
+  \cR^{\mu\nu}  \Delta(k + p_\un  ) -  \cR^{\mu\nu} \Delta(k - p_\deux  ) 
\biggr)  \\*
+ i  \int \frac{d^4k}{(2\pi)^4} \trace  X_\deux X_\un X_\trois  \biggl(    \Upsilon^\mu(k, k + p_\deux ) \Delta(k+p_\deux )  \Upsilon^\nu(k + p_\deux  , k  ) \Delta(k) \\*
-  \Upsilon^\mu(k, k - p_\un ) \Delta(k - p_\un  )    \Upsilon^\nu(k -  p_\un  , k  ) \Delta(k)   +  \cR^{\nu\mu} \Delta(k + p_\deux  )  - 
\cR^{\nu\mu}Ê \Delta(k - p_\un  ) \biggr) \label{DecomposAn} \; \; . 
\end{multline}
The commutator component $\propto \trace [ÊX_\un , X_\deux]  ÊX_\trois $ 
of  (\ref{DecomposAn})  gives rise to the vector field 
contribution to the vacuum expectation value of the insertion of two 
currents, as required by the current Ward identity
\begin{multline}  
i (p_{\un\, \sigma} + p_{\deux\, \sigma})  \Bigl< J^\mu(X_\un , p_\un ) J^\nu(X_\deux, p_\deux) J^\sigma(X_\trois, - p_\un - p_\deux ) \Bigr> \\* = i \Bigl< J^\mu(X_\un , p_\un ) J^\nu([ X_\deux, X_\trois] , - p_\un)  \Bigr>+ i  \Bigl< J^\mu([ X_\un , X_\trois ] , - p_\deux ) J^\nu(X_\deux, p_\deux) \Bigr> \; \; . 
\end{multline}

By contrast, the anticommutator component of  (\ref{DecomposAn}) is 
proportional to  $\trace J X_\un X_\deux X_\trois$, and reduces to the 
difference of divergent integrals with respect to a constant shift of 
the integration variable, as for the one-loop contribution of the fermion 
fields.\footnote{Because ghosts do not give rise to any term of type
 (\ref{rank3invJ}), they do not contribute to the anomaly.}
The two first lines in (\ref{DecomposAn}) give rise to the 
difference of linearly divergent integrals 
\bea 
&& \frac12 \, \trace \{X_\un, X_\deux \} X_\trois  
\int \frac{d^4k}{(2\pi)^4}  \Bigl( I^{\mu\nu} (k  , p_\un ) -   
I^{\mu\nu} (k-p_\un  , p_\un )  -  I^{\mu\nu} (k  ,  - p_\un ) +    
I^{\mu\nu} (k + p_\deux  , -  p_\deux )\Bigr)  \CR
&=&  \frac12 \, \trace \{X_\un , X_\deux \} X_\trois  
\int \frac{d^4k}{(2\pi)^4}  \left(  p_{\un\, \sigma}Ê   
\frac{ \partial I^{\mu\nu} (k , p_\un )}{\partial k_\sigma }Ê + 
p_{\deux\, \sigma}Ê  \frac{ \partial I^{\mu\nu} 
(k , - p_\deux )}{\partial k_\sigma } \right)  \nonumber \label{DifInt}\; \; , 
\eea
with 
\be 
\trace X^3 I^{\mu\nu}( k , p ) = \trace X^3 \,   \Upsilon^\mu(k + p  , k  ) 
\Delta(k) \Upsilon^\nu(k, k + p) \Delta(k+p ) \; \; , 
\ee
where $X$ can be any $SU(8)$ generator. Because there is no invariant  
symmetric tensor of rank three in the adjoint of $E_{7(7)}$ by (\ref{rank3inv})
the last anticommutator term of  (\ref{DecomposAn}) reduces to a double 
difference of quadratically divergent integrals
\bea
& & \frac{1}{2}Ê\trace  \{ X_\un  , X_\deux \} X_\trois \, \cR^{\mu\nu} 
\int \frac{d^4k}{(2\pi)^4}\biggl(  \Delta(k + p_\un  ) -   \Delta(k - p_\deux  ) - 
\Delta(k + p_\deux  ) +  \Delta(k - p_\un  ) \biggr) \CR
&=&   \frac{1}{2}Ê\trace  \{ X_\un  , X_\deux \} X_\trois \, \cR^{\mu\nu} 
( p_{\un \, \sigma} + p_{\deux\, \sigma} ) ( p_{\un\, \rho} - 
p_{\deux\, \rho} )   \int \frac{d^4k}{(2\pi)^4} \frac{ \partial^2 
\Delta(k)}{\partial {k_\sigma} \partial {k_\rho}} \label{DifDif} \; \; . 
\eea
In the above derivation, we have made use of standard formulas
\cite{Bertlmann,Weinberg} to express the integrals as surface integrals
which leads to the final expressions with first and second derivatives
on the integrands.

Although finite, these integrals are not absolutely convergent, and they 
are subject to ambiguities associated to the order of integration of the 
momentum components $k_\mu$. This ambiguity can be fixed in the conventional 
case (with fermions in the loop) by requiring Lorentz invariance. However, 
when photons run in the loop, the integrands are not Lorentz invariant and 
this prescription cannot be consistently defined.\footnote{We are aware
 that a consistent dimensional regularisation via an $SO(3)$ invariant 
 prescription has been used successfully in other contexts, such as the
 post-Newtonian approximation in general relativity, where there are
 no anomalies (T.~Damour, private communication). However, this prescription 
 appears to give inconsistent results in the present case.}Ê
This problem is in fact general in the theory. Indeed, because the Feynman 
rules are not manifestly Lorentz invariant, and because of the explicit
appearance of the Levi-Civit\`a tensor $\varepsilon^{\i\j\k}$, one cannot 
regularise the theory with the dimensional regularisation. Nevertheless, 
we will now explain how one can perform a consistent computation using 
Pauli--Villars regularisation. 

\subsection{Pauli--Villars regularisation}
\label{PauliVillarsReg}
The formulation of the theory is defined such that it is formally equivalent 
to the manifestly diffeomorphism covariant formulation, up to a Gaussian 
integration of the 28 vector fields $A^{\bar \m}_\i$ as in (\ref{PreGauss}, \ref{VaMo}, \ref{PosGauss}). Therefore, we will require the massive Pauli--Villars vector 
fields, to be defined through a local formulation after Gaussian integration. 
This is the case only if the vectors $A_\i^{\bar \m}$ appear in the mass 
term through $F_{\i\j}^{\bar \m}$ up to a total derivative. The only 
`sensible' possibility is therefore to introduce a symmetric tensor $\Gamma_{mn}$
which is off-diagonal in the Darboux basis 
(\ie $\Gamma_{\m\n} = 0 = \Gamma_{\bar \m \bar  \n} $)
\begin{multline}\label{quadvectM} 
\L_0(M) =  \frac{1}{2}Ê \Omega_{mn}  \varepsilon^{\i\j\k} \partial_\zero A_\i^m 
 \partial_\j A_\k^n + \frac{i}{2} \Gamma_{mn} \varepsilon^{\i\j\k} M  A_\i^m 
 \partial_\j A_\k^n  \\* - 
\frac{1}{2}G_{mn} ( \delta^{\i\k}\delta^{\j\l} - \delta^{\i\l} \delta^{\j\k}) 
\partial_\i A_\j^m \partial_\k A_\l^n + b_m \partial_\i A_\i^m \; \; . 
\end{multline} 
We will show next that this Lagrangian gives rise to the standard equations of motion for the $28$ Pauli--Villars vector fields in the Coulomb gauge. Before doing so, note that there is no necessity to modify the interaction terms in the Lagrangian (\ref{quadvectM}), and that the tensor $\Gamma_{mn}$ necessarily 
breaks $SU(8)$ to (at most) $SO(8)$. In fact, a manifestly $SU(8)$ 
regularisation would be in contradiction with the possible existence of 
chiral anomalies. We define $\Gamma_{mn}$ such that it  reads
\be \Gamma_{\m \bar \n } =   \Gamma_{\bar \n \m} = \delta_{\m\bar \n} \; \; , \ee
in the Darboux basis. Following the procedure of section \ref{Equivalence}, the manifestly covariant action for the Pauli--Villars vector fields is obtained after a Gaussian integration of the $28$ (dual) Pauli--Villars vector fields $A^{\bar \m}_\i$. This amounts to performing the replacement
\be\label{M} 
F_{\zero\i}^\m \rightarrow F_{\zero\i}^\m + M A_\i^\m \; \; , 
\ee
in all expressions. In particular, the equations of motion read 
\be 
\partial^\i \scal{ÊF^\m_{\zero\i} + M A_\i^\m }  = 0 \; \;  , \qquad 
\partial^\mu F_{\i\mu} + M \partial_\i A_\zero  - M^2 A_\i^\m = 0 \; \; ,  
\ee
and are manifestly gauge invariant with respect to the modified gauge 
transformations 
\be 
\delta A^\m_\zero = \partial_\zero c^\m + M c^\m \; \;  ,\qquad 
\delta A_\i^\m =   \partial_\i c^\m \; \; . 
\ee
In the Coulomb gauge $\partial^\i A_\i = 0$, they reduce to 
\be 
A^\m_\zero =  0 \ , \qquad \Box A^\m_\i + M^2 A^\m_\i = 0 \; \; . 
\ee
The substitution (\ref{M}) breaks diffeomorphism invariance manifestly,
which can therefore be restored only after the regulator is removed
(possibly with a non-Lorentz invariant local counterterm, see below).

With these replacements, the propagator is manifestly massive in the 
duality invariant formulation.  Indeed, one has
\be 
\Delta^{-1}(p,M) = 
\left( \begin{array}{cc}\ \Omega_{mn} \varepsilon^{\i\j\k}  
p_\zero p_\k + \Gamma_{mn}  \varepsilon^{\i\j\k}  
M  p_\k +  G_{mn} ( \delta^{\i\j} p^2- p^\i p^\j ) \ & \ \ i p^\i  
\delta^n_m \ \ \vspace{2mm} \\ -  i p^\j  \delta^m_n & 0 \end{array}\right) \; \; , 
\ee
and the propagator is 
\be
\Delta(p,M) =  \frac{1}{p^2}\left( \begin{array}{cc}\ 
\frac{\Omega^{mn} \varepsilon_{\i\j\k}  p_\zero p^\k + \Gamma^{mn} \varepsilon_{\i\j\k} M p^\k  - 
G^{mn} ( \delta_{\i\j} p^2 - p_\i p_\j )}{  {p_\zero}^2  - p^2 - M^2 + i \varepsilon }  
\ & \ \ i p_\i  \delta^m_n \ \ \vspace{3mm} \\ -  i p_\j  \delta^n_m & 0 
\end{array}\right) \label{PauPro} \; \; , 
\ee
where $\Gamma^{mn}$ is the inverse of $\Gamma_{mn}$ and satisfies
\be \Gamma_{mp} \Omega^{pn} = - \Omega_{mp} \Gamma^{pn} \ , \qquad  \Gamma_{mp} G^{pn} = G_{mp} \Gamma^{pn} \ , \qquad \Gamma_{mp} \Gamma^{pn} = \delta_m^n \; \;  . \ee 
Therefore
\be ( p_\zero \Omega_{mp} + M \Gamma_{mp} )( p_\zero \Omega^{pn} + M \Gamma^{pn} )  = ( - p_\zero{}^2 + M^2 ) \delta_m^n \; \; , \ee
which permits to check (\ref{PauPro}). 

To define the associated $SU(8)$-current vertex one must distinguish the 
vector and the axial components, respectively, corresponding to the 
decomposition $\bf{63}\rightarrow \bf{28} \oplus \bf{35}$ of the 
$\su(8)$ adjoint under its $\so(8)$ subalgebra. One can thus consider a 
manifestly $SO(8)$ invariant regularisation by considering the coupling 
of the $SO(8)$ current source $B_\mu^{\, m}{}_n$ to the mass term. 
So we consider the coupled Lagrangian 
\begin{multline} 
\L_0[B] = \frac{1}{2}Ê \Omega_{mn}  
\varepsilon^{\i\j\k} \scal{Ê\partial_\zero A_\i^m + 
B_\zero^{\, m}{}_p A_\i^p} \scal{ 
 \partial_\j A_\k^n + B_\j^{\, n}{}_q A^q_\k } +  \frac{i}{2} \Gamma_{mn} Ê\varepsilon^{\i\j\k} M   A_\i^m   \Scal{    \partial_\j A_\k^n +   B_\j^{\, n}{}_p A_\k^p  } Ê \\* -  
\frac{1}{2}G_{mn} ( \delta^{\i\k}\delta^{\j\l} - \delta^{\i\l} \delta^{\j\k}) 
\scal{Ê\partial_\i A_\j^m  + B_\i^{\, m}{}_p A_\j^p} 
\scal{Ê\partial_\k A_\l^n + B_\k^{\, n}{}_q A_\l^q }Ê+ b_m 
\scal{Ê\partial_\i A_\i^m + B_\i^{\, m}{}_n A_\i^n }  \; \; . 
\label{PVSourceLagrange}
\end{multline}
Note however that the mass term does only couple to the ${\bf 35}$ axial component of the source $B_\j^{\, n}{}_p$, because an axial generators $X_m{}^p$ satisfies
\be X_m{}^p \Gamma_{pn} - ÊX_n{}^p \Gamma_{mp} = 0 \ .  \label{AxialCom} \ee

For simplicity, we will focus on the contribution of the massive 
Pauli--Villars vector fields to the vacuum expectation value of three 
axial currents (the three of them in the ${\bf 35}$) for which the vertices $\Upsilon^\mu$ are still defined by (\ref{Upsilon}). Using (\ref{AxialCom}), 
one obtains that the latter is given by 
\begin{multline}\label{3pointM}  
\Bigl< J^\mu(X_\un , p_\un ) J^\nu(X_\deux, p_\deux) J^\sigma(X_\trois, 
- p_\un - p_\deux ) \Bigr>_{\rm PV} 
=  i \int \frac{d^4k}{(2\pi)^4} \trace  X_\un X_\deux X_\trois  
\biggl(  \\*\Delta(k+p_\un , M)  \Upsilon^\mu(k + p_\un  , k  ) 
\Delta(k,-M) \Upsilon^\nu(k, k - p_\deux ) \Delta(k - p_\deux ,M )  
\Upsilon^\sigma(k- p_\deux  , k + p_\un  )  \\* 
+  \Delta(k+p_\un ,M ) R^{\mu\nu}  \Delta(k - p_\deux ,M )  
\Upsilon^\sigma(k- p_\deux  , k + p_\un  ) +  \Delta(k+p_\un ,M )
\Upsilon^\mu(k + p_\un  , k  ) \Delta(k,-M) R^{\nu\sigma} \\* + 
R^{\sigma\mu}   \Delta(k,-M) \Upsilon^\nu(k, k - p_\deux ) 
\Delta(k - p_\deux ,M )  \biggr)   \\
 +  i  \int \frac{d^4k}{(2\pi)^4} \trace X_\deux X_\un X_\trois\biggl(    
\\* \Delta(k + p_\deux ,M ) \Upsilon^\nu(k + p_\deux  , k  ) 
\Delta(k,-M) \Upsilon^\mu(k, k - p_\un ) \Delta(k - p_\un ,M )  
\Upsilon^\sigma(k- p_\un  , k + p_\deux  )   \\*
+  \Delta(k+p_\deux ,M)   R^{\nu\mu}  \Delta(k - p_\un  ,M)  \Upsilon^\sigma(k- p_\un  , k + p_\deux  ) +  \Delta(k+p_\deux ,M) \Upsilon^\nu(k + p_\deux  , k  ) \Delta(k,-M) R^{\mu\sigma} \\* +  R^{\sigma\nu}   \Delta(k,-M) \Upsilon^\mu(k, k - p_\un ) \Delta(k - p_\un ,M )\biggr)
\end{multline}
where the propagator $\Delta(k,-M)$ gets an opposite mass through the 
commutation with the axial generators (similarly as in the standard fermion 
triangle). (\ref{3pointM}) is therefore the analogue of (\ref{3point1})
for $M\neq 0$. In addition to the traces (\ref{rank3inv}) and 
(\ref{rank3invJ}) there are two more types of traces, both of which 
give vanishing contribution because
\be 
\trace (\Omega \Gamma) X_\un X_\deux X_\trois =  0 
= \trace (G \Gamma) X_\un X_\deux X_\trois  \; \;  , 
\ee
Therefore the resulting integral is an {\em even} function of $M$. The 
massive generalisation of (\ref{FeynWard}) is
\be 
- i p_\mu \Upsilon^\mu(k+p,k)  =  \Delta^{-1}(k+p,M)  
- \Delta^{-1}(k,-M) - M \Upsilon_5(2k+p) \; \; , 
\ee
where 
\be
\Upsilon_5(p)  = 
\left( \begin{array}{cc}\   \Gamma_{mn}  \varepsilon^{\i\j\k}  
p_\k   \ & \ \ 0 \ \ \vspace{2mm} \\ 
 0 & 0 \end{array}\right)\; \;  ,
\ee
again indicating the formal similarity of our computation with the
usual fermionic triangle diagram.
Using the latter identity, one computes that 
\begin{multline}  
i (p_{\un\, \sigma} + p_{\deux\, \sigma})  \Bigl< J^\mu(X_\un , p_\un ) 
J^\nu(X_\deux, p_\deux) J^\sigma(X_\trois, - p_\un - p_\deux ) \Bigr>_{\rm PV}
  \\* =  i  \int \frac{d^4k}{(2\pi)^4} \trace  X_\un X_\deux X_\trois  
\biggl(  \Delta(k+p_\un,M )   \Upsilon^\mu(k + p_\un  , k  ) \Delta(k,-M) 
 \Upsilon^\nu(k, k + p_\un ) \\*
-   \Upsilon^\mu(k -  p_\deux  , k  ) \Delta(k,-M) 
 \Upsilon^\nu(k, k - p_\deux ) \Delta(k - p_\deux,M  ) 
+  \cR^{\mu\nu}  \Delta(k + p_\un )   -  \cR^{\mu\nu} \Delta(k - p_\deux ) 
\biggr)  \\*
 +  i \int \frac{d^4k}{(2\pi)^4} \trace  X_\deux X_\un X_\trois  \biggl(   
\Upsilon^\mu(k, k + p_\deux )   \Delta(k+p_\deux,-M )  
\Upsilon^\nu(k + p_\deux  , k  ) \Delta(k,M) \\*
- \Delta(k,M)   \Upsilon^\mu(k, k - p_\un ) \Delta(k - p_\un ,-M )    
\Upsilon^\nu(k -  p_\un  , k  )  +  \cR^{\nu\mu} \Delta(k + p_\deux  )  - 
\cR^{\nu\mu}Ê \Delta(k - p_\un  ) \biggr)\\*
 - i M    \int \frac{d^4k}{(2\pi)^4} \trace  X_\un X_\deux X_\trois  \biggl(  \Delta(k  + p_\un,M) \Upsilon^\mu(k + p_\un  , k  ) \Delta(k,-M) \Upsilon^\nu(k, k + p_\un ) \Delta(k+p_\un,M ) \\*
\times  \Upsilon_5 (2 k + p_\un - p_\deux) \;\; + \; \Delta(k + p_\un ,M) 
\cR^{\mu\nu}  \Delta(k - p_\deux,M  )  
\Upsilon_5 (2k + p_\un - p_\deux) \biggr)  \\*
- i M  \int \frac{d^4k}{(2\pi)^4} \trace  X_\deux X_\un X_\trois  \biggl(    \Delta(k+ p_\deux,M )  \Upsilon^\nu(k + p_\deux  , k  ) \Delta(k,-M)   \Upsilon^\mu(k, k - p_\un ) \Delta(k - p_\un,M )     \\*
 \times \Upsilon_5 (2k - p_\un + p_\deux) \; \; + \;  \Delta(k+ p_\deux ,M)
\cR^{\nu\mu}Ê  \Delta(k - p_\un,M  )   \Upsilon_5 (2k - p_\un + p_\deux)   \biggr) \label{DecomposAnPauli}
\end{multline}

\subsection{Computation of the anomaly coefficient}

To compute the anomaly we now follow the standard procedure by subtracting 
(\ref{DecomposAnPauli}) (indicated by the subscript ``$\rm{PV}$'') from
(\ref{DecomposAn}) (indicated by the subscript ``$\rm{vec}$''), and
then taking the limit $M\rightarrow\infty$. The first two 
 integrals in (\ref{DecomposAnPauli}) are 
very similar to the massless case (\ref{DecomposAn}), and their 
contribution to the anomaly reduces to a difference of linearly 
divergent integrals as well. Because such integrals only depend on the 
leading power in the momentum $k$, they do not depend on the mass and 
hence these contributions cancel precisely between (\ref{DecomposAn}) 
and (\ref{DecomposAnPauli}). It follows that the overall contribution 
of the vector fields to the anomaly is obtained by (minus) the sum of 
the two last integrals of (\ref{DecomposAnPauli}) in the limit 
$M\rightarrow\infty$ --- just like for the fermionic triangle in 
the standard computation with Pauli--Villars regulators, see \eg
\cite{Bertlmann}.

For the computation, it is straightforward to see that for the massive 
propagators and vertices, the relations (\ref{Vert}) remain 
unchanged, and that we have in addition,
\be\label{SimTran} 
\Delta(k,M)^T = \Delta(-k,M) \;\; , \quad 
\Upsilon_5(p)^T = \Upsilon_5(-p) \; \; . 
\ee
These relations permit to prove that the two last integrands in (\ref{DecomposAnPauli}) are invariant 
under the substitution  
\be 
k \leftrightarrow -k \ , \qquad p_\un \leftrightarrow p_\deux   
\ , \qquad \mu \leftrightarrow \nu \ . \label{SimPer} 
\ee
It follows that they are equal, and the anomaly being defined as
\begin{multline}  
\mathcal{A}_{\rm vec}^{\mu\nu}(p_\un, p_\deux) \, 
\trace J X_\un X_\deux X_\trois \equiv i (p_{\un\, \sigma} + p_{\deux\, \sigma})  \Bigl< J^\mu(X_\un , p_\un ) J^\nu(X_\deux, p_\deux) J^\sigma(X_\trois, - p_\un - p_\deux ) \Bigr>_{\rm vec+PV} \\* - i  \Bigl< J^\mu(X_\un , p_\un ) J^\nu([ X_\deux, X_\trois] , - p_\un)  \Bigr>_{\rm vec+PV} - i  \Bigl< J^\mu([ X_\un , X_\trois ] , - p_\deux ) J^\nu(X_\deux, p_\deux) \Bigr>_{\rm vec+PV} 
\end{multline} 
is given by 
\be 
\mathcal{A}_{\rm vec}^{\mu\nu}(p_\un, p_\deux)  =  
2 i \lim_{M \rightarrow +\infty}  \left[ M  \int \frac{d^4k}{(2\pi)^4} 
S^{\mu\nu}(p_\un , p_\deux , k )   \right]  \label{AnomalyPV} \; \; , 
\ee
where the integrand $S^{\mu\nu}(p_\un , p_\deux , k ) $ is
\begin{multline}    
S^{\mu\nu}(p_\un , p_\deux , k )  =  - \frac{1}{56} \trace J \biggl(  \Delta(k  + p_\un,M) \Upsilon^\mu(k + p_\un  , k  ) \Delta(k,-M) \Upsilon^\nu(k, k + p_\un ) \Delta(k+p_\un,M ) \\*
\times  \Upsilon_5 (2 k + p_\un - p_\deux) +  \Delta(k + p_\un ,M) 
\cR^{\mu\nu}  \Delta(k - p_\deux,M  )  \Upsilon_5 (2k + p_\un - p_\deux) 
\biggr) \; \; . 
\end{multline}
$S^{\mu\nu}$ includes three propagators, and takes the form 
\be 
S^{\mu\nu}(p_\un , p_\deux , k )  = \frac{ M  P^{\mu\nu} (p_\un , p_\deux , k ) }{{\scriptstyle  ( k + p_\un)^2} ( {Ê\scriptstyle (k_\zero + p_{\un\, \zero})^2 - ( k + p_\un)^2 - M^2 + i \varepsilon }){ \scriptstyle Êk^2 } ( { \scriptstyle k_\zero{}^2 - k^2 - M^2  + i \varepsilon } ) { \scriptstyle Ê( k - p_\deux)^2}Ê( { \scriptstyle (k_\zero - p_{\deux\, \zero} )^2 - ( k - p_\deux)^2 - M^2  + i \varepsilon  }) Ê} 
\ee
where $P^{\mu\nu}$ is a sum of monomials of order eight in $p_\un,\, p_\deux,\, k$ and $M$. One can neglect all terms of order three and higher in $p_\un$ and $p_\deux$, because they will not contribute to (\ref{AnomalyPV}). Moreover, for the terms of order two in  $p_\un$ and $p_\deux$, the denominator can be approximated as well by $k^{6} ( k_\zero{}^2 -  k^2 - M^2 + i \varepsilon)^{3}$ and one can use the usual simplifications 
\be k^{2n} k_{\zero}{}^{2m} k_\i k_\j \sim \frac{1}{3} \delta_{\i\j}  k^{2n+2} k_{\zero}^{2m} \ , \qquad k^{2n} k_\zero{}^{2m+1}Êk_\i \sim 0 \ , \ee
according to the standard integration rules. After a rather tedious
computation, we obtain
\bea
P^{\zero\i} &\sim& - k^4 ( k_\zero{}^2 -  k^2 - M^2 ) \varepsilon^{\i\j\k}Êk_\j ( 2 p_{\un\, \k} + 6 p_{\deux\, \k} ) +  \frac{1}{3} k^4 \scal{Ê4Êk^2 - 19  ( k_\zero{}^2 -  k^2 - M^2 ) }Ê\varepsilon^{\i\j\k} p_{\un\, \j} p_{\deux\, \k} \CR
P^{\i\j} &\sim& k^4 \varepsilon^{\i\j\k}Ê\biggl( 6 ( p_{\un\, \zero} + p_{\deux\, \zero} ) ( k_\zero{}^2 -  k^2 - M^2 ) k_\k - \frac{8}{3}( k_\zero{}^2 -  k^2 - M^2 ) Êk_\zero ( p_{\un\, \k} + p_{\deux\, \k} ) \CR
& & \quad  +\frac{1}{3}Ê ( k_\zero{}^2 -  k^2 - M^2 )  \Scal{Ê11  ( p_{\un\, \zero} + p_{\deux\, \zero} )  ( p_{\un\, \k} - p_{\deux\, \k} ) - 4 ( p_{\un\, \zero}p_{\deux\, \k} - p_{\deux\, \zero} p_{\un\, \k} ) }Ê\CR
&& \hspace{60mm}  - \frac{8}{3} k_\zero{}^2 ( p_{\un\, \zero}p_{\un\, \k} - p_{\deux\, \zero} p_{\deux\, \k} ) \biggr)
\eea
and $P^{\i\zero}Ê= - P^{\zero\i}$ using the symmetries (\ref{SimTran}, 
\ref{SimPer}). Using the formula~\footnote{Which itself follows from the
 standard formula \cite{Weinberg}
 $$
 \int_0^\infty \frac{x^{a-1} \, dx}{(x^2 + s^2)^b}
 = s^{a-2b} \, \frac{\Gamma(\frac{a}2)\Gamma(b-\frac{a}2)}{2\Gamma(b)} \;.
 $$}
\be   
\int \frac{d^4k}{(2\pi)^4}  \frac{k^{2n} (-k_\zero{}^2 )^{m}}{ 
( { \scriptstyle  Êk^2 - k_\zero^2 + M^2 -  i \varepsilon })^l} = 
- i \frac{\Gamma(\frac{1}{2} + m)  \Gamma(\frac{3}{2} + n) \Gamma(l-m-n-2)}{ 
( 2 \pi)^3 \Gamma(l) \, M^{2(l-m-n-2)} } \; \; , 
\ee
this leads to the integrals
\bea 
\mathcal{A}_{\rm vec}^{\zero\i}Ê&=&2M^2   
i \varepsilon^{\i\j\k} p_{\un\, \j} p_{\deux\, \k}  \int 
\frac{d^4k}{(2\pi)^4}  \left( \frac{4}{\scal{Êk^2 - k_\zero^2 + M^2 
- i \varepsilon }^3}- \frac{k^{-2} }{\scal{Êk^2 - k_\zero^2 + M^2 
- i \varepsilon }^2} Ê\right) \CR
&=& 0  \eea
and
\bea 
 \mathcal{A}_{\rm vec}^{\i\j}Ê&=& 2M^2 i   \varepsilon^{\i\j\k} ( p_{\un\, \zero}p_{\deux\, \k} - p_{\deux\, \zero} p_{\un\, \k} )   \int \frac{d^4k}{(2\pi)^4}  \left( \frac{ 8 k_{\zero}{}^2 k^{-2} }{ ( {Ê\scriptstyle  k_\zero{}^2 - Êk^2 -  M^2 + i \varepsilon })^3} - \frac{ \frac{4}{3} k^{-2}}{ ( {Ê\scriptstyle k^2 - k_\zero^2 + M^2 - i \varepsilon })^2 } \right) \CR
 && \hspace{-15mm}  + 2M^2 i   \varepsilon^{\i\j\k}  ( p_{\un\, \zero} + p_{\deux\, \zero} )  ( p_{\un\, \k} - p_{\deux\, \k} )   \int \frac{d^4k}{(2\pi)^4}  \left( \frac{ \frac{8}{3}  k_{\zero}{}^2 k^{-2} }{({ \scriptstyle k_\zero{}^2 - Êk^2 -  M^2 + i \varepsilon  })^3}   \right . \CR &&\hspace{60mm}Ê\left .-  \frac{4}{ ( {Ê\scriptstyle Êk^2 - k_\zero^2 + M^2 - i \varepsilon })^3}  -  \frac{ \frac{1}{3} k^{-2}}{( {Ê\scriptstyle Êk^2 - k_\zero^2 + M^2 - i \varepsilon })^2 } \right) \CR
 &=& \frac{1}{6 \pi^2} \varepsilon^{\i\j\k} ( p_{\un\, \zero}p_{\deux\, \k} - p_{\deux\, \zero} p_{\un\, \k} ) - \frac{1}{6 \pi^2}  \varepsilon^{\i\j\k}  ( p_{\un\, \zero} + p_{\deux\, \zero} )  ( p_{\un\, \k} - p_{\deux\, \k} ) \; \; \; . 
 \eea
 
The resulting anomaly is not Lorentz invariant, but this is not so surprising 
since we used a regulator that breaks Lorentz invariance. In order to restore
Lorentz invariance, one must renormalise the theory with a finite non-Lorentz 
invariant counterterm with the appropriate $\su(8)$ tensor structure. 
The only such $SO(3)$ invariant density is
\be 
\delta \L \, \, \propto\, \,   \varepsilon^{\i\j\k} J^m{}_n 
B_\zero^{\, n}{}_p B_\i^{\, p}{}_q \partial_\j  B_\k^{\, q}{}_m \label{CounterTerm} \; \; . 
\ee
It follows that the vacuum expectation value of three current insertions 
is only defined up to a shift
\be  
\delta  \Bigl< J^\i(X_\un , p_\un ) J^\j(X_\deux, p_\deux) J^\zero(X_\trois, - p_\un - p_\deux ) \Bigr> = - i a \, \varepsilon^{\i\j\k} (  p_{\un\, \k} - p_{\deux\, \k} )  \trace J  X_\un X_\deux X_\trois \ ,  \ee
and permutations. This shift affects the anomaly factor as
\be \delta  \mathcal{A}_{\rm vec}^{\zero\i} = - \delta  \mathcal{A}_{\rm vec}^{\i\zero} = a \,  \varepsilon^{\i\j\k} p_{\un\, \j}  p_{\deux\, \k}  \ , \qquad \delta  \mathcal{A}_{\rm vec}^{\i\j} = a  \, \varepsilon^{\i\j\k}  ( p_{\un\, \zero} + p_{\deux\, \zero} )  ( p_{\un\, \k} - p_{\deux\, \k} ) \ , 
\ee
and so, choosing $a = \frac{1}{6 \pi^2}$, one recovers the 
{\em Lorentz invariant anomaly} 
\be 
\mathcal{A}_{\rm vec}^{\mu\nu}(p_\un, p_\deux)  =  \frac{1}{6 \pi^2}
Ê\varepsilon^{\mu\nu\sigma\rho} p_{\un\,\sigma}Êp_{\deux\,\rho}    \; \;  . 
\ee
We have thus verified that the anomaly coefficient associated to the 
vector fields is, as predicted from the family's index theorem, $(-2)$ times 
the one associated to the Dirac fermion fields. Taking into account the 
fermionic contributions it follows that the total coefficient of the 
anomaly vanishes for $\N=8$ supergravity, in agreement with (\ref{Marcus}).
 
Within the path-integral formulation of the theory, the variation of the 
formal integration measure with respect to an infinitesimal $\su(8)$ local 
transformation gives rise to a local functional of the fields linear in 
the  $\su(8)$ parameter $C_\ka$ which defines a 1-cocycle over the space 
of $\su(8)$ gauge transformations. This factor can be compensated by a redefinition of 
the local action if this cocycle is trivial in local cohomology. The 
triviality of this cohomology class is equivalent, via a transgression 
operation, to the triviality of a 2-cocycle over the moduli space of framed 
$\su(8)$-connections $B$ identified modulo $\su(8)$-gauge transformations 
in local cohomology. The latter can be computed by means of the family's  
index theorem \cite{Singer}, as the Chern class of the vector bundle defined over an $S^2$ two parameters family of $\su(8)$-gauge orbits of $\su(8)$ framed connections with fibre the index of the chiral differential 
operators 
\be 
( 1+ i \gamma_{5}) \baaa D  \, \, Ê ,\qquad ( 1 + J \star ) d_B  \, \,  ,
\qquad ( 1+ i\gamma_{5} )  \star e^a_{\, \wedge}Ê \gamma_a d_B\, \,  , 
\ee 
acting on the fields of spin 1/2, 1, and 3/2, respectively. A similar construction applies to gravitational and mixed anomalies. According to the family's index theorem, the contribution of the fermion fields and the vectors has been computed in \cite{Singer}, and applied to various supergravity theories in \cite{Marcus}, giving for instance the cancelation (\ref{Marcus}) of the $\su(8)$ anomaly in $\N=8$ supergravity.

Let us now turn to the generalisation of these results to $E_{7(7)}$.
Unlike the linear $SU(8)$ anomaly, the full $E_{7(7)}$ current
and the non-linearly realised $E_{7(7)}$ symmetry give rise to an {\em infinite number} 
of potentially anomalous diagrams. Namely, for the
complete $\e_{7(7)}$ current Ward identities, one must also take into 
account the potential anomalies associated to the ${\bf 70}$ component
of the current, as well as diagrams with any number of scalar field 
insertions. We will write $X_\un, \, X_\deux$ for $\su(8)$ generators, 
and $Y_\un,\,  Y_\deux$ for generators in the ${\bf 70}$. Because 
there is no $\bf 63$ in the {\em symmetric} tensor product
$({\bf {70}\otimes \bf{70}})_{sym}$,~\footnote{The antisymmetric
  product just gives the usual contribution to the non-anomalous
  Ward identity.} the Ward identity associated to
\be 
\Bigl< J^\mu(Y_\un , p_\un ) J^\nu(Y_\deux, p_\deux) 
J^\sigma(X_\un, - p_\un - p_\deux ) \Bigr> 
\ee
cannot be anomalous (we are here using the same notation as in (\ref{JDeco}),
with $J^\mu(X)$ denoting the projection of the current $J^\mu$ along the Lie 
algebra element $X$). However, further anomalies can appear if one includes 
scalar field insertions, as \eg 
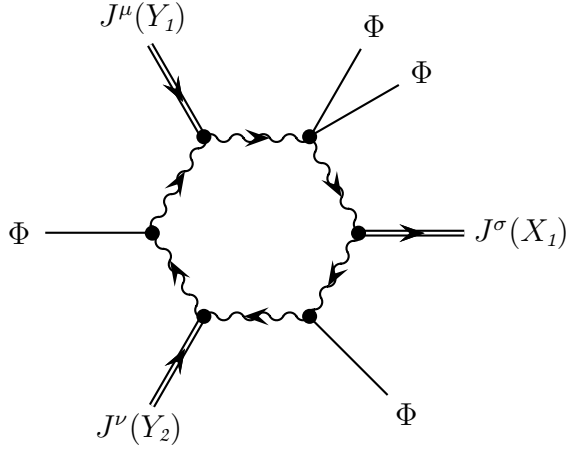
\begin{center}
\scalebox{1}{
\psset{xunit=5pt,yunit=5pt,linewidth=0.8pt,dotsize=2pt}
\def\zigzag{\psline[linearc=.1]{-}(0,0)(.5,1)(1.5,-1)(2,0)}
\begin{pspicture}(43,36)
\rput{300}(4,28.8)
{
\psline[doubleline=true]{-}(6,8)(10,8)
\psline[doubleline=true,arrowsize=0.24]{->}(2,8)(7,8)
\rput{*0}(0,8.5){$J^\mu(Y_\un)$}
}

\rput{60}(18,-2)
{
\psline[doubleline=true]{-}(6,8)(10,8)
\psline[doubleline=true,arrowsize=0.24]{->}(2,8)(7,8)
\rput{*0}(0,8){$J^\nu(Y_\deux)$}
}

\psline[doubleline=true]{-}(31.7,16.8)(35.7,16.8) 
\psline[doubleline=true,arrowsize=0.24]{->}(27.7,16.8)(32.7,16.8)
\rput{*0}(40,16.8){$J^\sigma(X_\un)$}

\multips(16,24)(2,0){4}{\zigzag}
\psline[arrowsize=0.24]{->}(20,24)(21,24)

\rput{300}(12,28.5)
{
\multips(10,8)(2,0){4}{\zigzag}
\psline[arrowsize=0.24]{->}(14,8)(15,8)
}

\rput{60}(25.8,-2.5)
{
\multips(10,8)(2,0){4}{\zigzag}
\psline[arrowsize=0.24]{->}(14,8)(13,8)
}

\multips(16,10.5)(2,0){4}{\zigzag}
\psline[arrowsize=0.24]{->}(20,10.5)(19,10.5)

\rput{300}(0,22)
{
\multips(10,8)(2,0){4}{\zigzag}
\psline[arrowsize=0.24]{->}(14,8)(13,8)
}

\rput{60}(14,4.5)
{
\multips(10,8)(2,0){4}{\zigzag}
\psline[arrowsize=0.24]{->}(14,8)(15,8)
}

\rput{0}(-13,8.03)
{
\psline{-}(17,8.8)(25,8.8) 
\rput{*0}(15,8.8){$\Phi$}
}

\rput{30}(13.5,7.9)
{
\psline{-}(17,8.8)(25,8.8) 
\rput{*0}(27,8.8){$\Phi$}
}

\rput{60}(23,4.7)
{
\psline{-}(17,8.8)(25,8.8) 
\rput{*0}(27,8.8){$\Phi$}
}

\rput{315}(6,16)
{
\psline{-}(17,8.8)(25,8.8) 
\rput{*0}(27,8.8){$\Phi$}
}

\qdisk(16,10.5){0.1}
\qdisk(16,24.1){0.1}
\qdisk(24,24.1){0.1}
\qdisk(24,10.5){0.1}
\qdisk(27.7,16.8){0.1}
\qdisk(12.1,16.8){0.1}
\end{pspicture}
}
\end{center}
This is because the insertion of one scalar field into the diagram does not only 
add one propagator, but also two derivatives, whence the degree of 
divergence of the diagram remains the same with any number of external 
scalar fields (the same is true for fermionic loops, where the insertion
of an extra fermionic propagator is accompanied by one derivative, as well as for the current vertex including scalar fields legs, which do not carry derivatives, but do not add propagators either).
As a first non-trivial example, consider 
the vacuum expectation value 
\be  
\Bigl< J^\mu(X_\un , p_\un ) J^\nu(X_\deux, p_\deux) 
J^\sigma(Y_\un, p_\trois ) \Phi(Y_\deux, -p_\un-p_\deux-p_\trois) \Bigr> \; \; . 
\ee
It satisfies the $\su(8)$ Ward identity
\begin{multline} 
- i p_{\deux\, \sigma} \Bigl< J^\mu(X_\un , p_\un ) J^\sigma(X_\deux, p_\deux) J^\nu(Y_\un, p_\trois ) \Phi(Y_\deux, -p_\un-p_\deux-p_\trois )\Bigr> \\*
= i   \Bigl< J^\mu([X_\un, X_\deux] , p_\un+p_\deux )  J^\sigma(Y_\un, p_\trois ) \Phi(Y_\deux, -p_\un-p_\deux-p_\trois) \Bigr>\\*  + i  \Bigl< J^\mu(X_\un , p_\un )J^\sigma([Y_\un, X_\deux], p_\deux  + p_\trois ) \Phi(Y_\deux, -p_\un-p_\deux-p_\trois) \Bigr> \\* + i  \Bigl< J^\mu(X_\un , p_\un )  J^\sigma(Y_\un, p_\trois ) \Phi([Y_\deux, X_\deux], -p_\un-p_\trois) \Bigr> \label{WI70} \; \; . 
\end{multline}
By contrast, the  $\e_{7(7)}$ Slavnov--Taylor identity takes a more
complicated form because the transformation is non-linear (see (\ref{Ward})
below for the derivation)
\begin{multline} 
- ip_{\trois\, \sigma} \Bigl< J^\mu(X_\un , p_\un ) J^\nu(X_\deux, p_\deux) J^\sigma(Y_\un, p_\trois ) \Phi(Y_\deux, -p_\un-p_\deux-p_\trois )\Bigr> \\*
=  i \Bigl< J^\mu([X_\un, Y_\un] , p_\un+p_\trois )  J^\nu(X_\deux, p_\deux)  \Phi(Y_\deux, -p_\un-p_\deux-p_\trois) \Bigr>\\*  + i  \Bigl< J^\mu(X_\un , p_\un ) J^\nu([X_\deux, Y_\un], p_\deux  + p_\trois ) \Phi(Y_\deux, -p_\un-p_\deux-p_\trois) \Bigr> \\* +  \Bigl< J^\mu(X_\un , p_\un )   J^\nu(X_\deux, p_\deux)  \Phi^A(p_\trois)   \, \Phi(Y_\deux, -p_\un-p_\deux- p_\trois) \Bigr>  \Bigl< \Bigl[ \frac{\Phi}{\tanh(\Phi)} \ast Y_\un \Bigr]_A\Bigr>  \\*
+ i  \Bigl< J^\mu(X_\un , p_\un )   J^\nu(X_\deux, p_\deux)  \Bigl[ \frac{\Phi}{\tanh(\Phi)}(-p_\un - p_\deux)  \ast Y_\un \Bigr]_A \Bigr>  \Bigl<  \Phi^A(p_\un + p_\deux + p_\trois)    \Phi(Y_\deux, -p_\un-p_\deux- p_\trois)  \Bigr> \\* 
+ i  \Bigl< J^\mu(X_\un , p_\un )     \Bigl[ \frac{\Phi}{\tanh(\Phi)}(-p_\un)  \ast Y_\un \Bigr]_A \Bigr>  \Bigl<  J^\nu(X_\deux, p_\deux) \Phi^A(p_\un + p_\trois) \Phi(Y_\deux, -p_\un-p_\deux- p_\trois)  \Bigr> \\* 
+   i\Bigl<   J^\nu(X_\deux, p_\deux)  \Bigl[ \frac{\Phi}{\tanh(\Phi)}(-p_\deux)  \ast Y_\un \Bigr]_A \Bigr>  \Bigl<  J^\mu(X_\un , p_\un )  \Phi^A(p_\deux + p_\trois)    \Phi(Y_\deux, -p_\un-p_\deux- p_\trois)  \Bigr>  \, \, . \label{ST70}Ê
\end{multline}
where the index $A = 1 $ to $70$  labels an orthonormal basis of the coset component of the $E_{7(7)}$ Lie algebra.

At one loop, there is a potential anomaly to these Ward identities 
of the form 
\be   \propto \varepsilon^{\mu\nu\sigma\rho} p_{\un\, \sigma} 
p_{\deux\, \rho}  \trace J X_\un X_\deux [ Y_\un , Y_\deux]Ê    \nn
\ee
for the $\su(8)$ Ward identity (\ref{WI70}), and an anomalous
contribution to the Slavnon-Taylor identity (\ref{ST70})
\be 
\propto \varepsilon^{\mu\nu\sigma\rho}  \scal{Ê3 p_{\un\, \sigma} 
p_{\deux\, \rho}  + (p_{\un\, \sigma} - 
p_{\deux\, \sigma}) p_{\trois \, \rho}} \trace J X_\un X_\deux 
[ Y_\un , Y_\deux]Ê \, \, . 
\ee
Similarly, the Ward identities associated to the vacuum expectations values 
\bea 
&&  \Bigl< J^\mu\Scal{ÊX_\un , - 
\sum_{m=1}^{2N+3} p_m }J^\nu(X_\deux, p_{2N+3} ) J^\sigma(Y_\un, p_\un ) 
\prod_{n=2}^{2+2N} \Phi(Y_n, p_n) \Bigr>\; \; ,  \CR
 &&\Bigl< J^\mu\Scal{ÊX_\un , - \sum_{m=1}^{2N+4} p_m }J^\nu(Y_\un, p_\un ) J^\sigma(Y_\deux, p_\deux ) \prod_{n=3}^{4+2N} \Phi(Y_n, p_n) \Bigr> \; \; , \CR
&& \Bigl< J^\mu\Scal{ÊY_{2N+6} , - \sum_{m=1}^{2N+5} p_m } 
J^\nu(Y_\un, p_\un ) J^\sigma(Y_\deux, p_\deux ) 
\prod_{n=3}^{5+2N} \Phi(Y_n, p_n) \Bigr> \; \; , 
\eea 
are potentially anomalous for all $N\ge 0$. 
Computing these anomalies explicitly would involve an infinite number of 
Feynman diagrams of increasing complexity. Fortunately, as we are going to 
see in the following section, the coefficients associated to these anomalies 
are determined by the Wess--Zumino consistency conditions in terms of the 
$\su(8)$ anomaly coefficient. It thus follows from the computation of 
this section that they all vanish.

What about higher loops? Remarkably, for strictly non-renormalisable
\footnote{\ `Strictly' in the sense that there are no coupling constant of dimension $\ge 0$.} 
theories the Adler--Bardeen Theorem is almost trivial in the following sense. 
By non-renormalisability and power counting higher loop anomalies 
would have a different form and dimension (involving more derivatives) 
from the one-loop anomaly studied above. However, such anomalies can be 
ruled out by the cohomology arguments given in the next section. In conclusion,
with the cancellations exhibited above {\em there are no $\su(8)$ or $\e_{7(7)}$ 
anomalies in $\N=8$ supergravity at any order in perturbation theory.}

\section{Wess--Zumino consistency condition}
\label{WessZumino}
The purpose of this section is to show that `non-linear' $\e_{7(7)}$ anomaly is 
completely determined by the `linear' $\su(8)$ anomaly. In this way the 
determination of an infinite number of potentially anomalous diagrams 
involving three currents and an arbitrary number of scalar field insertions 
can be reduced to the single diagram computed in section~\ref{Anomaly1loop}.
 As already
mentioned in the introduction, this result has its differential geometric 
roots in the homotopy equivalence (\ref{E7homotopy}).

\subsection{The $\e_{7(7)}$ master equation}
 
The `non-linear' $\e_{7(7)}$ Ward identities are Slavnov--Taylor identities, 
which can be summarised in a master equation for the 1PI generating 
functional $\Gamma$. To simplify the discussion, we will postpone the 
discussion of the ghost sector and the compatibility with the BRST master 
equation to the next section. 

Because the discussion of this section does not rely on particular 
properties of $E_{7(7)}$ and applies equally to other supergravity theories
 coupled to abelian vector fields and scalar fields parametrising a 
symmetric space, we keep it general by considering a Lie algebra $\g$ 
with decomposition
\be
\g \cong  \ka \oplus \pa \; \; , 
\ee
with maximal `compact subalgebra' $\ka$ and the `non-compact' part $\pa$,
and the usual commutation relations
\be\label{comm}
[\ka,\ka]\subset \ka \;,\;\;
[\ka,\pa]\subset \pa \;,\;\;
[\pa,\pa]\subset \ka  \; \; . 
\ee
As explained in section~\ref{SymmetricGauge}, the transformation $\delta^\g$ in the
non-linear realisation acts on the scalar fields as 
\be\label{scalartrafo} 
\delta^\g \Phi \equiv \delta^\ka \Phi + \delta^\pa  \Phi
= -  [ÊC_\ka ,  \Phi ]Ê +  \frac{\Phi}{\tanh\Phi} * C_\pa  \; \; ,  
\ee
where the compact subalgebra $\ka$ acts linearly with parameter 
$C_\ka$, while the remaining transformations $\delta^\pa $ with parameter 
$C_\pa$ are realised {\em non-linearly}. With regard to our previous 
discussion of these transformations in section~\ref{SymmetricGauge}, we note two important 
differences:
\begin{enumerate}
\item As we wish to treat the theory within the `BRST formalism', we 
      will from now on take the transformation parameters $C_\ka$ and 
      $C_\pa$ as {\em anti-commuting} (which is the reason why we use 
      the letter $C$ rather than $\Lambda$ for the transformation parameters).
\item Although the $\g$ symmetry acts {\em rigidly}, we will nevertheless
      take $C$ to be a {\em local} parameter, \ie to depend on $x$. The
      corresponding source fields $B\equiv B_\ka + B_\pa$
      coupling to the conserved $G$ Noether current consequently 
      transform as (non-abelian) gauge fields with these parameters.
\end{enumerate}
With regard to the second point we emphasise that the introduction of
an artificial local $G$ invariance here is merely a formal device (well 
known to specialists) which will enable us to derive current Ward identities 
for $\g$. The sources $B$ are {\em external fields}, which are not part 
of any supermultiplet and are not integrated over in the path integral. 
Hence, the symmetries of the physical degrees of freedom of $\N=8$ 
supergravity and their interactions are still the same as before.
Similarly, $C_\ka$ and $C_\pa$, though $x$-dependent, are not quantum 
fields. Readers might nevertheless find it convenient to consider them as 
ghosts for the fictitious local $G$ symmetry, when the $G$ current is 
coupled to the sources $B_\pa$ and $B_\ka$. For instance, we will shortly 
consider a grading that corresponds to the order of the functional in 
these parameters, and that can be thought of as a ghost number (although 
it must not be confused with the true ghost number associated to the BRST 
operator which implements the gauge symmetries of the theory).

With these comments, the action of the transformations (\ref{scalartrafo}) 
on the other fields is straightforward to describe. On the fermionic fields 
(as well as on the supersymmetry ghost or superghosts) the transformations 
act via an induced $\ka$ transformation with parameter
\be  
C_\ka + \tanh( \Phi / 2 ) * C_\pa 
\ee
while on the vector fields and their ghosts the variations act 
linearly with parameter $C$ in the corresponding representation
(the $\bf{56}$ of $E_{7(7)}$ for $\N=8$ supergravity). Finally, writing
$C\equiv C_\ka + C_\pa$, we have 
\be\label{deltaB} 
\delta^\g B = - d C - \{ÊB , C \}Ê\ , \qquad \delta^\g C = - C^2 
\ee
on the current source $B\equiv B_\ka + B_\pa$ and on the parameter itself,
both of which transform in the adjoint of $\g$ (that is, the $\bf{133}$ of
$E_{7(7)}$ for $\N=8$ supergravity). The anticommutator in this formula 
appears because $\delta^\g$ anticommutes with forms of odd degree. 

In summary, on all the fields (but $C_\ka$) the differential $\delta^\g$ 
decomposes into a $\ka$ transformation of parameter $C_\ka$, which we 
will denote $\delta^\ka$, and the remaining (coset) transformation 
$\delta^\pa $ with parameter $C_\pa$ 
\be 
\delta^\g = \delta^\ka + \delta^\pa  \ , \qquad 
\delta^\ka \equiv \delta^{\ka}(C_\ka) \, , \  
\delta^\pa  \equiv \delta^\pa (C_\pa)  
\ee
For instance, and as a consequence, (\ref{deltaB}) splits as follows 
\begin{eqnarray}
\delta^\ka B_\ka &=& - dC_\ka - \{ B_\ka , C_\ka \} \;\; , \quad
\delta^\ka B_\pa = - \{ B_\pa , C_\ka \}   \; \; ,  \nn\\
\delta^\pa B_\ka &=& - \{ B_\pa , C_\pa \} \;\; , \quad
\delta^\pa B_\pa = - dC_\pa - \{ B_\ka , C_\ka\}  \; \; ,  \CR
\delta^\ka C_\pa &=& - \{ C_\ka , C_\pa \} \;\; , \quad
\delta^\pa C_\pa = 0 \; \;  , 
\end{eqnarray}
$\delta^\ka$ is a nilpotent differential defined on all the fields, 
including $C_\ka$ with 
\be \delta^\ka C_\ka = - C_\ka{}^2 \ .\ee
By contrast, $\delta^\pa$ makes sense only on expressions which do not depend
on $C_\ka$. If such expressions are moreover $\ka$-invariant, the operator
$\delta^\pa$ is nilpotent as a consequence of  (\ref{comm}), \ie
\be\label{cohomp}
\delta^\pa (C_\pa) \circ \delta^\pa  (C_\pa) =
\delta^\ka (C_\pa^2) \approx 0 \ .
\ee
We will refer to such $\ka$-invariant expressions which do not depend of $C_\ka$ as `{\em $\ka$-basic}'; and the cohomology of the nonlinear operator $\delta^\pa $ on the complex of  
$\ka$-basic expressions, as the {\em equivariant cohomology} $\mathcal{H}_K^\bullet(\delta^\pa)$ (see for example \cite{Equivariant} for a mathematical definition). 

We will write $S[\varphi,B]$ for the classical action coupled to $G$-current 
sources $B$, where by $\varphi^\Afi$  we designate {\em all} the fields of 
the theory including ghosts. For each field $\varphi^\Afi$ we introduce 
a source $\varphi^\g_\Afi$ for the non-linear $\g$ transformation 
$\delta^\pa(C_\pa)$ of the field $\varphi^\Afi$ of anti-commuting 
parameter $C_\pa$. We define the action coupled to sources by
\be 
\Sigma\big[\varphi,\varphi^\g,B,C\big]Ê\equiv \frac{1}{\kappa^2}ÊS[ 
\varphi,B]Ê- \int d^4x \sum_\Afi (-1)^\Afi \varphi^\g_\Afi 
\,  \delta^\pa(C_\pa) \varphi^\Afi \; \;  , 
\ee
where $(-1)^\Afi$ is $\pm1$ depending of the Grassmann parity of the 
field $\varphi^\Afi$, and the letter $\Afi$ labels {\em all} the fields
of the theory. Of course, the parity of the antifields is the reverse of the corresponding fields, such that the action $\Sigma$ is bosonic and of zero ghost number.
$\Sigma[\varphi,\varphi^\g,B,C]$ satisfies the linear functional identity
\begin{multline}  
\delta^\ka \Sigma = 
\int d^4 x \left( \sum_\Afi \delta^\ka(C_\ka) \varphi^\Afi \, 
\frac{\delta^L \Sigma}{\delta \varphi^\Afi} + \sum_\Afi \delta^\ka(C_\ka) 
\varphi^\g_\Afi \, \frac{\delta^L \Sigma}{\delta \varphi^\g_\Afi} \right .\\* 
\left .   Ê-  \scal{Êd C_\ka + \{ÊB_\ka , C_\ka \} } \cdot \frac{\delta^L 
\Sigma}{\delta B_\ka} -   \{Ê C_\ka , B_\pa \}  \cdot 
\frac{\delta^L \Sigma}{\delta B_\pa} -  \{ÊC_\ka , C_\pa \}Ê \cdot \frac{\delta^L \Sigma}{\delta C_\pa} \right) = 0 \ ,  
\end{multline}
associated to the $\ka$-current Ward identities, and the bilinear 
functional identity 
\begin{multline}Ê 
\int d^4 x \left( \sum_\Afi \frac{ \delta^R \Sigma}{\delta \varphi^\g_\Afi}\frac{\delta^L \Sigma}{\delta \varphi^\Afi} Ê-  \scal{Êd C_\pa + \{ÊB_\ka , C_\pa \} } \cdot \frac{\delta^L \Sigma}{\delta B_\pa}   \right . \\* \left . -    \{Ê C_\pa , B_\pa \}  \cdot \frac{\delta^L \Sigma}{\delta B_\ka}    - \sum_\Afi \varphi^\g_\Afi \delta^\ka(C_\pa{}^2) \varphi^\Afi  \right) = 0 \ ,  
\end{multline}
associated to the $\pa$-current Slavnov--Taylor identities, where the dots 
stand for the appropriately normalised $K$-invariant scalar products. 

Here we disentangled the linear and the non-linear Ward identities, however, 
in order to discuss possible anomalies it will be more convenient to combine
both of them into a single bilinear {\em $G$ master equation} 
\be  \bigl( \, \Sigma , \Sigma \, \bigr)_\g = 0 \  ,  \ee 
which can be obtained by introducing sources for the sources $B$ and the 
parameter $C$ \cite{PiguetSorella}. In the absence of anomalies, the 
above master equation can be elevated to a $G$ master equation for the
full effective action, \ie the 1PI generating functional $\Gamma$ 
\be 
\bigl( \, \Gamma , \Gamma \, \bigr)_\g = 0 \ . \label{MasterG}Ê
\ee 
This, then, is the equation which encapsulates the $\g$ invariance of the 
theory up to any given order in perturbation theory. 

Before discussing the anomalies, let us give an example of Slavnov--Taylor 
identities that can be obtained from the (to be proved to be) non-anomalous 
master equation (\ref{MasterG}). For example, one can consider correlation 
functions involving scalar fields only, with
\be
\left(  \prod_{n \in I} X_n \cdot \frac{ \delta \hspace{5mm}}{\delta \Phi(x_n)} \, X \cdot  \frac{\delta\hspace{5mm}\, }{\delta C(x)}  \,  \bigl( \, \Gamma , \Gamma \, \bigr)_\g  \right)\Bigg|_0  = 0 \; \;  , 
\ee
where the notation $|_0$ means that we set all the classical field 
$\varphi^\Afi$ and sources to zero after differentiation. This 
gives the Ward identity 
\begin{multline}  
\frac{\partial\, \, }{\partial x^\mu} \Bigl< J^\mu(X,x) \prod_{n\in I} \Phi(X_n,x_n) \Bigr> \\* = \sum_{J \subset I}Ê\Bigl< \Phi^A(x) \prod_{m\in J}  \Phi(X_m,x_m) \Bigr> \Bigl< \left[Ê\frac{\Phi}{\tanh(\Phi)}(x) \ast X \right]_A  \prod_{n\in I \setminus J}  \Phi(X_n,x_n) \Bigr> \ ,  \label{nlWard}
\end{multline}
where the sum over $J \subset I$ is the sum over all odd subsets of indices 
$J$ inside the odd set of indices $I$. In the same way (\ref{ST70}) is the 
Fourier transform of 
\be\label{Ward} 
\left( X_\un \cdot \frac{ \delta \hspace{5mm}}{\delta B_{\ka\, \mu}
    (x_\un)}\right)\left( X_\deux \cdot \frac{ \delta 
  \hspace{5mm}}{\delta B_{\ka\, \nu}(x_\deux)}\right)\left( 
  Y_\un \cdot  \frac{\delta\hspace{5mm}\, }{\delta C(y_\un)}\right) 
  \left(Y_\deux 
\cdot  \frac{\delta\hspace{5mm}\, }{\delta \Phi(y_\deux)}\right)  \,  
\bigl( \, \Gamma , \Gamma \, \bigr)_\g \Bigg|_0  = 0 \ .  
\ee

Let us first briefly recall why the existence of anomalies is equivalent 
to a cohomology problem. It is well known that the master equation 
(\ref{MasterG}) can in principle be broken by the renormalisation process 
at each order $n$ in perturbation theory, such that 
\be  
\bigl( \, \Gamma_n , \Gamma_n \, \bigr)_\g =  \hbar^n Ê\mathcal{A}_nÊ+ 
\mathcal{O}(\hbar^{n+1}) \ , 
\ee 
where $\Gamma_n\equiv \sum_{p\leq n} \hbar^p \, \Gamma^\ord{p}$ is the
$n$-loop renormalised 1PI generating functional, and $\mathcal{A}_n$ is 
a local functional of the fields and antifields 
linear in $C$. Because of the `anti-Jacobi' functional identity 
\be 
\bigl( \, \Gamma ,  \bigl( \, \Gamma , \Gamma \, \bigr)_\g \, \bigr)_\g = 0 
\ee
the anomaly nevertheless satisfies the Wess--Zumino consistency condition 
\be  
\bigl( \, \Gamma_n  ,  \mathcal{A}_nÊ\, \bigr)_\g = \mathcal{O}(\hbar)  \, 
\ee
and therefore 
\be \bigl( \, \Sigma ,  \mathcal{A}_nÊ\, \bigr)_\g = 0 \; \; , \ee
where $\Sigma$ is the classical action. If $\mathcal{A}_n$ satisfies 
\be 
\mathcal{A}_n=  \bigl( \, \Sigma , \Sigma^\ord{\flat\, n}Ê\, \bigr)_\g \; \; , 
\ee
for a local functional of the fields  $\Sigma^\ord{\flat\, n}$, the anomaly 
is trivial, because one can simply add it to the bare action in order to 
define a 1PI generating functional which is not anomalous at this order (as we did for example in the last section with the counterterm (\ref{CounterTerm}) in order to restore Lorentz invariance). 
The  existence of an anomaly therefore requires that the cohomology of the 
linearised Slavnov--Taylor operator ${\cal{F}}\rightarrow
(\Sigma ,{\cal{F}} )$ on the set of local functionals $\{{\cal{F}}\}$
of the fields is non-trivial. This cohomology is equivalent to the 
cohomology $\mathcal{H}^1(\delta^\g)$ of the differential operator 
$\delta^\g$ which generates the non-linear $\e_{7(7)}$ action on the 
set of local functionals of the fields identified modulo the equations 
of motion \cite{HenneauxCoh}. 

As we already pointed out, the property that $\mathcal{A}$ is a local 
functional is known as the quantum action principle \cite{PiguetSorella}. 
This principle holds true generally for any well defined regularisation 
scheme. Because of the rather non-standard character of the duality invariant 
formulation of the theory we are using, it is important to show that such 
consistent  regularisation scheme exists for the theory. Although a fully
rigorous proof of the validity of the quantum action principle within the 
Pauli--Villars regularisation scheme defined in the preceding section is 
beyond the scope of the present paper, the one-loop computation of the 
preceding section provides a strong indication for its validity.

\subsection{The $SU(8)$-equivariant cohomology of $\e_{7(7)}$}

To investigate the general structure of anomalies, we need a basis of 
local functionals. For this purpose it is convenient to consider 
functions of the fields and their covariant derivatives, defined as 
\be 
d_B \Phi \equiv d \Phi + [\, B_\ka , \Phi \, ]Ê - 
\frac{\Phi}{\tanh\Phi} * B_\pa \label{CovBDer}  
\ee
for the scalars, and similarly for the other fields. Keeping in mind
that $\delta^\g$ and the exterior derivative {\em anti-}commute, we 
then have
\be\label{dBPhi} 
\delta^\g \ \scal{Êd_B \Phi }Ê =  -  \{Ê\, C_\ka , \, d_B \Phi\,  \}Ê-  
d_B \Scal{  \frac{\Phi}{\tanh\Phi} } * C_\pa \ . 
\ee
In deriving this formula, we make use of the closure property (\ref{E7comm})
in the form
\be
\delta^\pa(C_\pa) \left( \frac{\Phi}{\tanh\Phi}\right) * B_\pa +
\delta^\pa(B_\pa) \left( \frac{\Phi}{\tanh \Phi}\right) * C_\pa
= - \big\{ B_\pa , C_\pa \big\} * \Phi \ , 
\ee
which allows us to trade one expression (the variation of $\Phi/\tanh \Phi$)
which we cannot write in closed-form in terms of another (the $B_\pa$
covariantisation of the last term in (\ref{dBPhi})) which we also cannot
write in closed-form.

Given a basis of local functionals, a potential anomaly $\mathcal{A}$ 
decomposes into a term linear in $C_\ka$ and a term linear in $C_\pa$ 
\be 
\mathcal{A}Ê= \int \Scal{  \mathcal{F} \cdot C_\ka +  
\mathcal{G} \cdot C_\pa } \ , 
\ee
with two local functionals $\mathcal{F}$ and $\mathcal{G}$ of the fields, 
the current sources and their derivatives. $\mathcal{F}$ and $\mathcal{G}$
take values in $\ka$ and $\pa$, respectively. Accordingly, the Wess--Zumino 
consistency condition $\delta^\g  \mathcal{A}Ê = 0$ decomposes into three 
components 
\bea  \label{WZconcon}
\int  \delta^\ka  \scal{  \mathcal{F}Ê \cdot C_\ka }Ê&=& 0 \; \; , \CR
\int \Scal{Ê\delta^\pa Ê \mathcal{F}  \cdot C_\ka +  \delta^\ka \scal{Ê\mathcal{G}Ê \cdot C_\pa }Ê} &=& 0 \; \; , \CR
\int \Scal{Ê-  \mathcal{F}  \cdot {C_\pa}^2 +  \delta^\pa Ê \mathcal{G}Ê \cdot C_\pa } &=& 0 \; \; , 
\eea
corresponding to the coefficients of $C_\ka^2, C_\ka C_\pa$ and $C_\pa^2$,
respectively. The first equation is the condition that $\int \mathcal{F} 
\cdot C_\ka$ defines a consistent anomaly for the $\ka$ current Ward 
identity. {\em A priori}, there are therefore two kinds of anomalies, 
the ones associated to the linearly realised subgroup $K$ and determined by 
$\int \mathcal{F} \cdot C_\ka$ in $\mathcal{H}^1(\delta^\ka)$,\footnote{The
  superscript on $\mathcal{H}$ here refers to the `ghost number'.} which 
would have to be extended to the non-linear representation by an 
 appropriate $\int \mathcal{G} \cdot C_\pa$; and {\em `genuinely non-linear 
anomalies'}, with $\int \mathcal{F} \cdot C_\ka = 0$,  associated to the 
non-linear transformations only and given by $\int \mathcal{G} \cdot C_\pa$. 
The latter expression is then a $\ka$-invariant functional of the fields 
and the current sources  which is $\delta^\pa $ closed by (\ref{WZconcon}). 
If it can be written as the $\delta^\g$ variation of a functional of the 
fields, 
the latter must be $K$ invariant, and the action of $\delta^\g$ and 
$\delta^\pa $ on it are identical. Such a  functional $\int \mathcal{G} \cdot C_\pa$, if non-trivial, defines 
by definition a cocycle representative of the equivariant cohomology 
$\mathcal{H}^1_K(\delta^\pa )$ of $\delta^\g$ with respect to $K$. 
This property can be summarised in the following exact sequence 
\be  
0  \hookrightarrow  \mathcal{H}^1_K(\delta^\pa )\stackrel{\iota}{\longrightarrow}  \mathcal{H}^1(\delta^\g) \stackrel{\pi}{\longrightarrow}  \mathcal{H}^1(\delta^\ka)  \ , 
\ee
which states that to each element of  $ \mathcal{H}^1_K(\delta^\pa )$ 
there corresponds one element of $ \mathcal{H}^1(\delta^\g) $, and that 
all the other elements of  $ \mathcal{H}^1(\delta^\g) $ correspond to 
elements of  $\mathcal{H}^1(\delta^\ka)$ (although $\pi$ is not necessarily 
surjective {\it a priori}). 

Let us first consider the non-trivial anomalies associated to $\ka$ 
current anomalies. The anomalies associated to a linearly realised group 
are well known, and are classified by symmetric Casimirs. A nice way to 
derive such anomalies is by means of the `Russian formula' \cite{DixonR,Russian,BaulieuR,Stora}
\be\label{Russian}
( d + \delta^\ka ) \scal{ÊB_\ka + C_\ka }Ê+  \scal{ÊB_\ka + C_\ka }^2 
= F^\ord{0}_\ka \equiv d B_\ka + B_\ka{}^2  \; \; , 
\ee
to derive a $(d+ \delta^\ka)$-cocycle from any symmetric Casimirs by use of 
the Cartan homotopy formula. In four dimensions, the relevant Casimir is 
the symmetric tensor of rank three, and the Cartan homotopy formula gives 
\be 
( d + \delta^\ka )  \trace \Scal{\tilde{B}_\ka {F}^\ord{0}_\ka{}^2 - \frac{1}{2} \tilde{B}_\ka{}^3  {F}^\ord{0}_\ka + \frac{1}{10}  \tilde{B}_\ka{}^5 } = \trace  {F}^\ord{0}_\ka{}^3 = 0 \label{CHF} \ , 
\ee
where we define the {\em extended connection} (always indicated by a
tilde) as
\be 
\tilde{B}_\ka \equiv B_\ka + C_\ka \ , 
\ee
and the trace is taken in the complex representation of $\ka$. 
Picking the component of the Chern--Simons function of form degree four, 
one obtains from this equation the conventional non-abelian 
Adler--Bardeen anomaly 
\be 
\mathcal{A}_\ka = \int \trace d C_\ka \Scal{ B_\ka F^\ord{0}_\ka 
- \frac{1}{2} B_\ka{}^3 }Ê\ , 
\ee
which satisfies 
\be  \delta^\ka \, \mathcal{A}_\ka= 0 \ . \ee

Here we are specifically interested in the case when the rigid symmetry 
group $G$ does {\em not} admit an invariant tensor of rank three, as for
$G=E_{7(7)}$. In this case the trace must be taken in a {\em complex
representation} of the subgroup $K$, and (\ref{CHF}) cannot be defined 
from the straightforward extension of the `linear' Russian formula 
(\ref{Russian}) to the linear formula 
\be 
( d + \delta^\g ) \scal{ÊB + C }Ê+  \scal{ÊB + C }^2 = F_\g = d B + B^2 \ , 
\ee
for the full Lie algebra $\g$, because this formula would only make sense 
in a linear representation of $E_{7(7)}$. Instead we must now look for a 
{\em non-linear variant of the Russian formula.} To this aim, we first 
observe that the closure of the non-linear representation of $\g$ 
on the fermion fields implies 
\be 
\delta^\g  \scal{ C_\ka + \tanh( \Phi / 2 ) * C_\pa }Ê+  
\scal{ C_\ka + \tanh( \Phi / 2 ) * C_\pa }^2 = 0  \ . \label{CCdonneC}Ê
\ee
in the given complex representation of $\ka$. This formula (which 
is a non-linear analogue of the usual BRST variation, cf. second
formula in (\ref{deltaB})) suggests that one can define a non-linear 
Russian formula for the $\g$ symmetry in the fundamental representation 
of $\ka$. The most natural guess for the (extended) `non-linear 
$\g$ connection' is
\be 
\tilde{B} \equiv B_\ka + C_\ka + \tanh(\Phi/2) \ast (   B_\pa + C_\pa) \ , 
\ee
which is indeed valued in the Lie subalgebra $\ka$. In turn, this motivates 
the following definition of the (extended) $\g$ field strength, {\it viz.}
\be \tilde{F}_\g \equiv ( d + \delta^\g ) \tilde{B} + \tilde{B}^2 \  , \ee
which one then computes (using the extended version of (\ref{CCdonneC}) to $B + C$) to be 
\be 
\tilde{F}_\g =  F_\ka   +  \tanh(\Phi/2) \ast F_\pa + 
d_B \scal{Ê\tanh(\Phi/2) } \ast  ( B_\pa +  C_\pa ) \ , \label{ERF}
\ee
with
\be\label{nonlinF} 
F_\ka \equiv d B_\ka + B_\ka{}^2 + B_\pa{}^2 \ , \qquad F_\pa \equiv
d B_\pa + \{ÊB_\ka , B_\pa \}Ê\ , \ee
and  $d_B \tanh(\Phi/2)$ defined in terms of the covariant derivative (\ref{CovBDer}) similarly as in (\ref{dBPhi}). In contradistinction to the conventional Russian formula, the extended field-strength (\ref{ERF}) is not only the `horizontal' two-form curvature, but has an extra component linear in the parameter $C_\pa$.  Nevertheless, it does not depend on $C_{\ka}$, and transforms covariantly with respect to $\ka$ in the adjoint representation $\ka$, 
\be 
\delta^\ka\tilde{F}_\g =  - \bigl[\, ÊC_\ka \, , \, \tilde{F}_\g \, \bigr]Ê \ .
\ee
With these definitions, the Cartan homotopy formula 
\be 
( d + \delta^\g )  \trace \Scal{\tilde{B}_\g \tilde{F}_\g{}^2 - 
\frac{1}{2} \tilde{B}_\g{}^3  \tilde{F}_\g + 
\frac{1}{10}  \tilde{B}_\g{}^5 } = \trace  \tilde{F}_\g{}^3  \label{ECHF} 
\ee
therefore admits a non-vanishing right-hand-side (whereas for a linear
representation of $E_{7(7)}$, the right hand side of (\ref{ECHF}) would
simply vanish). But because it is independent of $C_\ka$ and 
$\ka$-invariant, and hence $\ka$-basic, it defines a cocycle of the 
equivariant cohomology  $\mathcal{H}_K^2(\delta^\pa )$. Note that it is 
a cocycle of  `ghost number' $2$,  because the associated $4$-form component 
is of `ghost number' $2$. We here tacitly use the corollary of the algebraic 
Poincar\'e lemma, \ie all $d$-closed functions of the fields and their 
derivatives of form-degree $p\le 3$ are $d$-exact, whence the cohomology of 
a differential $\delta^\pa $ in the complex of  local functionals of 
`ghost number' $n$ is isomorphic to the cohomology of the extended 
differential $d + \delta^\pa $ in the complex of functions of the 
fields of form-degree plus `ghost number' $4+n$ \cite{HenneauxBook}. 

If this cocycle is trivial in $\mathcal{H}_K^2(\delta^\pa )$, \ie if 
there exists a $\ka$-basic function $\tilde{M}$ such that 
\be 
( d + \delta^\pa ) \tilde{M} = \trace  \tilde{F}_\g{}^3  \ , 
\ee
one can extend the $\ka$ Adler--Bardeen anomaly to a $\g$ anomaly by 
considering the integral of the 4-form component 
\be 
\mathcal{A}Ê=  \int  \biggl( \trace \Scal{\tilde{B}_\g \tilde{F}_\g{}^2 - 
\frac{1}{2} \tilde{B}_\g{}^3  \tilde{F}_\g + \frac{1}{10}  
\tilde{B}_\g{}^5 }\Big|_{\gra{4}{1}} - M_\gra{4}{1} \biggr) 
\ee
It follows that the only possible obstruction to extend a $\ka$ anomaly 
in $\mathcal{H}^1(\delta^\ka)$ to a full $\g$ anomaly in 
$\mathcal{H}^1(\delta^\g)$ are defined by cohomology classes of the 
second equivariant cohomology group $\mathcal{H}_K^2(\delta^\pa )$. 

One can summarise these properties into the exact sequence 
\be\label{exactS}
  0  \hookrightarrow  \mathcal{H}^1_K(\delta^\pa)\stackrel{\iota}{\longrightarrow}  \mathcal{H}^1(\delta^\g) \stackrel{\pi}{\longrightarrow}  \mathcal{H}^1(\delta^\ka)  \longrightarrow \mathcal{H}_K^2(\delta^\pa) \ee
which states that the  $\mathcal{H}^1(\delta^\g) $ and $ \mathcal{H}^1(\delta^\ka) $ only differ by cocycles associated to equivariant cohomology classes. The last arrow is the map which associates to any $K$ consistent anomaly, the corresponding invariant polynomial in $\tilde{F}_\g$. 

Now that we have motivated our interest in the equivariant cohomology, we 
are going to prove that it is trivial. The intuitive idea is the following, the equivariant cohomology on the set of local functional of the fields is closely related to the equivariant cohomology on the set of functions of the scalars only, and the latter is homomorphic to the De Rham cohomology of the coset space $G / K \cong \IR^n $ which is trivial \cite{Equivariant}.

In order to carry out this program, 
it will turn out to be useful to introduce a filtration in terms of the order 
of the functional in naked scalar fields $\Phi$, (considering $d_B \Phi$ as 
independent). The expansion of the variation of $\Phi$ and its covariant 
derivative are 
\bea 
\delta^\pa \Phi &=&  C_\pa +  \frac{1}{3}[\Phi , [\Phi , C_\pa ]]- \frac{1}{45}[\Phi, [\Phi, [\Phi, [ \Phi, C_\pa]]]] + \mathcal{O}(\Phi^6) \; \; ,   \CR
d_B \Phi &=& - B_\pa + d_{B_\ka} \Phi  - \frac{1}{3} [\Phi , [\Phi , B_\pa]] +  \frac{1}{45} [\Phi, [\Phi, [ \Phi , [ \Phi, B_\pa]]]] + \mathcal{O}(\Phi^6) \; \; ,  \CR
\delta^\g_{C_\ka} \scal{Êd_{B_\ka} \Phi }Ê &=& - d_{B_\ka} C_\pa + [\Phi , \{ B_\pa , C_\pa\} ] - \frac{1}{3} d_{B_\ka} [\Phi , [ \Phi , C_\pa ] ] + \mathcal{ O}(\Phi^4)  \; \; . 
\eea
The first order in $\Phi$ of the equivariant differential only acts on $\Phi$ itself as $\delta^{\g\, \ord{-1}}_{C_\ka} \Phi =  C_\pa$. Any $SU(8)$-invariant local function of the fields admits an expansion 
\be X = \sum_{k \in \IN} X^\ord{n+k} \ee
and 
\be  \delta^\pa  X = 0 \quad \Rightarrow \quad\delta^{\pa\, \ord{-1}} X^\ord{n} = 0 \; \; . \ee
If $X^\ord{n}$ depends non-trivially on $C_\pa$ or $\Phi$, then there exist a function $Y^\ord{n+1}$  such that $X^\ord{n} =\delta^{\pa\, \ord{-1}} Y^\ord{n+1}$ \cite{HenneauxBook}. To see this, let us define the trivialising homotopy $\sigma$, which acts trivially on all fields, but $C_\pa$
\be  \sigma C_\pa =  \Phi\; \; ,  \hspace{10mm} \{ \sigma , \delta^{\pa\, \ord{-1}} \} X =  N X \equiv \int d^4x \Scal{  C_\pa \frac{\delta\, }{\delta C_\pa} + \Phi \frac{\delta\, }{\delta \Phi} } X \; \; ,  \ee
and 
\be [N , \delta^{\pa\, \ord{-1}}] = [N , \sigma ]= 0 \; \; .  \ee
If $X^\ord{n}$ depends non-trivially on $C_\pa$ or $\Phi$, $N^{-1} X^\ord{n}$ exists and 
\be X^\ord{n} = \{\sigma , \delta^{\pa\, \ord{-1}}\}N^{-1} X^\ord{n} = \delta^{\pa\, \ord{-1}} \, \sigma N^{-1} X^\ord{n} \ee
Now with $Y^\ord{n+1} = \sigma N^{-1} X^\ord{n}$,\footnote{Note that although $\delta^{\pa \, \ord{0}}$ vanishes on $\Phi$ and the fermion fields, it acts non-trivially on the electromagnetic fields and their ghost, and so $ \delta^{\pa\, \ord{0}}  Y^\ord{n+1} $ does not vanish in general.} 
\be X - \delta^\pa  Y^\ord{n+1} = X^\ord{n+1} -  \delta^{\pa\, \ord{0}}  Y^\ord{n+1} + \mathcal{O}(\Phi^{n+2}) \; \; . \ee
Using the trivialising homotopy, one proves in the same way that $Y^\ord{n+2}$ exists such that 
\be  X - \delta^\pa  \scal{Y^\ord{n+1} + Y^\ord{n+2} }= X^\ord{n+2} -  \delta^{\pa\, \ord{1}} Y^\ord{n+1} - \delta^{\pa\, \ord{0}}  Y^\ord{n+2} +  \mathcal{O}(\Phi^{n+3}) \; \; . \ee
Iteratively one proves that there exist a formal power series 
\be 
Y = \sum_{k \in \IN} Y^\ord{n+1+k} \ee
 in $\Phi$ that trivialises $X$,
\be X =  \delta^\pa  Y \; \; . 
\ee
This proof extends trivially to functionals \cite{HenneauxBook}, and therefore 
\be  
\mathcal{H}^n_K(\delta^\pa) \cong 0  \qquad \mbox{for } n \ge1 \; \; . 
\ee 
As a direct consequence, the exact sequence (\ref{exactS}) implies the isomorphism 
\be  
\mathcal{H}^1(\delta^\g) \cong \mathcal{H}^1(\delta^\ka) \; \; . 
\ee
The equivalence of these two cohomology groups is a main result of this 
paper: it states that the $\e_{7(7)}$ consistent anomalies are in 
one-to-one correspondence with the $\su(8)$ consistent anomalies. 
In particular, it follows that their coefficients are the same, establishing 
as a corollary that the absence of anomalies for the $\su(8)$ current 
Ward identities implies the absence of anomalies for the non-linear 
$\e_{7(7)}$ Ward identities. This statement completes our proof that the rigid $E_{7(7)}$ symmetry of $d=4$ $\N=8$ supergravity is not anomalous in perturbation theory.

In the remaining part of this section, we want to illustrate in some more detail how a potential $\su(8)$ Adler--Bardeen anomaly would generalise to an $\e_{7(7)}$ anomaly. The three-form component 
$\trace \tilde{F}_\ka{}^3 |_{\gra{3}{3}}$ is cubic in $C_\pa$, and being 
$  \delta^\pa$-closed by construction, there exists an $SU(8)$-invariant 
function $M_\gra{3}{2}$ of the fields quadratic in $C_\pa$ such that 
\be 
\trace \tilde{F}_\ka{}^3 |_{\gra{3}{3}} = \delta^\g M(F_\ka, F_\pa, B_\pa, 
d_B \Phi, \Phi , C_\pa)_\gra{3}{2}  \; \; . 
\ee
Then $ \trace \tilde{F}_\ka{}^3 |_{\gra{4}{2}}  - d M_\gra{3}{2}$ is itself $  \delta^\pa$-closed because of the Bianchi identity,
\be \delta^\pa  \scal{\trace \tilde{F}_\ka{}^3 |_{\gra{4}{2}}  - d M_\gra{3}{2} } = - d \, \trace \tilde{F}_\ka{}^3 |_{\gra{3}{3}} + d\,   \delta^\pa  M_\gra{3}{2} = 0 \; \; , \ee
 and being quadratic in $C_\pa$, there exists a $K$-invariant function $M_\gra{4}{1}$ of the fields linear in $C_\pa$ such that 
\be \trace \tilde{F}_\ka{}^3 |_{\gra{4}{2}} = \delta^\g M(F_\ka, F_\pa, B_\pa, d_B \Phi, \Phi , C_\pa)_\gra{4}{1} + d M(F_\ka, F_\pa, B_\pa, d_B \Phi, \Phi , C_\pa)_\gra{3}{2} \; \; .   \ee
The consistent $E_{7(7)}$ anomaly is defined as 
\be \int \left( \trace \Scal{\tilde{B}_\ka \tilde{F}_\ka{}^2 - \frac{1}{2} \tilde{B}_\ka{}^3  \tilde{F}_\ka + \frac{1}{10}  \tilde{B}_\ka{}^5 }\Big|_{\gra{4}{1}} - M(F_\ka, F_\pa, B_\pa, d_B \Phi, \Phi , C_\pa)_\gra{4}{1} \right)  \ee
where 
\begin{multline}  \trace \Scal{\tilde{B}_\ka \tilde{F}_\ka{}^2 - \frac{1}{2} \tilde{B}_\ka{}^3  \tilde{F}_\ka + \frac{1}{10}  \tilde{B}_\ka{}^5 }\Big|_{\gra{4}{1}}  \\* \hspace{-15mm} = \trace \Scal{ C_\ka + \tanh(\Phi / 2) *    C_\pa }  d \biggl( \Scal{ B_\ka + \tanh(\Phi / 2) *    B_\pa }
d \Scal{ B_\ka + \tanh(\Phi / 2) *    B_\pa } \\* \hspace{80mm} + \frac{1}{2} \Scal{ B_\ka + \tanh(\Phi / 2) *    B_\pa }^3 \biggr) \\* \hspace{-20mm}  + \trace d_B \scal{Ê \tanh(\Phi / 2)} *   C_\pa  \biggl( \Scal{ B_\ka + \tanh(\Phi / 2) *    B_\pa } d \Scal{ B_\ka + \tanh(\Phi / 2) *    B_\pa } \biggr . \\* \biggl . + d \Scal{ B_\ka + \tanh(\Phi / 2) *    B_\pa }\, \Scal{ B_\ka + \tanh(\Phi / 2) *    B_\pa } +  \frac{3}{2} \Scal{ B_\ka + \tanh(\Phi / 2) *    B_\pa }^3 \biggr) \; \; . 
\end{multline}
One computes perturbatively that 
\begin{multline}  M_\gra{3}{2} = \frac{1}{8} \trace \biggl( -  [\Phi, B_\pa ] \{C_\pa , B_\pa \}^2 + \frac{1}{2} \Scal{3 \{C_\pa , d_{B_\ka} \Phi \}- [\Phi , d_{B_\ka} C_\pa ] } \Bigl\{[\Phi , B_\pa ] , \{C_\pa , B_\pa \} \Bigr\} \biggr)\\*  +\; \;  \mathcal{O}(\Phi^3) \; \; ,  \end{multline}
and 
\begin{multline}  M_\gra{4}{1} =  \frac{3}{8} \trace  \Bigl\{ \{C_\pa , B_\pa \}- 2 \{C_\pa , d_{B_\ka} \Phi\} + [\Phi , d_{B_\ka} C_\pa ] , [\Phi , B_\pa ]\Bigr\}\scal{F_\ka - B_\pa{}^2 } \\* 
+ \frac{1}{4} \trace \Bigl\{ [\Phi, B_\pa ], \{C_\pa , B_\pa \}\Bigr\}\Scal{ [\Phi , F_\pa ]+ \{B_\pa , d_{B_\ka} \Phi \}} + \mathcal{O}(\Phi^3) \; \; . \end{multline} 
There is no difficulty in computing higher order terms in $\Phi$, but the complete solution is not obvious. Anyway, the important property is that it exists, at least as a formal power series in $\Phi$ (the issue of convergence being irrelevant in perturbative theory). 

The results of this section extend straightforwardly to any consistent 
$K$ anomaly for any supergravity theory in arbitrary dimensions. For 
example, the solution for $\trace \tilde{F}_\ka $ is rather trivial
when $K$ admits a $U(1)$ factor, as for lower $\N$-extended supergravity 
theories. In that case, $\trace \tilde{F}_{\g\, \,  \wedge}  \, R^{ab}_{\, \, \, \, \wedge} R_{ab}= 0 $ 
and one has the anomaly 
\be 
\mathcal{A}_{\mathfrak{u}(1)} = \int \trace \Scal{C_\ka + \tanh(\Phi/2) *  
    C_\pa } R^{ab}_{\, \, \, \, \wedge} R_{ab}\label{U1anomaly}  \; \; . 
\ee
In particular, this anomaly does not vanish when $C_\pa$ is constant, and 
the current sources are set to zero. It follows that the rigid Ward 
identities are anomalous at one-loop if the coefficient does not vanish, 
as is the case for minimal $\N=4$ supergravity with duality group
$SL(2,\IR)$, and more generally for $\N \le 4$ supergravities.

More specifically, for $G=SL(2,\IR)$, we can spell out the above
formulas in explicit detail. In this case, the second relation
in (\ref{deltaB}) becomes
\be 
\delta^{\sl_2}  \alpha = i \w \bar \w \; \; , \qquad \delta^{\sl_2}\w 
= - 2 i \alpha \w \; \; , 
\ee
where $\alpha\equiv C_\ka$ and $\w \equiv C_\pa$ are real and complex
anticommuting numbers, respectively. If we denote by $\phi$ the complex 
scalar parametrising the coset $SL(2,\IR)/SO(2)$ (the analogue of $\Phi$)
the formula (\ref{scalartrafo}) can be worked out as
\be 
\delta^{\sl_2}  \phi =  \Scal{ \frac{1}{2} + \frac{|\phi|}{\tanh 2 | \phi| }} 
  \w - 2 i \alpha\phi + \Scal{ \frac{ 1}{2 |\phi|^2} - \frac{ 1}{|\phi| 
 \tanh 2 |\phi|} } \bar \w \phi^2  
\ee 
and the anomaly (\ref{U1anomaly}) reads
\be  
\mathcal{A}_{\mathfrak{u}(1)} = \int\Scal{\alpha  -  
  \frac{i \tanh | \phi| }{2  |\phi|}  \scal{\bar \phi \w - \phi \bar \w }} \,
 R^{ab}_{\, \, \, \, \wedge} R_{ab} \label{slAnomaly} 
\ee
Equivalently, within the triangular gauge parametrisation of $SL(2,\IR)/SO(2)$ by the complex modulus 
$\tau =  \tau_\un + i \tau_\deux $, and the (hopefully self explanatory)
notation
\be 
C \, = \,  \left(\begin{array}{cc} \, \, C_h\, \,  & \, \, C_e\, \,  
\vspace{2mm} \\ \, \, C_f\, \,  & - C_h\, \,  \end{array}\right) \; \; , 
\ee
the algebra reads
\begin{gather}  
\delta^{\sl_2}  C_h = - C_e C_f \; \; ,  \qquad 
\delta^{\sl_2}C_e = - 2 C_h C_e\; \; ,  \qquad  
\delta^{\sl_2}C_f = 2 C_h C_f \; \; ,  \CR
\delta^{\sl_2} \tau = - C_e - 2 C_h \tau  + C_f \tau^2 \; \; . 
\end{gather}
The consistent anomaly (\ref{U1anomaly}) then becomes ($C_f \tau_\deux$ being the parameter of the compensating $\mathfrak{u}(1)$ transformation in the triangular gauge) 
\be 
\mathcal{A}_{\mathfrak{u}(1)} =   
\int   C_f \tau_\deux    \,R^{ab}{}_\wedge R_{ab} \; \; . 
\ee
Indeed, explicit computation shows that 
\be  
\delta^{\sl_2}  \big( C_f \tau_\deux \big) = 0 \; \; , 
\ee
but $C_f \tau_\deux$ itself cannot be written as $ \delta^{\sl_2} \cF(\tau)$: 
indeed, the vanishing of the $C_e$ component of $ \delta^{\sl_2}  \cF(\tau)$ 
implies that $\cF$ is a function of $\tau_\deux$ only, and the vanishing 
of the $C_h$ component then entails that $\cF$ must be constant. 

In the conventional formulation of $\N=4$ supergravity, and similarly in $\N=2$ supergravity with a semi-simple duality group $SL(2,\IR) \times SO(2,n)$, we see that the non-linearly realised generator $\bf f$ of $\sl_2$ is anomalous at one-loop. More generally, in $\N=2$ supergravity theories with vector multiplets scalar fields parametrising a symmetric special K\"{a}hler manifold, the duality group will be anomalous at one-loop. Indeed, one computes similarly as in \cite{Marcus} that the addition of matter multiplets does not permit to cancel the $U(1)$ gravitational anomaly, the anomaly coefficient of (\ref{U1anomaly}) being proportional to $24 + 12 n_{\scriptscriptstyle \rm V}$ in $\N=4$ supergravity coupled to $n_{\scriptscriptstyle \rm V}$ vector multiplets, and to $102 + 10n_{\scriptscriptstyle \rm V} + 3 n_{\scriptscriptstyle \rm H}$ in $\N=2$ supergravity coupled to $n_{\scriptscriptstyle \rm V}$ vector multiplets and $n_{\scriptscriptstyle \rm H}$ hypermultiplets.

\section{Compatibility of $E_{7(7)}$ with gauge invariance}

Up to this point we have discussed the properties of the $E_{7(7)}$ symmetry and its possible anomalies, irrespectively of its compatibility with gauge invariance. We now extend this 
discussion to the full quantum theory, with the aim of deriving the Ward identities 
associated to the conservation of the $E_{7(7)}$ Noether current, thereby
corroborating our main claim that the non-linear $E_{7(7)}$ symmetry is
compatible with all gauge symmetries of the theory, in the sense that
it can be implemented order by order in a loop expansion of the full 
effective action. To this aim, we have to make use of the BRST formalism
(see e.g. \cite{Weinberg2,HenneauxBook}). Because the algebra of gauge 
transformations is `open',\footnote{That is, the gauge algebra closes 
  only modulo the equations of motion.} we have to go one step further 
by including higher order ghost interactions \cite{DeWitVanHolten},
and ultimately bring in the full machinery of the Batalin--Vilkovisky 
formalism \cite{BV}. 
In addition to the usual ghosts and antighosts (anti-commuting for the 
bosonic transformations, and commuting for the supersymmetry transformations) 
this requires introducing `antifields' for all fields and ghost fields 
of the theory. The compatibility of the of $E_{7(7)}$ with the BRST
symmetry is then encoded into two corresponding mutually compatible
`master equations'.

Now, a complete treatment of our $E_{7(7)}$ invariant formulation 
of maximal supergravity along these lines would be 
very involved and cumbersome, and certainly unsuitable for practical
computations of the type performed in \cite{Bern,Bern2}. Instead, we
here focus on the specific features of the duality invariant 
formulation in comparison with the conventional formulation of the 
theory, and the fact that the cancellation of $E_{7(7)}$ anomalies,
together with the well admitted absence of diffeomorphism and supersymmetry
anomalies in four space-time dimensions, eliminates any obstruction
towards implementing the BRST and $E_{7(7)}$ master equations at any
order in perturbation theory. We emphasise again that these results 
do not preclude the appearance of divergent counterterms, but
they ensure that potential divergences must respect the full
$E_{7(7)}$ symmetry of the theory.

\subsection{Batalin--Vilkovisky formalism}

Following \cite{Weinberg2} we will designate by $e^{\dd\, \mu}_a$, 
$A_m^{\dd\, \i}$, $\psi^{\dd\, \mu}_i$ and $\chi^\dd_{ijk}$ 
the antifields associated to the vierbein, the vector 
fields, the gravitino and the Dirac fields, respectively, and by 
$\xi^\mu$, $\Omega^a{}_b$ and $c^m$ the anticommuting ghost fields 
associated to diffeomorphism invariance, Lorentz invariance, and 
abelian gauge invariance, respectively; the commuting supersymmetry 
ghost is $\epsilon^i$. In addition we also need antifields $\xi^\dd,
\Omega^{\dd\, ab}, c^{\dd\,m}$ and $\epsilon^{\dd\,i}$ for these ghost
fields.

Regarding gauge-fixing, the $\e_{7(7)}$ Ward identities can be implemented 
without further ado as long as the gauge-fixing manifestly preserves
$E_{7(7)}$ invariance. Of course, this is trivially the case for any 
sensible gauge choice for diffeomorphism and Lorentz invariance, and 
it is also true for the Coulomb gauge we are using. An $E_{7(7)}$-invariant 
gauge choice for local supersymmetry can be achieved in terms of the 
$SU(8)$-covariant derivative 
\be 
D_\mu \psi^i_\nu = \partial_\mu \psi^i  - \frac{1}{3} \Scal{u_{jk}{}^{IJ} 
\partial_\mu u^{ik}{}_{IJ} - v_{jkIJ} \partial_\mu v^{ikIJ }} \psi^j_\nu \ .
\ee
(for instance by setting $D^\mu \psi^i_\mu$ = 0), with the extra proviso 
that the supersymmetry antighost and the Nielsen--Kallosh field transform 
in the non-linear representation of $E_{7(7)}$ conjugate to the one 
of $\psi^i$ and $\epsilon^i$. 

In the conventional formulation of the theory, the supersymmetry algebra 
closes on the fermionic 
 fields only modulo terms linear in the fermionic 
equations of motion. Within the Batalin--Vilkovisky approach, this problem 
is cured by introducing terms quadratic in the fermion antifields in the 
action. The functional form of the BRST operator then includes these terms 
in the BRST transformation of the fermions. Let us briefly recall how 
this works. Collectively designating the fields and ghosts as $\varphi^\Afi$, 
and their Grassmann parity as $(-1)^\Afi$, we have
\be 
s^2 \varphi^\Afi = \sum_\Bfi K(\varphi)^{\Afi\Bfi}Ê
\frac{ \delta^L \Sigma}{\delta \varphi^\Bfi} \ . \label{DefK}  
\ee
This equation simply expresses the fact that algebra closes  (that is,
$s^2 \approx 0$) only if the equations of motion are imposed. 
Introducing antifields 
$\varphi^\dd_\Afi$, the action $\Sigma[\varphi,\varphi^\dd]$ reads
\be 
\Sigma (\varphi^\Afi, \varphi^\dd_\Afi) = 
\frac{1}{\kappa^2} S[\varphi] -\int d^4 x \left(Ê \sum_\Afi  
 (-1)^\Afi  \varphi^\dd_\Afi s \varphi^\Afi + 
 \frac{\kappa^2 }{2}Ê \sum_{\Afi \Bfi}Ê \varphi^\dd_\Afi K^{\Afi \Bfi}
 (\varphi)  \varphi^\dd_\Bfi \right) \ , \label{Sigma1}   
\ee
The symmetry of the action and the closure of the algebra can then be
combined into a single {\em BRST master equation}~\footnote{See e.g.
  \cite{HenneauxBook} for further information.} (indexed by a $\dd$ to 
distinguish it from the $E_{7(7)}$ master equation to be introduced
below)
\be \bigl( \Sigma , \Sigma \bigr)_\dd \equiv \sum_\Afi \int d^4 x 
  \frac{ \delta^R \Sigma}{\delta \varphi^\dd_\Afi} 
  \frac{ \delta^L \Sigma}{\delta \varphi^\Afi} = 0 \ .  \label{BRSTM} 
\ee
This equation requires in addition that 
\bea\label{sK} 
s K^{\Afi\Bfi} + \frac{1}{2}Ê\sum_\Cfi \left( K^{\Afi\Cfi}Ê\frac{\partial^L s \varphi^{\Bfi}}{\partial \varphi^\Cfi} + (-1)^{(\Afi + 1)(\Bfi + 1)}  K^{\Bfi\Cfi}Ê\frac{\partial^L s \varphi^{\Afi}}{\partial \varphi^\Cfi} \right) &=& 0   \ , \\*
\sum_\Dfi \left( K^{\Cfi\Dfi}Ê\frac{\partial^L K^{\Afi\Bfi}}{\partial \varphi^\Dfi} +  (-1)^{\Cfi( \Afi + \Bfi)} K^{\Afi\Dfi}Ê\frac{\partial^L K^{\Bfi\Cfi}}{\partial \varphi^\Dfi} + (-1)^{\Bfi( \Cfi + \Afi)} K^{\Bfi\Dfi}Ê\frac{\partial^L K^{\Cfi\Afi}}{\partial \varphi^\Dfi} \right) &=& 0 \ . \label{sK2}
\eea
These identities are automatically satisfied modulo the equations of motion 
by integrability of definition (\ref{DefK}). For $\N=8$ supergravity in the conventional formulation, the
term quadratic in the antifields in (\ref{Sigma1}) only involves the fermionic antifields $\psi_\mu^{*\,i}$ and $\chi^{*\, ijk}$ in a first approximation (\ie neglecting the antifield dependent terms in (\ref{DefK})).
 In supergravity, such $K^{\Afi\Bfi}$ components associated to the fermions are bilinear in the superghosts $\epsilon^i$ (and depend as well on the vierbeine and the scalar fields), and the validity of the identity (\ref{sK}) is ensured by certain cubic Fierz 
identities in $\epsilon^i$. (\ref{sK2}) is trivially satisfied because $K^{\Afi\Bfi}$ only depends on fields on which the algebra is satisfied off-shell. Nevertheless, this modification of the 
BRST transformation of the fermions also affects the closure of the 
algebra on the bosons, such that the BRST transformation of the Lorentz 
ghost $\Omega^a{}_b$ must include terms linear in the fermion antifields 
as well. This entails terms quadratic in the antifields involving the 
Lorentz ghost antifield $\Omega_{ab}^\dd$ as well. Considering these 
terms in the nilpotency of the linearised Slavnov--Taylor operator 
$( \Sigma , \cdot \, )$ on the vierbeine, \ie 
\be 
\bigl( \Sigma , \bigl(  \Sigma , e_\mu^a  \bigr)_\dd  \bigr)_\dd  = 
\kappa^2 \sum_\Afi \Bigl(  \bar \epsilon_i \gamma^a K_\mu^{i\, \Afi}(\varphi) 
+   \bar \epsilon^i \gamma^a K_{\mu i}{}^{\Afi}(\varphi)  + 
K^a{}_b{}^\Afi(\varphi)  e^b_\mu \Bigr) \varphi^\dd_\Afi \ \ , 
\ee
where $K_\mu^{i\, \Afi},\, K_{\mu i}{}^{\Afi}$ and $K^a{}_b{}^\Afi$ are 
the components of $K^{\Afi\Bfi}$ to be contracted with $\bar \psi_i^{\dd\mu},\,
\bar\psi^{\dd \mu i}$ and $\Omega^\dd_a{}^b$, respectively, one observes 
that $K^a{}_b{}^\Afi$ is determined in function of $K_\mu^{i\, \Afi},
\, K_{\nu i}{}^{\Afi}$, such that the modification of the action to be 
carried out amounts to replacing the gravitino antifields appearing 
in the term quadratic in the fermion antifields by 
\be 
\psi^{\dd\, \mu}_i \, \rightarrow \psi^{\dd\, \mu}_i - e^\mu_a \gamma_b 
\Omega^{\dd\, ab} \epsilon_i \ . 
\ee

In our manifestly $E_{7(7)}$-invariant formulation, the situation is 
the same with regard to the fermion fields, but now the vector fields 
are also governed by a first order Lagrangian (first order in the time 
derivative), and hence the algebra of gauge transformations on the 
vectors likewise involves the equations of motion. It is important 
that the equation of motion of the vector fields here appears as  
(\ref{EOM1}), and not in its integrated form (\ref{EOM3}), as required
for the consistency of the Batalin--Vilkovisky formalism. We have checked
that the diffeomorphism transformations do close among themselves, and that local supersymmetry closes on the vector fields. However, their commutator on the vector fields 
close modulo the equations of motion of the fermion fields, {\it viz.} 
\begin{multline}
Ês^2 A_\i^{IJ} = \frac{N \xi^\zero}{\sqrt{h}} e^{\zero a} 
e_\i^b \left[ u_{ij}{}^{IJ} \left(
\bar{\epsilon}^{i}\gamma_\mu \gamma_{ab}\frac{\delta^L S}{\delta\bar{\psi}_{\mu \, j}}  +12\bar{\epsilon}_k\gamma_{ab}\frac{\delta^L S}{\delta\bar{\chi}_{ijk}}
\right)  \right . \\* \left . - v^{ijIJ} \left(
\bar{\epsilon}_{i}\gamma_\mu\gamma_{ab}\frac{\delta^L S}{\delta\bar{\psi}_\mu^{j}}  +12\bar{\epsilon}^k\gamma_{ab}\frac{\delta^L S}{\delta\bar{\chi}^{ijk}} \right) \right] \ . 
\end{multline} 
To remain consistent with the basic symmetry property $ K^{\Afi\Bfi} = 
- (-1)^{\Afi\Bfi} K^{\Bfi\Afi}$, the closure of diffeomorphisms with 
local supersymmetry on the fermion fields then requires correspondingly the equations of motion of the vector fields. We checked that this is indeed the case, and that the fermion equations of motion are not involved (as follows trivially from Lorentz invariance). The quadratic terms in the fermion antifields are also modified by non-manifestly diffeomorphism invariant terms, such that they are 
manifestly duality invariant (and so do not depend on the scalar fields). 

The quadratic terms in the antifields of the gauge fields are responsible 
for the quartic terms in the ghosts that appear in supergravity 
\cite{BV,QuarticGhost}. It follows that in the duality 
invariant formulation, we will also have quartic terms depending on the 
diffeomorphism ghost $\xi^\zero$, the supersymmetry ghost $\epsilon^i$, 
the abelian antighost $\bar c_m$ and the supersymmetry antighost $\eta_i$, 
which in a flat Landau-type gauge for local supersymmetry like 
$ D^\mu \psi_\mu^i \approx 0 $,  would for example be of the form
\be 
\frac{N \xi^\zero}{\sqrt{h}}Êe^{\zero a} e^b_\i \partial_\i \bar c_{IJ} 
\Scal{Êu_{ij}{}^{IJ} \bar\epsilon^i \gamma^\mu \gamma_{ab} D_\mu  
\eta^j - v^{ijIJ}Ê\bar\epsilon_i \gamma^\mu \gamma_{ab} D_\mu  \eta_j }Ê+ 
\mbox{c.c.} \ . 
\ee
In general, such vertices do not contribute to amplitudes of physical fields. 
However, the renormalisation of the theory in the absence of regularisation 
preserving all gauge symmetries requires the renormalisation of the composite 
BRST transformations. In consequence, the correlation  functions involving 
the insertion of the BRST transformation of the vector fields do involve 
such vertices.

Note that one can obtain the solution $\Sigma$ of the master equation in the covariant formulation form the duality invariant one, by carrying out the Gaussian integration of the momentum variable $\Pi^{\m\, \i}$ for the complete action $\Sigma$ with antifields, similarly as in the second section. Considering for example the terms 
\begin{multline}   
S_{\scriptscriptstyle \rm vec}^{\scriptscriptstyle \rm ghost} =  
\int d^4x \biggl( A^{*\, \i}_m \Scal{ - \scal{\xi^\j + N^\j \xi^\zero} 
F_{\i\j}^m + \frac{N}{2\sqrt{h}} \xi^\zero h_{\i\j} \varepsilon^{\j\k\l} 
J^m{}_n F^n_{\k\l} }  \\* -   c^*_m  \Scal{\frac{1}{2} \xi^\i \xi^\j 
F_{\i\j}^m + \xi^\zero \xi^\i \scal{- N^\j F^m_{\i\j}  + 
\frac{N}{2\sqrt{h}}  h_{\i\j} \varepsilon^{\j\k\l} J^m{}_n F^n_{\k\l} }}  
+ \cdots \biggr) 
\end{multline}
and the $F^m_{\i\j}$ dependent terms that appear in the supersymmetry 
transformations of the fermions, as well as the gauge-fixing terms, one sees that the vector fields only appear through their field strength $F^m_{\i\j}$. It 
is therefore important (and true!) that the whole quantum action can be 
treated in the way described in the second section. The $\Re[ F_{\i\j}^{IJ}]$ dependent terms 
in the supersymmetry variation of the fermions and of the $\sl_2 (\IC)$ ghost 
$\Omega^a{}_b$, as well as the ones in the diffeomorphism variation of the vector fields and 
their ghost $c^{IJ}$, are replaced upon Gaussian integration of the momentum 
variables $\Pi^{IJ\, \i}$ by the solution of $\Pi^{IJ\, \i}$ according to 
 their equation of motion. This step will restore manifest diffeomorphism 
invariance. All these terms will also produce quadratic terms in the 
sources, which define the required equations of motion in order to close 
the gauge algebra in the $E_{7(7)}$ invariant formulation of the theory. 
This way one obtains that the only terms quadratic in $\Im[ A^{*\, \i}_{IJ}]$ 
involve the diffeomorphism ghost $\xi^\mu$, and that they vanish once
one puts the source $\Re[ A^{*\, \i}_{IJ}]$ equal to zero, in agreement 
with the explicit computation in the formalism.

\subsection{BRST extended current}
\label{BRSTCurrent}
Having set up the BRST transformations and the Batalin--Vilkovisky
framework, the next task is to define the $\e_{7(7)}$ current Ward 
identities in such a way that their mutual consistency is preserved also
in perturbative quantisation. To this aim, one must in principle couple the 
whole chain of operators associated to the current via the BRST descent 
equations, that is, extend the current constructed in section~\ref{ClassicalCurrent} by 
appropriate ghost and antifeld terms. Because the classical current 
defines a physical Noether charge which is BRST invariant, one has 
\be 
s J_\gra{3}{0}(\Lambda)   = - d J_\gra{2}{1}(\Lambda) \; \; , 
\ee
where we now write $J_\gra{3}{0}\equiv J$, indicating the form
degree and ghost number.
Considering the functional BRST operator $s$ acting on both fields 
and antifields, the conservation of the current reads 
\be 
d  J_\gra{3}{0} (\Lambda)  = - s J_\gra{4}{-1} (\Lambda) \; \; , 
\ee
where $J_\gra{4}{-1}$ is the composite operator linear in the 
antifields
\be J_\gra{4}{-1}(\Lambda) \equiv  \sum_\Afi \varphi^\dd_\Afi \sept(\Lambda) 
\varphi^\Afi \; \; . 
\ee
Then
\be 
d J_\gra{3}{0}(\Lambda) =   \sum_\Afi \left( \sept(\Lambda) \varphi^\Afi \, 
\frac{\delta^L \Sigma}{\delta \varphi^\Afi}  + \sept(\Lambda) \varphi^\dd_\Afi \, 
\frac{\delta^L \Sigma}{\delta \varphi^\dd_\Afi} \right)  \, \, ,    
\ee
as defined by the Noether procedure on the complete gauge 
fixed action $\Sigma[\varphi,\varphi^\dd]$, and where $\varphi^\dd_\Afi$ 
transforms with respect to $E_{7(7)}$ in the representation conjugate 
to $\varphi^\Afi$. 

The whole chain of operators appearing in the descent equations defines 
an extended form $\tilde{J}$ which is a cocycle of the extended 
differential $d+s$ \cite{HenneauxCoh},
\be 
( d + s ) \scal{ J_\gra{4}{-1} + J_\gra{3}{0} + J_\gra{2}{1} + 
J_\gra{1}{2} + J_\gra{0}{3} } = 0 \; \; . 
\ee
The complete form of the extended current $\tilde{J}$ which now also
depends on the ghosts and antifields is again very complicated, and its
explicit form would not be very illuminating. Let us nevertheless 
discuss some salient features of this extended current, neglecting 
terms depending on the antifields and terms linear in the equations 
of motion. With these assumptions we can take $J_\gra{4}{-1} $ to vanish, 
and $J_\gra{3}{0}$ can be identified with the Gaillard--Zumino current
constructed in section~\ref{ClassicalCurrent}, where we also disregard the `curl component' 
leading to a trivial cocycle. Let us first rewrite the components of 
$J_\gra{3}{0}$ in terms of differential forms, cf. (\ref{J1F}, \ref{Rmu})
\bea 
R^i{}_j &=& - 2 i e^a_{\, \wedge}Ê\Scal{Ê \bar{\psi}^{i}_{\, \wedge}Ê\gamma_a 
\psi_{j} - \frac{1}{8}Ê\delta^i_j  \bar{\psi}^{k}_{\, \wedge}Ê\gamma_a 
\psi_{k} } -   \frac{1}{48} \varepsilon_{abcd} e^{b}_{\, \wedge}Êe^c_{\, \wedge}Êe^d \Scal{Ê  \bar{\chi}^{ikl} \gamma^a \chi_{jkl} - \frac{1}{8} \delta^i_j   \bar{\chi}^{klp} \gamma^a \chi_{klp} }\; \; , \CR
R_{ijkl} &= &- \star \hat{\mathcal{A}}_{ijkl} + \frac{i}{2} e^a_{\, \wedge} e^b_{\, \wedge}Ê \Scal{Ê \bar{\chi}_{[ijk} \gamma_{ab}  \psi_{l]}  + \frac{1}{4!} \varepsilon_{ijklpmnpq}  \bar{\chi}^{mnp} \gamma_{ab}  \psi^{q} }  \label{RinForm}\; \; , 
\eea
and 
\be  J_{\scriptscriptstyle \rm GZ}^{(2)}(\Lambda) = - 
\frac{1}{2} \, ÊA^m_{\, \, \wedge} F^n \, 
\Lambda_m{}^p \Omega_{pn}\, \, .
\ee
Conveniently, the extended current $\tilde{J}$ takes a form similar
to the current constructed in section~\ref{ClassicalCurrent} 
\be 
\tilde{J} (\Lambda) = - \frac{1}{24}Êe^{i_\xi} \mbox{tr}Ê\scal{Ê\V^{-1}Ê\tilde{R} \V \Lambda }Ê+  \tilde{J}_{\scriptscriptstyle \rm GZ}^{(2)}(\Lambda) \ . \label{ExtendedCurrent}
\ee
so we only need to explain how to obtain the `tilded' version of the above
currents. The operator $i_\xi$ is the (commuting) Cartan contraction with 
respect to the anti-commuting vector $\xi^\mu$; its exponentiated action 
takes care automatically of all modifications involving the diffeomorphism 
ghost fields $\xi^\mu$ \cite{Baulieu}. In order to understand how to extend 
the remaining piece $J_{\scriptscriptstyle \rm GZ}^{(2)}(\Lambda) $ to $\tilde{J}_{\scriptscriptstyle \rm GZ}^{(2)}
(\Lambda) $, it is again convenient to write a Russian formula 
\be 
( d + s ) \scal{ÊA^m + c^m } = e^{i_\xi}Ê\tilde{F}^m \ , 
\ee
where the extended curvature $\tilde{F}$ is defined as 
\begin{multline}Ê
\tilde{F}^{IJ} \equiv  F^{IJ} + u_{ij}{}^{IJ} \Scal{ \frac{1}{4}Ê 
\bar{\epsilon}_k e^a \gamma_a \chi^{ijk} + 2 \bar{\epsilon}^i \psi^j  +  
\bar{\epsilon}^i \epsilon^j  } - v^{ijIJ} \Scal{ \frac{1}{4}Ê \bar{\epsilon}^k
 e^a \gamma_a \chi_{ijk}   + 2 \bar{\epsilon}_i \psi_j  +   
\bar{\epsilon}_i \epsilon_j  } \\*
\hspace{-45mm}Ê= \hat{F}^{IJ} + u_{ij}{}^{IJ} \Scal{ \frac{1}{4}Ê \overline{[\psi + \epsilon]}_k e^a 
\gamma_a \chi^{ijk} +  \overline{[\psi+ \epsilon]}^i [\psi+ \epsilon] ^j  } 
\\* -  v^{ijIJ} \Scal{ \frac{1}{4}Ê \overline{[\psi+ \epsilon]}^k e^a \gamma_a \chi_{ijk}   
+ \overline{[\psi+ \epsilon]}_i [\psi+ \epsilon]_j } \label{ExtendedF}
 \; \; . 
\end{multline}
The gravitinos here appear only in the supercovariantisation of $F^{IJ}$ or through the combination $\psi^i + \epsilon^i$. In addition we need the nilpotent extended differential \cite{Baulieu}
\be 
\tilde{d} \equiv  e^{-i_\xi} ( d + s ) e^{i_\xi}  = d + s - 
\L_\xi + i_{( \bar \epsilon \gamma \epsilon)} \, \, , 
\ee
where $i_{( \bar \epsilon \gamma \epsilon)}$ is the Cartan contraction 
with respect to the vector $\bar \epsilon_i \gamma^\mu \epsilon^i$. 
Defining 
\be 
\tilde{A}^mÊ\equiv  A^m + c^m - i_\xi A^m \ , 
\ee
it is obvious that 
\be   
( d + s) \, e^{i_\xi}  \,  \Big( Ê\tilde{A}^m_{\, \, \wedge} \tilde{F}^n \, 
\Lambda_m{}^p \Omega_{pn} \Big)\, \,  = e^{i_\xi} \, 
\Big( \tilde{F}^m_{\, \, \wedge}Ê\tilde{F}^n \, \Lambda_m{}^p \Omega_{pn}
\Big) \, \, . 
\ee
The right-hand-side being gauge-invariant, the extended form  $\tilde{R}$ 
can be obtained from the equation 
\be 
\tilde{d}  \,\left(  \frac{1}{24}Ê \mbox{tr}Ê\scal{Ê\V^{-1}Ê\tilde{R} 
\V \Lambda }\right) = 
\, \frac{1}{4} \, \tilde{F}^m_{\, \, \wedge}Ê\tilde{F}^n \, 
\Lambda_m{}^p \Omega_{pn}\, \, . \label{ExtGZ} 
\ee
which is an extended version of the Gaillard-Zumino construction.
For any gauge invariant extended form, such as $\tilde{R}$ or $\tilde{F}^m$, 
 supersymmetry covariance implies that the gravitino field $\psi^i$ only 
appears in supercovariant forms, or `naked', through the wedge product 
of $\psi^i+ \epsilon^i$ with supercovariant forms. It follows that 
$\tilde{R}$ is simply obtained 
from $R$ by performing the replacement $\psi^i\rightarrow
\psi^i+ \epsilon^i$ everywhere inside (\ref{RinForm}). 

The $(4,0)$ component of  (\ref{ExtGZ}) is simply the current conservation. 
To see that   (\ref{ExtGZ}) is indeed satisfied for the other components, 
let us consider the $(0,4)$ component of this equation. From (\ref{ExtendedF})
we see that the right hand side is the $\e_{7(7)}$ component of the square 
of $u_{ij}{}^{IJ} \bar\epsilon^i \epsilon^j - v^{ijIJ}Ê \bar\epsilon_i 
\epsilon_j $. By $E_{7(7)}$ covariance, the scalar fields dependence then 
reduces to a similarity transformation with respect to $\V$ (as the left 
hand side), and one can concentrate on the $\e_{7(7)}$ element quadratic 
in $\bar\epsilon^i \epsilon^j $. Because (for commuting spinors)
\be 
\bar\epsilon^{[i} \epsilon^j\bar\epsilon^k \epsilon^{l]} = 0 \ ,  
\ee
this term only contributes in the $\su(8)$ component $i Ê( Ê\bar \epsilon^i \epsilon^k) (  \bar \epsilon_j \epsilon_k ) - 
\frac{i}{8} \delta^i_j ( Ê\bar \epsilon^k \epsilon^l) 
(  \bar \epsilon_k \epsilon_l ) $. Because $\tilde{R}$ has a vanishing $(0,3)$ component, the left hand side is the Cartan contraction of its $(1,2)$ component $-2i e^a  \scal{Ê\bar \epsilon^i \gamma_a \epsilon_j 
- \frac{1}{8}Ê\delta^i_j \bar \epsilon^l \gamma_a \epsilon_l }$  with the vector $\epsilon_i \gamma^\mu \epsilon^i$.  Using the Fierz identity 
\be 
Ê( Ê\bar \epsilon^i \epsilon^k) (  \bar \epsilon_j \epsilon_k ) - 
\frac{1}{8} \delta^i_j ( Ê\bar \epsilon^k \epsilon^l) 
(  \bar \epsilon_k \epsilon_l ) Ê= -  \frac{1}{2}Ê ( \bar 
\epsilon_k \gamma^a \epsilon^k ) \Scal{Ê\bar \epsilon^i \gamma_a \epsilon_j 
- \frac{1}{8}Ê\delta^i_j \bar \epsilon^l \gamma_a \epsilon_l }Ê\ ,
\ee
one obtains the validity of the $(0,4)$ component of (\ref{RinForm}). 

Considering the complete antifield dependent extended current $\tilde{J}$,~\footnote{We have computed the complete $\xi^\mu$ dependent part of $\tilde{J}$ including the antifields to check that the non-manifest Lorentz invariance does not give rise to extra difficulties. Nevertheless, its exhibition would not shed much light in this discussion. However we have not computed explicitly the $\epsilon^i$ dependent terms that would involve the quadratic terms in the antifields of the solution $\Sigma$ of the master equation.} one can 
couple the $E_{7(7)}$ current to the action in a way fully consistent with 
BRST invariance. Indeed, considering sources $B_\gra{p}{1-p}$ for each 
component of the current $\tilde{J}$, one obtains that 
\be 
\Sigma[B] = \Sigma + \int \tilde{B}_{\, \wedge} \tilde{J} \ , 
\ee
where we use the Berezin notation
\be 
\int \trace  \tilde{B}_{\, \wedge} \tilde{J} = 
\int \trace \Scal{ B_\gra{0}{1} J_\gra{4}{-1} + B_{\gra{1}{0}\, \wedge} 
J_\gra{3}{0} + B_{\gra{2}{-1}\, \wedge} J_\gra{2}{1} + 
B_{\gra{3}{-2}\, \wedge} J_\gra{1}{2} + B_{\gra{4}{-3}} J_\gra{0}{3} } \ , 
\ee
satisfies the master equation 
\be 
\scal{ \Sigma , \Sigma }_\dd - \int d \tilde{B}_{\, \wedge} 
\frac{ \delta \Sigma}{\delta \tilde{B} }Ê= 0  \ . 
\ee
This formal notation means that 
\be 
( d + s ) \tilde{B} = 0  \ .  
\ee
This would be enough for insertions of one single current in a BRST invariant 
way, but consistency with $E_{7(7)}$ will require the consideration of
higher order terms in $\tilde{B}$ in $\Sigma$, such that these equations are then only valid up to quadratic terms in the sources $B_\gra{p}{1-p}$.

Introducing a source for the $E_{7(7)}$ current, the rigid $\e_{7(7)}$ 
Ward identity is promoted to a local $\e_{7(7)}$ Ward identity expressing
the conservation of the $E_{7(7)}$ current, such that 
\be 
\sept \tilde{B} = -  d C -   \{ \tilde{B} , C \} \; \; . 
\ee 
All the components of $\tilde{B}$ thus transform in the adjoint 
representation, and $B_\gra{1}{0}$ transforms as an $\e_{7(7)}$ gauge field. 
In order for the current Ward identity to be satisfied, each derivative 
in the action must be replaced by an $\e_{7(7)}$ covariant derivative 
with respect to the gauge field $B_\gra{1}{0}$. It follows that the 
linear component is defined as $ \int \trace  B_{\gra{1}{0}\, \wedge} 
J_\gra{3}{0}  $, by definition of the Noether current. The kinetic terms 
of the scalar fields, the Maxwell fields, their ghosts, and the 
supersymmetry ghost being quadratic in derivatives, they give rise to 
bilinear terms in $B_\gra{1}{0}$ in the action. The compatibility 
with BRST invariance therefore requires to also add  quadratic terms 
in the other sources defining $\tilde{B}$. 

In order to ensure that $\delta^\g$ anticommutes with $s$, one must 
then define the BRST transformation of $\tilde{B}$ such that 
\be ( d + s ) \tilde{B}  + \tilde{B}^2 = 0 \ . \ee
In this way one has the consistent `very extended' Russian formula
\be 
( d + s + \delta^\g ) \scal{Ê\tilde{B} + C }Ê  + 
\scal{Ê\tilde{B} + C}^2 = 0 \ ,  
\ee 
and
\be 
s B_\gra{0}{1}  = - {B_\gra{0}{1}}^2 \qquad s B_\gra{1}{0} = - d   
{B_\gra{0}{1}} - [B_\gra{1}{0} , B_\gra{0}{1} ] \ .  
\ee
The master equation for the completed $\Sigma[\tilde{B}]$ (including 
quadratic couplings in $\tilde{B}$ is therefore 
\be 
\scal{ \Sigma , \Sigma }_\dd - \int ( d \tilde{B} + \tilde{B}^2)_{\, \wedge} 
\cdot \frac{ \delta \Sigma}{\delta \tilde{B} }Ê= 0  \ , 
\ee
It is straightforward to compute the solution $\Sigma[\tilde{B}]$ for
a non-linear sigma model coupled to gravity, but the derivation of the 
complete solution in the case of $\N=8$ supergravity is beyond the scope 
of this paper. Nevertheless, one can say that this solution can be written as 
\be  
\Sigma[\tilde{B}] = \frac{1}{\kappa^2} S[\varphi,\tilde{B}] -
\int d^4 x Ê\left(  
\sum_\Afi  (-1)^\Afi  Ê\varphi^\dd_\Afi s_{\tilde B} \varphi^\Afi + 
\frac{\kappa^2 }{2} \sum_{\Afi \Bfi}Ê Ê\varphi^\dd_
\Afi K^{\Afi \Bfi}(\varphi)  Ê\varphi^\dd_\Bfi \right) \ , 
\label{BRSTMasterSol} 
\ee
such that $s_{\tilde B}$ defines a differential operator which is nilpotent 
modulo the equations of motion of $S[\varphi, \tilde{B}]$ satisfying
\be 
s_{\tilde{B}} S[\varphi, \tilde{B}] = 0 \, \,  , 
\ee
and which anti-commutes with $\delta^\g(C)$ for a $x$ dependent 
parameter $C$.~\footnote{Whereas the BRST operator $s$ anti-commutes with $\delta^\g(C)$ 
 only for constant parameter $C$.}   We emphasise that this is not 
equivalent to gauging the theory with respect to a local $E_{7(7)}$ 
symmetry, because the components of $\tilde{B}$ are classical sources 
and do not constitute part of a supermultiplet in the conventional sense. 

In order to arrive at a consistent definition of the BRST master equation 
(\ref{BRSTM}) and the $\e_{7(7)}$ master equation (\ref{MasterG}), one has 
to introduce sources $\varphi^\g_\Afi$ for the non-linear symmetry, 
sources (or antifields) for the BRST transformations, as well as sources 
$\varphi^{\dd\g}_\Afi$ for the non-linear transformations of the 
BRST transformations \cite{Shadow}, which all transform with respect 
to $E_{7(7)}$ in the representation conjugate to the one of the corresponding 
fields. Given the $E_{7(7)}$ invariant solution (\ref{BRSTMasterSol}) to 
the BRST master equation, one computes that the complete action \footnote{The 
  only sources $\varphi^{\dd\g}_\Afi$  that are involved quadratically in 
  the action are $\psi^{\dd\g\, \mu}_i,\, \chi^{\dd\g}_{ijk}$, 
  $A^{\dd\g\, \i}_{m}$, and $- \delta^\pa(C_\pa)$ is defined as a 
  linear $\e_{7(7)}$ transformation on $A^{\dd\g\, \i}_m$, and as an 
  $\su(8)$ transformation of parameter $ \tanh(\Phi / 2) \ast C_\pa$ on 
  $\psi^{\dd\g\, \mu}_i$ and $\chi^{\dd\g}_{ijk}$.}
\begin{multline}  
\Sigma = \frac{1}{\kappa^2} S[\varphi,\tilde{B}] -\int d^4 x Ê \sum_\Afi  (-1)^\Afi  \Scal{Ê\varphi^\dd_\Afi s_{\tilde B} \varphi^\Afi +   
\varphi^\g_\Afi \delta^\pa(C_\pa) \varphi^\Afi  +   
\varphi^{\dd\g}_\Afi \delta^\pa(C_\pa) s_{\tilde B}  \varphi^\Afi } \\* - 
\frac{\kappa^2 }{2}Ê\int d^4 x  \sum_{\Afi \Bfi}Ê \Scal{Ê  \varphi^\dd_\Afi 
- (-1)^\Afi \delta^\pa(C_\pa) \varphi^{*\g}_\Afi }ÊÊK^{\Afi \Bfi}(\varphi)  
\Scal{Ê  \varphi^\dd_\Afi - (-1)^\Bfi \delta^\pa(C_\pa) \varphi^{\dd\g}_\Bfi }
\ ,  
\end{multline}
yields a consistent solution of the BRST master equation 
\be
Ê\int d^4 x \sum_\Afi \left( \frac{ \delta^R \Sigma }{\delta \varphi^\dd_\Afi}Ê\frac{ \delta^L \Sigma}{\delta \varphi^\Afi}Ê-  (-1)^\Afi \varphi^\g_\Afi \frac{Ê\delta^L \Sigma }{\delta \varphi^{\dd\g}_\Afi}Ê\right) - \int ( d \tilde{B} 
+ \tilde{B}^2)_{\, \wedge} \cdot \frac{ \delta \Sigma}{\delta \tilde{B} }Ê= 0
 \ , 
\ee
the linear $\su(8)$ Ward identity 
\begin{multline}  
\int d^4 x \sum_\Afi \left( \delta^\ka(C_\ka) \varphi^\Afi \, \frac{\delta^L \Sigma}{\delta \varphi^\Afi} +  \delta^\ka(C_\ka) \varphi^\g_\Afi \, \frac{\delta^L \Sigma}{\delta \varphi^\g_\Afi} +  \delta^\ka(C_\ka) \varphi^\dd_\Afi \, \frac{\delta^L \Sigma}{\delta \varphi^\dd_\Afi} + \delta^\ka(C_\ka) \varphi^{\dd\g}_\Afi \, \frac{\delta^L \Sigma}{\delta \varphi^{*\g}_\Afi} \right)\\*    Ê- \int  \left(  \scal{Êd C_\ka + \{Ê\tilde{B}_\ka , C_\ka \} }{}_{\wedge} \cdot \frac{\delta^L \Sigma}{\delta \tilde{B}_\ka}  +    \{Ê C_\ka , \tilde{B}_\pa \}{}_{\wedge}   \cdot \frac{\delta^L \Sigma}{\delta \tilde{B}_\pa}  + \{ÊC_\ka , C_\pa \}Ê \cdot \frac{\delta^L \Sigma}{\delta C_\pa} \right) = 0 \ ,  
\end{multline}
and the $E_{7(7)}$ master equation  
\begin{multline}Ê 
\int d^4 x \sum_\Afi \left(  \frac{ \delta^R \Sigma}{\delta \varphi^\g_\Afi}\frac{\delta^L \Sigma}{\delta \varphi^\Afi} Ê+ (-1)^\Afi \varphi^\dd_\Afi \frac{Ê\delta^L \Sigma }{\delta \varphi^{\dd\g}_\Afi}Ê
 +  (-1)^\Afi \delta^\ka(C_\pa{}^2) \varphi^{\dd\g}_\Afi  \frac{Ê\delta^L \Sigma }{\delta \varphi^{\dd}_\Afi}Ê
- \varphi^\g_\Afi \delta^\ka(C_\pa{}^2) \varphi^\Afi  \right)  \\*
- \int \left(  \scal{Êd C_\pa + \{Ê\tilde{B}_\ka , C_\pa \} }{}_{\wedge}  \cdot \frac{\delta^L \Sigma}{\delta \tilde{B}_\pa}  +     \{Ê C_\pa , \tilde{B}_\pa \}{}_{\wedge}Ê  \cdot \frac{\delta^L \Sigma}{\delta \tilde{B}_\ka}   \right) = 0 \ .   
\end{multline}
According to the quantum action principle \cite{PiguetSorella}, these 
functional identities are satisfied by the $n$-loop 1PI generating functional 
$\Gamma_n$, modulo possible anomalies defined by local functionals 
$\mathcal{A}^\g_n$ and $\mathcal{A}^\dd_n$. We have established in this 
paper that there is no non-trivial anomaly for the non-linear $E_{7(7)}$ 
master equation. It is commonly admitted (although no general proof exists 
to our knowledge) that there is no non-trivial anomaly to the BRST master 
equation in four dimensions (that is, diffeomorphisms and local supersymmetry
are non-anomalous in four space-time dimensions). Once one has enforced the 
$E_{7(7)}$ master equation, the cohomology of the BRST operator of ghost 
number one associated to the possible anomalies to the BRST symmetry must 
be defined on the complex of $E_{7(7)}$ invariant functionals. Nevertheless, 
it rather obvious that the a BRST antecedent of an $E_{7(7)}$ invariant 
solution to the BRST Wess--Zumino consistency condition can always be 
chosen to be $E_{7(7)}$ invariant. We therefore conclude that there 
exists a renormalisation scheme such that these three functional identities 
are satisfied by the 1PI generating functional $\Gamma$ to all orders 
in perturbation theory. 

The Pauli--Villars regularisation employed in this paper breaks all these 
Ward identities, and so the determination of the non-invariant finite 
counterterms would require checking their validity in each order of 
perturbation theory. In principle, preserving $E_{7(7)}$ invariance 
requires testing the $\e_{7(7)}$ Ward identities separately, and local 
supersymmetry will not be enough. As an example,  the 
three-loop supersymmetry invariant starting as the square of the 
Bel--Robinson tensor does not preserve $E_{7(7)}$ invariance \cite{BroedelDixon,Elvang}. 
Therefore, the supersymmetry master equation does not determine its coefficient 
in the bare action, independently of the property that there is no 
logarithmic divergence at this order, and one must use the $E_{7(7)}$ 
master equation to determine its value. $L=3$ is therefore the first 
loop order at which a renormalisation prescription may fail to preserve 
$E_{7(7)}$ invariance. One would expect that the prescription used in
\cite{Bern,Bern2} to compute $\N=8$ on-shell amplitudes should satisfy 
the $\e_{7(7)}$ Slavnov--Taylor identities, but this needs to be checked.

The BRST master equation and the $E_{7(7)}$ master equation are 
more constraining than the requirement of local supersymmetry and 
rigid $E_{7(7)}$ invariance. For this reason it would be interesting 
to see if the prospective divergent counterterms at 7 and 8 loop could 
possibly be ruled out by these master equations.

\subsection{Energy Coulomb divergences}
\label{Coulomb}
There is still one subtlety concerning the Coulomb gauge which we have
not yet addressed. It is well known that the Coulomb gauge in non-abelian 
gauge theories gives rise to energy divergences which are not easily 
dealt with in the renormalisation program \cite{Coulomb,Coulomb2}. Because 
the ghost `kinetic' term does not involve a time derivative, any ghost loop 
contribution is the energy integral of a function independent of the 
energy $k_\zero$, which diverges linearly. However, in the flat Coulomb 
gauge we use the ghost field $c^m$ only appear in its free `kinetic' term
\be\label{cddc} 
- \bar c_m \partial_\i \partial_\i c^m \ . \ee
Therefore, although the antighost $\bar c_m$ couples to the other fields 
via the diffeomorphism ghosts $\xi^\mu$  and the supersymmetry ghosts 
$\epsilon^i$, and so `ghost particles' can decay, they cannot be created, 
and there is no closed loops involving the ghost $c^m$. It follows that 
the Coulomb energy divergences do not appear in the loop corrections to 
amplitudes. It is in fact very important that the Coulomb gauge we use 
is field independent for this property to be true. For instance, a metric 
dependent gauge such as $ \partial_\i (\sqrt{h} h^{\i\j}ÊA_\j)$ would 
give rise to the ghost Lagrangian
\be 
- \bar c_m \partial_\i \scal{ \sqrt{h} h^{\i\j} \partial_\j c^m } \ ,  
\ee
whence perturbation theory would involve energy divergences through the 
couplings to the metric. Although BRST invariance in principle guarantees 
that these energy divergences should cancel with the energy divergences 
involving vector fields, the compensating process might be difficult 
to exhibit. 

Even within the `free Coulomb gauge', the energy divergences do not 
disappear when one considers insertions of non-gauge-invariant composite 
operators, and in particular when one considers insertions of the 
$E_{7(7)}$ current, since the latter couples to the ghosts in a way very 
similar as in non-abelian gauge theory, in such a way that (\ref{cddc})
is replaced by
\be\label{cDDc}
-\bar{c}_m D_\i D_\i c^m 
\ee
with the $E_{7(7)}$ covariant derivative $D_\i c^m \equiv \partial_\i c^m 
+ B_\i^m{}_n c^n$. For all (and only for) the correlation functions 
of $N$ $E_{7(7)}$ currents, there is one one-loop diagram associated to 
a `ghost particle' interacting with each of the currents for each 
ordering of the currents, which gives an integral of the form
\begin{multline}  \Bigl< \prod_{a=1}^N J^{\i_a}(X_a,p_a) \Bigl>_{\rm \scriptscriptstyle ghost} =  -   2 \sum_{\varsigma} \trace \left( \prod_{\varsigma(a)=1}^N X_{\varsigma(a)} \right) \times \\*   \int \frac{d^4 k}{(2\pi)^4} \frac{ \left( 2 k^{\i_N}Ê- \sum_{c=1}^{N-1} p_{c}^{\i_N} \right)   \prod_{a=1}^{N-1} \Scal{Ê2 \left(  k^{\i_{a}}   + \sum_{b=1}^{a-1} p_{b}^{\i_{a}}  \right) + p_{a}^{\i_{a}} }  }{Ê\prod_{a=1}^N \left( k + \sum_{b=1}^{a-1}Êp_{b} \right)^2}   + \mbox{C.T.}\ , \end{multline}
where the sum over $\varsigma$ is the sum over non-cyclic permutations, 
(\ie the permutations identified modulo cyclic ones), and C.T. correspond to the diagrams involving contact terms.

The contributions of the vector fields to such insertion is given at one-loop by 
\begin{multline}  \Bigl< \prod_{a=1}^N J^{\i_a}(X_a, p_a) \Bigr>_{\rm \scriptscriptstyle vec} = (-i)^N \sum_{\varsigma} \trace \prod_{a=1}^N  X_{\varsigma(a)}  Ê\int \frac{d^4 k}{(2\pi)^4}Ê\prod_{b=1}^N \Bigl( \Delta(k_{\varsigma,b}) \Upsilon^{\i_{\varsigma(b)}}(k_{\varsigma,b},k_{\varsigma,b}+p_{\varsigma(b)}) \Bigr)\\*  + \mbox{C.T.}  \ , \label{Ncurrent} \end{multline}
where $k_{\varsigma,a} = k + \sum_{\varsigma(b)=1}^{\varsigma(a)-1} p_{\varsigma(b)}$ and the sum over $\varsigma$ is the sum over non-cyclic permutations. The leading order in  $\k_\zero$ in the limit $k_\zero {}^2 \rightarrow + \infty $ of the product 
\bea && \Delta(k) \Upsilon^\k(k,k+p)  = \CR
  &&\hspace{-15mm} \frac{1}{k^2}Ê
\left( \begin{array}{cc}\ i \delta^m_n \scal{Ê\delta_\i^\j k^\k - \delta_\i^\k k^\j + k_\i \delta^{\k\j}Ê} + \mathcal{O}(k_\zero{}^{-1})  \ & \ \Omega^{mn}Ê\varepsilon^\k{}_{\i\l} k^\l k_\zero{}^{-1} + \mathcal{O}(k_\zero{}^{-2}) 
  \ \ \vspace{2mm} \\ \Omega_{mn} \varepsilon^{\j\k\l} k_\l k_\zero + \mathcal{O}(1)   &  i k^\k \delta^m_n  \end{array}\right) \label{LeadTrunc} 
\eea
is such that 
\be \Bigl< \prod_{a=1}^N J^{\i_a}(X_a, p_a) \Bigr>_{\rm \scriptscriptstyle vec} =  \sum_{\varsigma} {\rm tr} \prod_{a=1}^N  X_{\varsigma(a)}  Ê\int \frac{d^4 k}{(2\pi)^4} {\rm tr}Ê\left(  \prod_{b=1}^N K^{\i_{\varsigma(b)}} (k_{\varsigma,b})  + \mathcal{O}(k_\zero{}^{-1}) \right)  \ , \ee
with
\be K^\k(k) = \frac{1}{k^2} \left( \begin{array}{cc}\ \delta_\i^\j k^\k - \delta_\i^\k k^\j + k_\i \delta^{\k\j}Ê  \ & \ Ê\varepsilon^\k{}_{\i\l} k^\l  
  \ \ \vspace{2mm} \\ \varepsilon^{\j\k\l} k_\l   &  k^\k  \end{array}\right) \ , \ee
where we used the property that the trace is invariant with respect to 
inverse rescalings of the two off-diagonal components,\footnote{This 
  can easily be proved using a similarity transformation of the form 
  $K \rightarrow S^{-1} K S$ with \be S = \left( \begin{array}{cc}\ 
  \delta_\i^\j k_\zero{}^\frac{1}{2}   \ & \ Ê0   \ \ 
 \vspace{2mm} \\ 0    & k_\zero{}^{-\frac{1}{2}}   \end{array}\right) 
 \nonumber \ee} and the property that the contact terms are subleading in $k_\zero$ because
 \be  \Delta(k) \cR^{\i\j} = \left( \begin{array}{cc}\  \mathcal{O}(k_\zero{}^{-1})  \ & \ 0  
  \  \vspace{2mm} \\  \mathcal{O}(1)   &  0  \end{array}\right)  \ . \ee
We observe that this matrix can be written 
\be 
K^\k(k) =  \frac{k_\i  \sigma^\i }{k^2} \sigma^\k \ , 
\ee
where the $\sigma^\i$ are the $4\times 4$ pure imaginary Pauli matrices,
\be 
\sigma^\k \equiv   i \left( \begin{array}{cc}\ \varepsilon_\i{}^{\j\k}  \ & \ \delta^\k_\i 
  \ \ \vspace{2mm} \\ - \delta^{\k\j}   &  0  \end{array}\right) \ , 
\ee
satisfying 
\be 
\sigma^\i \sigma^\j =  \delta^{\i\j} - i  \varepsilon^{\i\j}{}^\k 
\sigma^\k \ . 
\ee
Rewriting the `leading' vector field contribution to the $N$ $\su(8)$ 
currents insertion in this way,
\be 
\Bigl< \prod_{a=1}^N J^{\i_a}(X_a, p_a) \Bigr>_{\rm \scriptscriptstyle vec} = \int \frac{d k_\zero}{2\pi}  \sum_{\varsigma} {\rm tr} \prod_{a=1}^N  X_{\varsigma(a)}  Ê\int \frac{d^3 k}{(2\pi)^3} {\rm tr} \left(  \prod_{b=1}^N \frac{1}{ \ba k_{\varsigma,b}} \sigma^{\i_{\varsigma(b)}}  + \mathcal{O}(k_\zero{}^{-1}) \right)  \ ,  
\ee
one recognises that the integrand 
\be  
\sum_{\varsigma} {\rm tr}\prod_{a=1}^N X_{\varsigma(a)}  \int \frac{d^3 k}{(2\pi)^3}    {\rm tr} \prod_{b=1}^N \frac{1}{ \ba k_{\varsigma,b}} \sigma^{\i_{\varsigma(b)}}   \ ,  
\ee
is the one-loop $N$ $\su(8)$-currents insertion in a three-dimensional 
theory of free bosonic spinor fields. 

It follows that the contribution to the $N$ $\su(8)$-current insertions 
responsible for energy divergences can be computed in an Euclidean 
three-dimensional effective theory, with 56 doublets of anti-commuting 
scalar fields $\bar c_m ,\, c^m$ and 56 Dirac spinor fields $\lambda^m$, 
understood as $SU(2)$ ${\bf 2} \oplus \bar {\bf 2}$ real spinors with 
\be 
\bar \lambda_m = \lambda^{n\, T} G_{nm} \ , 
\ee
coupled to an external $\su(8)$-current as 
\be 
S^{\rm 3D}Ê=  \int d^3 x \Bigl(  \frac{1}{2} \bar \lambda_m \baaa D 
\lambda^m - \bar c_m D_\i D^\i c^m \Bigr) \ . 
\ee
The corresponding contributions to the $N$ $\su(8)$-currents insertions are
\be 
\exp({-\Gamma[B]}) = \frac{{\rm Det}[D_\i D^\i]}{{\rm Det}[\baaa D]^\frac{1}{2}}  \ . \label{DetRatio} 
\ee
and therefore do not vanish. Nevertheless, they can be compensated by the contribution of a trivial free-theory. Consider the fermionic fields $\theta_\i^m ,\,  \bar \theta_m$ and the bosonic fields $L^m,\,  \bar L_m$, with BRST transformations 
\be s \theta_\i^m = \partial_\i L^m \  , \qquad s L^m = 0 \  , \qquad s \bar L_m  = \bar \theta_m \ , \qquad s \bar \theta_m = 0 \ .  \ee
The BRST invariant Lagrangian 
\be \frac{1}{2} \Omega_{mn} \varepsilon^{\i\j\k}Ê \theta^m_\i \partial_\j \theta_\k^n + s  \, \Scal{Ê \bar L_m  \partial_\i \theta_\i^m }Ê=  \frac{1}{2} \Omega_{mn} \varepsilon^{\i\j\k}Ê \theta^m_\i \partial_\j \theta_\k^n +    \bar \theta_m  \partial_\i \theta_\i^m  +  \bar L_m  \partial_\i \partial_\i L ^m \ , \ee
is a fermionic equivalent of the abelian Chern--Simons Lagrangian. The coupling of this theory to the current gives rise to a contribution to the $N$ $\su(8)$-current insertions which cancels the ratio of determinants (\ref{DetRatio}). One can therefore disregard the energy divergences without affecting the BRST symmetry, although the extended current (\ref{ExtendedCurrent}) is modified by a non-trivial BRST cocycle
\be \tilde{J}(\Lambda)_{\scriptscriptstyle \rm C} \approxÊ \frac{1}{2} dt_{\, \wedge}  \scal{Êdx^\i  \theta^m_\i  +  L^m }{}_{\wedge} \scal{Êdx^\j   \theta^n_\j +  L^n } \Omega_{np}Ê\Lambda_m{}^p  \ . \ee
Nevertheless, this term vanishes when the equations of motion are imposed with the appropriate boundary conditions, 
\be \partial_{[\i} \theta_{\j]}^m = 0  \; ,  \quad \partial_\i \theta_\i^m = 0 \quad   \Rightarrow \quad \theta_\i^m = 0 \; . \ee

This contribution to the energy divergences is reproduced by the Pauli--Villars fields, within the  prescription for the vector fields defined in section~\ref{PauliVillarsReg}, and the prescription for the ghosts that their Pauli--Villars Lagrangian is mass-independent. For the ghosts, this implies that their contribution is entirely eliminated by their Pauli--Villars `partners', and one simply omits them at one-loop. This prescription is rather natural, since it preserves the BRST symmetry associated to the abelian gauge invariance of the Pauli--Villars vector fields (the mass term in (\ref{quadvectM}) being $M \Gamma_{mn}  \varepsilon^{\i\j\k} A_\i^m F^n_{\j\k}$). The leading $k_\zero$ independent integrand in  (\ref{Ncurrent})  is mass-independent for the Pauli--Villars vector field Feynman rules as well, and that is why their contribution cancel precisely the vector fields energy Coulomb divergences.

By property of the Pauli--Villars regularisation, the regularised divergences in $M$ can be computed by expending the integrant in powers of the external momenta (since $p^2 \ll M^2$ and $p_\zero^{\; 2} \ll M^2 $), and no non-local divergent contribution can be produced. The energy divergences are therefore consistently eliminated within the Pauli--Villars regularisation. One computes indeed that the divergent contribution to the regularised two-points function is
\be
\Bigl< J^\i(X_\un,p) J^\j(X_\deux, -p) \Bigl>_{\rm \scriptscriptstyle vec + PV} \sim \frac{i}{48 \pi^2 }Ê\trace \scal{ÊX_\un X_\deux }Ê \, \biggl(a M^2 -   \Scal{Ê\delta^{\i\j} \scal{Ê p^2 - p_\zero^{\; 2} } - p^\i p^\j } \ln M  \biggr)  \ ,   \label{VecCD}Ê 
\ee
similarly as for the Dirac fermion contribution. In particular, we see that the Coulomb energy divergence
\be \Bigl< J^\i(X_\un,p) J^\j(X_\deux, -p) \Bigl>_{\rm \scriptscriptstyle ghost + \lambda \bar \lambda} =  \int \frac{ d k_\zero}{4\pi}\;   \trace \scal{ÊX_\un X_\deux }  \; \frac{1}{|p|}Ê \scal{Ê\delta^{\i\j} p^2 - p^\i p^\j }\; , \ee
{\em does not} require a `catastrophic' non-local renormalisation
\be \propto   \int \frac{d^4 p}{(2\pi)^4} \frac{M}{|p|} \scal{Ê\delta^{\i\j} p^2 - p^\i p^\j }  \trace B_\i(p) B_\j(-p) \; , \ee
within the prescription. The coefficient $a$ depends on the axial / vector character of the elements $X_\un$ and $X_\deux$, and is not unambiguously determined within the prescription, because it diverges logarithmically in the UV (\ie at $\alpha \rightarrow 0$)
\bea a_{\scriptscriptstyle \rm A} &=& \int_0^{\infty} d\alpha \left( \frac{5}{3}ÊM^{-2} \alpha^{-2} \Scal{Êe^{-\alpha M^2}Ê- 1}Ê + \Scal{Ê 3 \alpha^{-1} + 2 M^2 } e^{-\alpha M^2} \right)  \CR
a_{\scriptscriptstyle \rm V} &=& \int_0^{\infty} d\alpha \left( \frac{5}{3}ÊM^{-2} \alpha^{-2} \Scal{Êe^{-\alpha M^2}Ê- 1}Ê + \frac{1}{3}Ê \alpha^{-1}  e^{-\alpha M^2} \right) \ . 
\eea
This difficulty is not associated to the Coulomb divergences, but to the general property that the Pauli--Villars regularisation does not permit to regularise divergences behaving like  $\sim M^2 \ln M$. For example, the same problem appears in the Dirac fermion contribution to the two-point function when $X_\un$ and $X_\deux$ are axial. These divergences are irrelevant anyway, since they do not affect the renormalised correlation functions at higher orders. 

\section{Conclusions}

We have exhibited in this paper the consistency of the duality invariant 
formulation of $\N=8$ supergravity in perturbation theory. The non-standard
non-manifestly Lorentz invariant Feynman rules turn out to satisfy the quantum
action principle, and diffeomorphism invariance can therefore be maintained 
through appropriate renormalisations. The theory can be gauge-fixed within 
the Batalin--Vilkovisky formalism, and although the abelian ghosts exhibit 
Coulomb energy divergences in insertions of the $E_{7(7)}$ current, these divergences are consistently removed within the Pauli--Villars regularisation.   

Furthermore, we have solved the Wess--Zumino consistency conditions for 
the anomaly associated to the non-linear $\e_{7(7)}$-current Ward idendities, 
and shown that these solutions are uniquely determined in terms of the 
corresponding solutions to the Wess--Zumino consistency condition associated 
to the linear $\su(8)$-current Ward identity. It follows that any non-linear
$E_{7(7)}$ anomaly in perturbation theory is entirely determined by the 
one-loop coefficient of the linear $\su(8)$ anomaly. In particular, we have 
explicitly computed the one-loop contribution of the vector fields to the 
anomaly, establishing the validity of the family's index prediction, and 
therefore the vanishing of the anomaly at one-loop. 

The main result of the paper is that the non-linear Slavnov--Taylor 
identities associated to the $\e_{7(7)}$ Ward identities are maintained 
at all orders in perturbation theory, if one renormalises the theory 
appropriately. Although we proved this theorem within the symmetric gauge, 
it remains in principle valid within the $SU(8)$ gauge invariant formulation \cite{deWit}. 

What are the implications of the non-linear $E_{7(7)}$ symmetry for possible 
logarithmic divergences of the theory? Regarding the definition of 
BPS supersymmetric invariants which cannot be written as full superspace 
integrals (but as integrals over subspaces of superspace classified
by their BPS degree), the linear approximation suggests that they cannot 
be duality invariant. Indeed, the BPS invariants are defined in the linearised approximation as 
partial superspace integrals of functions of the scalar superfield 
$W_{ijkl}(x,\theta) = \phi_{ijkl} + {\cal{O}}(\theta)$, but there is no 
$E_{7(7)}$ invariant function that can be built out of these scalar fields 
in any $SU(8)$ representation. It is therefore hard to see how such
supersymmetric invariants (\ie the supersymmetrisations of the Bel--Robinson square $R^4,\, \partial^4 R^4$ and $\partial^6 R^4$) could be made invariant under the full non-linear
duality symmetry. Nevertheless, this argument may not be entirely 
`watertight', as a similar argument appears to fail in higher dimensions,
where, however, the duality groups are non-exceptional. For instance,
the logarithmic divergences found in dimensions $\geq 6$ imply that 
there must exist an $SO(5,5)$ invariant 1/8 BPS counterterm in six 
dimensions, an  $SL(5,\IR)$ invariant 1/4 BPS counterterm in seven 
dimensions, and an $SL(2,\IR) \times SL(3,\IR)$ invariant 1/2 BPS 
counterterm in eight dimensions. Nevertheless, \cite{BroedelDixon,Elvang} 
exhibited that the 1/2 BPS invariant is not $E_{7(7)}$ invariant, which implies that 
the absence of logarithmic divergence at 3-loop is a consequence of the $\e_{7(7)}$ 
Ward identities. 

The duality invariance may therefore entail various non-renormalisation 
theorems, which might explain the absence of logarithmic divergences 
in maximal supergravity in five dimensions at four loops \cite{Bern2},
and in maximal supergravity in four dimensions at three, five and 
six loops.  A similar argument would lead to the conclusion that 
$\N=6$ supergravity admits its first logarithmic  divergence at 
five loops, and $\N=5$ supergravity at four loops.  However, 
establishing such non-renormalisation theorems will require further 
investigation of BPS invariants in supergravity. 

As another application, the $\e_{7(7)}$ Slavnov--Taylor identities such 
as (\ref{nlWard}) may imply special identities among the on-shell amplitudes 
in the `multi-soft-momenta limit',  generalising the ones derived in 
\cite{ArkaniHamed} at all orders in perturbation theory.

As shown by several examples  (see \eg \cite{BHS}), the study of supersymmetric counterterms 
is not enough to reach definite conclusions regarding the appearance
of certain logarithmic divergences in supersymmetric theories. The non-linear 
$\e_{7(7)}$ Ward identities may therefore imply more stringent restrictions than 
one would deduce from the existence of $E_{7(7)}$ invariant supersymmetric counterterms. 

\subsection*{Acknowledgements}
We would like to thank Ido Adam, Thibault Damour, Paul Howe, Ilarion Melnikov, 
Pierre Ramond, Hidehiko Shimada, Kelly Stelle, Pierre Vanhove and Bernard de Wit for discussions related to this work. We are grateful to the referee for suggesting several 
improvements in the original version of this article.


 \end{document}